\documentclass[twocolumn,aps,pra,superscriptaddress,nofootinbib,floatfix]{revtex4-2}
\usepackage{bm,physics,amsmath,embedfile}

\usepackage[normalem]{ulem}

\usepackage[T1]{fontenc}
\usepackage{tikz}

\usepackage[utf8]{inputenc}
\usepackage{xcolor}
\usepackage{lipsum}
\usepackage{mathtools} 
\usepackage{amssymb}   
\usepackage{appendix}
\usepackage{braket}
\usepackage{graphicx}
\usepackage{hyperref}
\usepackage{multirow}
\usepackage{hhline}
\usepackage{float}

\usepackage{breqn}

\begin{document}

\title{Comparison of renormalized interactions using one-dimensional few-body systems as a testbed
}

	\author{Fabian Brauneis}
	\affiliation{Technische Universit\"{a}t Darmstadt$,$ Department of Physics$,$ 64289 Darmstadt$,$ Germany}
	
	\author{Hans-Werner Hammer}
	\affiliation{Technische Universit\"{a}t Darmstadt$,$ Department of Physics$,$ 64289 Darmstadt$,$ Germany}
	\affiliation{ExtreMe Matter Institute EMMI and Helmholtz Forschungsakademie
  Hessen f\"ur FAIR (HFHF)$,$ GSI Helmholtzzentrum f\"ur Schwerionenforschung GmbH$,$ 64291 Darmstadt$,$ Germany}
	
	\author{Stephanie M. Reimann}
	\affiliation{Division of Mathematical Physics and NanoLund$,$ Lund University$,$ SE-221 00 Lund$,$ Sweden}
	
	\author{Artem G. Volosniev}
	\affiliation{Institute of Science and Technology Austria (ISTA)$,$ Am Campus 1$,$ 3400 Klosterneuburg$,$ Austria} 
  \affiliation{Department of Physics and Astronomy$,$ Aarhus University$ ,$ Ny Munkegade 120$,$ DK-8000 Aarhus C$,$ Denmark}

\begin{abstract}
Even though the one-dimensional contact interaction requires no regularization, renormalization methods have been shown to improve the convergence of numerical {\it ab initio} calculations considerably. In this work, we compare and contrast these methods: `the running coupling constant' where the two-body ground state energy is used as a renormalization condition, and two effective interaction approaches that include information about the ground as well as excited states.  
In particular, we calculate the energies and densities of few-fermion systems in a harmonic oscillator with the configuration interaction method, and compare the results based upon renormalized and bare interactions.
We find that the use of the running coupling constant instead of the bare interaction improves convergence significantly. A comparison with an effective interaction, which is designed to reproduce the relative part of the energy spectrum of two particles, showed a similar improvement.
The effective interaction provides an additional improvement if the center-of-mass excitations are included in the construction. Finally, we discuss the transformation of observables alongside the renormalization of the potential, and demonstrate that this might be an essential ingredient for accurate numerical calculations.
\end{abstract}

\maketitle

\section{Introduction}
\label{sec:Introduction}

Experimental advances in cold-atom physics allowed one to create and study (quasi) one-dimensional (1D) systems of a few (up to 10) particles in a highly controllable setting~\cite{serwane2011deterministic, Wenz2013, Zurn2013}. 
This rapid progress fueled studies of the corresponding Schr{\"o}dinger equation~\cite{Sowinski2019,mistakidis2022cold}. Thanks to the small number of particles, this equation can be solved accurately using numerical methods, such as the multi-layer multi-configuration time-dependent Hartree
method for atomic mixtures~\cite{kronke2013non, cao2017unified,Lode2020}, flow equation approaches~\cite{brauneis2021impurities, brauneis2022artificial, brauneis2023emergence}, coupled cluster methods~\cite{Cederbaum2006,grining2015many,Grining2015PRA}, or exact diagonalization methods, also known as the configuration interaction method (CI)~\cite{Harshman2012,Gharashi2013, DAmico_2014, Pecak2016, Wlodzinsky2022, Rammelmueller2023, rontani2017renormalization, Bjerlin2016, Christensson2009} and its extension, the importance truncated CI~\cite{Roth2009, chergui2023superfluid}.
These powerful numerical techniques, however, usually face problems for strong interactions (see, e.g.,~\cite{DAmico_2014,grining2015many}).

This difficulty is well-known in the nuclear physics community. There, the renormalization of nuclear potentials using so-called ``Lee-Suzuki'' transformations~\cite{suzuki1980} or renormalization group methods~\cite{bogner2001, Bogner2007} has enabled rapid progress in nuclear structure calculations over the last twenty years~\cite{BOGNER2010}. Similar ``effective interactions'' were used lately in the cold-atom community~\cite{rotureau2013, Lindgren2014, rammelmuller2023magnetic,AnhTai2023}, see also Refs.~\cite{Ernst2011, KOSCIK2018, Jeszenszki2018, Rojo-Francas2022} for other related approaches.

Zero-range interactions require no regularization in one-dimensional geometries (unlike in 2D and 3D), however, Refs.~\cite{rotureau2013, Lindgren2014, rammelmuller2023magnetic,AnhTai2023,Ernst2011, KOSCIK2018, Jeszenszki2018, Rojo-Francas2022} demonstrate that renormalized interactions significantly boost the convergence of numerical simulations, enabling calculations with larger particle numbers and stronger interactions. However, to the best of our knowledge, these potentials were only benchmarked against the bare interaction. A direct comparison between different renormalization methods is still lacking and this paper aims to fill this gap. Additionally, we discuss a unitary transformation of observables together with the potential. This procedure is well-known in nuclear physics, but not in one-dimensional few-body problems. We will showcase that such a transformation can be necessary to improve the convergence of observables other than the energy.

This work provides a brief introduction to renormalized interactions, and a detailed discussion of corresponding convergence patterns. This should help the reader to decide on the most suitable method for the problem at hand. The simplest method (the running coupling constant) fixes the two-body ground state energy in a truncated Hilbert space and can be a go-to method for qualitative calculations. At the same time, we demonstrate that calculations of the energy as well as other observables are more accurate with an effective interaction that reproduces the {\it full} low-energy spectrum of the two-body problem. 
This interaction has not been explored before in few-body models of cold atoms (to the best of our knowledge). It outperforms effective interactions introduced in previous studies because it captures both the center-of-mass and relative degrees of freedom.

Our testbed for illustrating differences between different renormalization methods is a system of harmonically trapped fermions interacting via short range interactions. This system is well-studied (see, e.g.,~\cite{Harshman2012,Sowinski2013,Gharashi2013, DAmico_2014,grining2015many,Koutentakis2019}) and features slow convergence of observables~\cite{grining2015many}, making it an ideal testbed for renormalized interactions and an optimal candidate for comparative studies.
Furthermore, the corresponding two-body problem is analytically solvable~\cite{avakian1987spectroscopy, busch1998}, which will facilitate construction of effective interactions in our calculations. 

We calculate properties of this system using the CI method. It is an ``exact'' method in the sense that no approximation besides the truncated Hilbert space needs to be performed.  Even though CI calculations for more than just a handful of particles are nearly impossible, the main characteristics of renormalized interactions become apparent even in small systems. 
The CI method will allow us to study excited states and dynamical observables like the transition rate between different states; something out of reach for most other methods. 

Although the main focus of our study is on one-dimensional systems, our results can be extended to 2D systems, which have also been been realized experimentally~\cite{bayha2020observing, Holten2021Pauli, holten2022observation}, see
the outlook for a brief comparison of renormalization schemes used in theoretical calculations~\cite{Bjerlin2016, Christensson2009, rontani2017renormalization,Bengtsson2020}. 3D systems have been discussed mostly in the context of nuclear physics, see Ref.~\cite{BOGNER2010} and references therein, but some works also studied cold-atom systems~\cite{rotureau2013}.

We start in Sec.~\ref{sec:System} by introducing the system of our study and by briefly explaining the CI method. Then, in Sec.~\ref{sec:Interaction} we review the basic ideas behind running coupling in momentum space. After that, we introduce the different renormalization schemes used in this work: The running coupling constant where the two-body ground state energy is used as a renormalization condition, an effective interaction that reproduces the relative energy spectrum of two particles, and an effective interaction approach, which includes information about all (relative and center-of-mass) excitations. Using these interactions, we calculate energies and densities, and compare them in Sec.~\ref{sec:Results}. Additionally, we discuss the transformation of observables alongside the renormalization of the interaction for one-body observables (the one-body part of the Hamiltonian: kinetic energy and harmonic confiment) and for a two-body observable (the transition matrix elements for a periodic modulation of the interaction). Finally, in Sec.~\ref{sec:Conclusion} we summarize our findings and provide a short outlook into future research topics. In particular, we present preliminary results for 2D systems.

\section{Formalism}
\label{sec:System}

\subsection{System}
\label{subsec:System}

\begin{figure}
    \centering
    \includegraphics{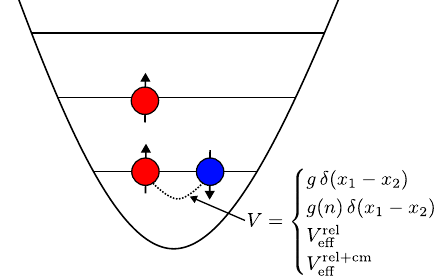} 
    \caption{We consider two-component harmonically-trapped fermions with four different two-body interactions: bare contact interaction, $g\delta(x_1-x_2)$; contact interaction with a running coupling constant, $g(n)\delta(x_1-x_2)$, where $n$ is the one-body cutoff in our calculations; effective interaction defined in the relative coordinates, $V_{\mathrm{eff}}^{\mathrm{rel}}$; effective interaction that includes relative and center-of-mass excitations, $V_{\mathrm{eff}}^{\mathrm{rel+cm}}$ (see the text for more details). The goal of the paper is to compare and contrast convergence of numerical calculations based upon these different interactions.}
    \label{fig:sketch}
\end{figure}

{\it Hamiltonian.} We consider a one-dimensional system of spin-$1/2$ fermions in a harmonic trap (see Fig.~\ref{fig:sketch}). The Hamiltonian of the system reads:
\begin{equation}
    H=T+W+\sum\limits_{i< j} V(x_i, x_j),
    \label{eq:Hamiltonian}
\end{equation}
where $x_i$ and $x_j$ are the coordinates of particles $i$ and $j$  in the laboratory frame, and $V(x_i,x_j)$ is their interaction potential.
The operators of the kinetic energy $T$ and the external trapping potential $W$
are given by
\begin{equation}
\label{eq:kinetic_trapped}
T=-\sum\limits_{i=1}^N\frac{\hbar^2}{2m}\frac{\partial^2}{\partial x_i^2}\; \qquad W=\sum\limits_{i=1}^N\frac{m\omega^2}{2}x_i^2,
\end{equation}
where $N=N_\uparrow+N_\downarrow$ is the number of fermions. We assume that the spin of a particle is conserved so that $N_\uparrow$ and  $N_\downarrow$ are fixed. As we are interested in low-energy scattering of atoms, we assume that $V$ is a  zero-range (contact) potential~\cite{BRAATEN2006259, Bloch2008}. 
This implies that only fermions of different spins can interact as the Pauli exclusion principle forbids two fermions with equal spin to occupy the same position.  

As was mentioned above, the Hamiltonian in Eq.~(\ref{eq:Hamiltonian}) is well-studied theoretically~\cite{Sowinski2019,Minguzzi2022,mistakidis2022cold}, in particular, due to the possibility to realize such systems experimentally~\cite{serwane2011deterministic, Zurn2013, Wenz2013}. The plethora of existing results makes the system ideal for comparing and contrasting different renormalization schemes in one dimension, which is the overall goal of this study. 
In what follows we shall use the system of units in which \mbox{$\hbar=m=\omega=1$}. 

{\it Potential}. We employ four different potentials as sketched in Fig.~\ref{fig:sketch}. The `bare potential' $V(x_1, x_2)=g\delta(x_1-x_2)$ serves as a base line for presenting our results. Further, we use three `renormalized' interactions: (i) A potential with a running coupling constant $V(x_1, x_2)=g(n)\delta(x_1-x_2)$, where $n$ is the cutoff on the basis states included in our calculations (see Sec.~\ref{sec:Runningcoupling}); (ii) an effective interaction, $V_{\mathrm{eff}}^{\mathrm{rel}}$, defined in relative coordinates using the corresponding two-body energy spectrum (see Sec.~\ref{sec:Veff_rel}); (iii) an effective interaction, $V_{\mathrm{eff}}^{\mathrm{rel+cm}}$, based upon the complete two-body energy spectrum (see Sec.~\ref{sec:Veff_relcm}).

\subsection{Numerical method: Configuration interaction method}
We employ numerical calculations based upon the configuration interaction (CI) method~\cite{CremonPhDThesis, BjerlinPhdThesis}, which diagonalizes the Hamiltonian in the formalism of second quantization
\begin{equation}
\label{eq:hamiltonian_second}
    H=\sum\limits_{i, j}A_{ij}a_i^\dagger a_j+\sum\limits_{i,j,k,l}V_{ijkl}a_i^\dagger a_j^\dagger a_l a_k.
\end{equation}
Here, $a_i^\dagger$ ($a_i$) are fermionic creation (annihilation) operators. The index $i$ includes information about the oscillator state and the spin quantum number of the fermion. $A_{ij}$ and $V_{ijkl}$ are one- and two-body matrix elements written in the basis of the non-interacting system, i.e., the eigenfunctions of the harmonic oscillator:
\begin{equation}
    \Phi_{l}(z)=\frac{1}{\sqrt{2^l l!}}\left(\frac{1}{\pi}\right)^{\frac{1}{4}}e^{-\frac{1}{2}z^2}H_{l}(z),
\end{equation}
where $H_{l}$ is the $l$th Hermite polynomial. In the CI method, we first construct antisymmetric $N-$particle basis states and then use them to build the Hamiltonian matrix, which is subsequently diagonalized using the Arnoldi/Lanczos algorithm~\cite{golubmatrix}.

To truncate the Hilbert space we work with the $n$ lowest one-body basis states\footnote{Note that in one-dimensional systems the number of one-body basis states, $n$, is equivalent to a one-body energy cutoff of $n+1/2$.}. It is worth noting that there exist more sophisticated truncation schemes, which further restrict the many-body basis states. For example, one can fix a maximal many-body energy as, e.g., in Ref.~\cite{RojoFrancs2020,Bjerlin2016}, use selection rules imposed by symmetries~\cite{Harshman2012} or employ importance truncation as in Refs.~\cite{Roth2009, chergui2023superfluid, Bengtsson2020}. As our focus is  on studying different renormalization schemes, it is not necessary in this context to optimize the truncation scheme, and we use the most general one.

The main problem of solving a system with $N_\uparrow$ spin-up and $N_\downarrow$ spin-down fermions with the CI method is that the size of the Hilbert space grows as
\begin{equation}
    \mathrm{dim}\,\mathcal{H}=\binom{n}{N_\uparrow}\binom{n}{N_\downarrow},
\end{equation}
where $\binom{n}{k}=n!/(k!(n-k)!)$ is the binomial coefficient.
Rapid increase of $\mathrm{dim}\,\mathcal{H}$ means that numerical solution of a few-body problem is only possible if the size of the one-body basis states, $n$, is sufficiently small. 

\section{Renormalized interaction}
\label{sec:Interaction}

In general, strong interactions imply that `many' one-body basis states are needed to achieve accurate results, see, e.g, Sec.~\ref{sec:Results}. This can make the investigation of strongly interacting systems hard if not impossible. However, the numerical value of `many' can be drastically decreased by properly renormalizing the interactions enabling the study of strongly interacting few-body systems.
The basic idea of renormalized interactions is to fix physical parameters (such as the scattering length or two-body binding energy) in a truncated Hilbert space. One possible interpretation of this procedure is as follows. By truncating a Hilbert space we effectively change properties of the interaction, and thus we need to renormalize the interaction to describe actual physical quantities. 
This section introduces the commonly used renormalization schemes in one-dimensional systems, see Refs.~\cite{Lindgren2014, Rammelmueller2023, Ernst2011, Jeszenszki2018, Rojo-Francas2022}.

Three renormalizations of the contact interaction, i)-iii) are mentioned above in Sec.~\ref{subsec:System} (see also Fig.~\ref{fig:sketch}).
We will benchmark the convergence of the different regularization methods against the bare contact interaction and then compare them with each other. 

Before we proceed, let us briefly review the relevant literature. The running coupling constant approach [i)] is closely related to
Ref.~\cite{Rojo-Francas2022} where an analytical expression for the running coupling constant was derived by truncating the exact solution of two interacting particles in a harmonic trap. In practice, however, we explicitly fit the two-body ground state energy (see Sec.~\ref{sec:Runningcoupling}), which allows us to easily extend our approach also to different traps or dimensions, something not possible following Ref.~\cite{Rojo-Francas2022} as it is based upon the availability of an exact solution.
Another study that introduced a running coupling constant used the energy of strongly-repulsive bosons for the renormalization~\cite{Ernst2011}. This approach works well in one-dimensional systems where the energy of infinitely-strongly repulsive bosons is that of non-interacting fermions~\cite{Girardeau1960,Minguzzi2022}.

The effective potentials  $V_{\mathrm{eff}}^{\mathrm{rel+cm}}$ and $V_{\mathrm{eff}}^{\mathrm{rel}}$ 
are based on the same idea, but the former takes into account the full two-body spectrum. The application of this idea to cold atoms is discussed in Refs.~\cite{rotureau2013,Lindgren2014}; its connection to renormalization group frameworks can be understood from Ref.~\cite{bogner2001}.
However, their performance for one-dimensional systems has not been benchmarked against each other; $V_{\mathrm{eff}}^{\mathrm{rel+cm}}$ has not been explored at all in this context, to the best of our knowledge. 
Although setting up these effective potentials involves a few more steps than establishing a running coupling constant, their performance might be superior to that of $g(n)\delta$ (as demonstrated in Sec.~\ref{sec:Results}).

\subsection{Running coupling constant}
\label{sec:Runningcoupling}

To set the stage, we first illustrate the renormalization of a contact interaction in an unconfined system adapting the approach presented for 2D systems in Ref.~\cite{nyeo2000regularization}. This allows us to introduce a running coupling constant in a simple way using momentum space. Afterwards, we connect this procedure to our CI-calculations in a trapped system.

{\it Unconfined system.} To define a two-body interaction potential, it is sufficient to solve the two-body problem. Therefore, we consider the two-body Schr{\"o}dinger equation in relative coordinates $x=(x_1-x_2)/\sqrt{2}$, 
\begin{equation}
    \left(-\frac{1}{2}\frac{\partial^2}{\partial x^2}+\frac{g}{\sqrt{2}}\delta(x)\right)\Psi(x)=E_{1+1}^{\mathrm{unconf.}}\Psi(x),
    \label{eq:homogeneous_two_body}
\end{equation}
where $E_{1+1}^{\mathrm{unconf.}}$ is the two-body energy of the unconfined system. For now, we discuss only $g<0$ because we want to use the energy of a bound system with $E_{1+1}^{\mathrm{unconf.}}<0$ as the guiding condition for renormalization. Note that in 1D, there is always a bound state if $g<0$~\cite{griffiths2018introduction}. 

By solving Eq.~(\ref{eq:homogeneous_two_body}) in momentum space, we arrive at the following equation for the energy (see Sec. IV the supplementary material~\cite{SuppMat} for details):
\begin{equation}
\label{eq:rcoup}
    -\frac{1}{g}=\frac{1}{\sqrt{2}\pi}\int_{-\infty}^\infty\frac{dk}{k^2-2E_{1+1}^{\mathrm{unconf.}}}.
\end{equation}
The integral on the right-hand-side of this equation converges\footnote{This is a property of one-dimensional geometries. In higher spatial dimensions the corresponding integrals diverge and one always needs to regularize the zero-range potential. See for example Refs.~\cite{BRAATEN2006259, nyeo2000regularization,rontani2017renormalization} for possible regularization schemes in two and three spatial dimensions.} for all $E_{1+1}^{\mathrm{unconf.}}<0$. Introducing a momentum space cutoff $\lambda$
restricts the Hilbert space (as mandatory in numerical calculations).  

We fix the two-body energy $E_{1+1}^{\mathrm{unconf.}}$ for each  $\lambda$ by introducing a running coupling constant $g_{\mathrm{unconf.}}(\lambda)$ defined by Eq.~\eqref{eq:rcoup} with
the $k$-integration restricted to the interval $[-\lambda,\lambda]$:
\begin{equation}
\label{eq:gRG}
    -\frac{1}{g_{\mathrm{unconf.}}(\lambda)}=\frac{1}{\pi\sqrt{|E_{1+1}^{\mathrm{unconf.}}|}}\mathrm{arctan}\left[\frac{\lambda}{\sqrt{|2E_{1+1}^{\mathrm{unconf.}}|}}\right].
\end{equation}
This equation renormalizes the interaction in a finite Hilbert space, in the sense, that the ground state energy $E_{1+1}^{\mathrm{unconf.}}$ becomes independent of the size of the Hilbert space. Instead of $g$, the two-body energy $E_{1+1}^{\mathrm{unconf.}}$ now determines the physical properties of the system. 

In the limit of large $\lambda$, we 
derive from Eq.~(\ref{eq:gRG})
\begin{equation}
-\frac{1}{g_{\mathrm{unconf.}}(\lambda)}\simeq \frac{1}{2\sqrt{|E_{1+1}^{\mathrm{unconf.}}|}}-\frac{\sqrt{2}}{\pi\lambda}.
\label{eq:large_lambda}
\end{equation}
With this we retrieve the well-known expression $g_{\mathrm{unconf.}}(\lambda\to\infty)=-2\sqrt{|E_{1+1}^{\mathrm{unconf.}}|}$ for two particles interacting via a contact interaction potential~\cite{griffiths2018introduction}. Note also that the second term in Eq.~(\ref{eq:large_lambda}) is independent of the physical properties of the system just like in three spatial dimensions~\cite{BRAATEN2006259}. 
For $g>0$, a similar renormalization can be performed by fixing the phase shift of two scattered atoms for a given momentum space cutoff.   

Finally, we note that Eq.~(\ref{eq:large_lambda}) allows one also to understand the convergence of the energy $E_{1+1}^{\mathrm{unconf.}}(\lambda)$ as a function of the Hilbert space size in calculations based upon bare contact potentials, $g \delta(x_1-x_2)$. In this case the second term on the right-hand-side of Eq.~(\ref{eq:large_lambda}) can be treated perturbatively.
This leads to (see App.~\ref{App:ConvergenceEnergy} for details)
\begin{equation}
E_{1+1}^{\mathrm{unconf.}}(\lambda)\simeq E_{1+1}^{\mathrm{unconf.}}(\lambda\to\infty)+\frac{\sqrt{2} C}{\pi\lambda},
\label{eq:E11_convergence}
\end{equation}
where $C=g^2\braket{\Psi|\delta(x)|\Psi}$ is the contact parameter~\cite{Barth2011}. The functional behavior of Eq.~(\ref{eq:E11_convergence}) is in agreement with Refs.~\cite{volosniev2017flow,Jeszenszki2018}. We will see that, in general, this behavior depends on the trapping potential.  In an unconfined system the cutoff $\lambda$ is a continuous parameter while for the trapped system the cutoff $n$ is discrete. For large quantum numbers\footnote{Large means that the energy of an excited state is much larger than the excitation energy to a neighboring state of a higher energy. For the harmonic oscillator, which we discuss below, this means that $n\gg 1$. A different line of argument with the same conclusion can be based on the asymptotic approximation of Hermite polynomials, see, e.g., Ref.~\cite{BOYD1984382}.}, we connect $\lambda$ and $n$ below using the semi-classical approximation.

\begin{figure}
    \centering
    \includegraphics[width=1\linewidth]{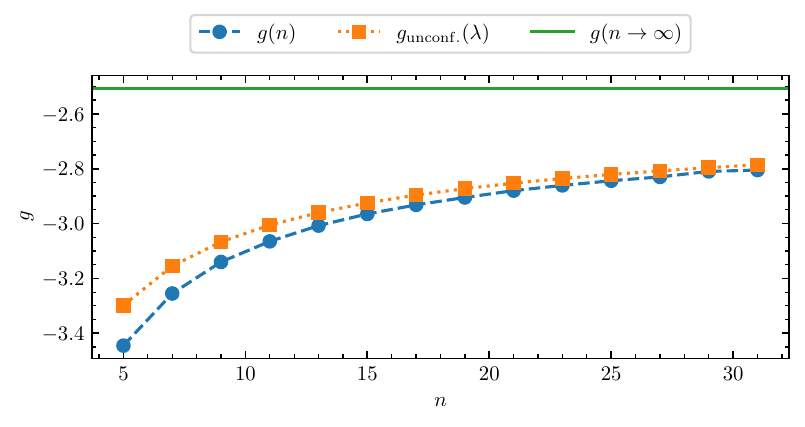}
    \caption{Running coupling constant for $g(n\to\infty)=-2.5067$ (corresponding to $E_{1+1}=-1$, the two-body energy in the trapped system). Dots show $g(n)$ calculated by explicitly fixing the two-body ground state energy of two fermions in a harmonic trap for each single-body cutoff $n$. Squares show $g_{\mathrm{unconf}.}$ calculated by integrating out high momentum contributions of two fermions in free space, Eq.~\eqref{eq:gRG}. The parameter $\lambda$ is related to the single-body cutoff $n$ via Eq.~\eqref{eq:ConnectionLambdan}. The solid line represents the value of the coupling constant for an infinitely large Hilbert space, $g(n\to\infty)=-2.5067$.
    }
    \label{fig:running}
\end{figure}

{\it 1D Harmonic oscillator.} By analogy to the discussion above, we introduce here a running coupling constant for a trapped system via
\begin{equation}
    V=g(n)\delta(x_1-x_2),
\end{equation}
and use the two-body ground state energy for renormalization, i.e., we tune $g(n)$ for each cutoff such that the two-body ground state energy is fixed\footnote{Note that unlike the unconfined system, both repulsive and attractive interactions can be renormalized using the two-body ground-state energy because the state is always bound due to the harmonic confinement.}. 

Let us now relate the running coupling constant $g(n)$ to $g_{\mathrm{unconf.}}(\lambda)$ presented in  Eq.~\eqref{eq:gRG}. First, we connect the momentum space cutoff $\lambda$ with $n$. To this end, we use the Wilson–Sommerfeld quantization condition $
    \oint_{H(p, x)=E} p\,dx=2\pi n$, which for a one-dimensional harmonic oscillator leads to
\begin{equation}
\label{eq:ConnectionLambdan}
    \lambda=2\sqrt{n}.
\end{equation}
For a derivation of this equation and an extension to general polynomial potentials, see Sec. V in the supplementary material~\cite{SuppMat}. We see that convergence is relatively slow (because $1/\lambda = 1/(2\sqrt{n})$) for a harmonic potential (see also Ref.~\cite{grining2015many}), making the use of renormalized interactions for trapped systems more important than in an unconfined case.

To compare $g(n)$ with $g_{\mathrm{unconf.}}(2\sqrt{n})$, we fix the value of $g(n\to\infty)$ and insert the expression $E_{1+1}^{\mathrm{unconf.}}=-g(n\to\infty)^2/4$ in Eq.~\eqref{eq:gRG} to calculate $g_{\mathrm{unconf.}}$.  Then, we compute the two-body ground state energy of the trapped system (see Eq.~\eqref{eq:ExactEnergy}) and tune $g(n)$ such that this energy is reproduced in a truncated Hilbert space for each cutoff.
In Fig.~\ref{fig:running} we show the calculated running coupling constants $g(n)$ and $g_{\mathrm{unconf.}}(2\sqrt{n})$  [$g(n\to\infty)=-2.5067$ corresponding to $E_{1+1}=-1$, the two-body energy in the trapped system]. The behavior of $g(n)$ resembles that of $g_{\mathrm{unconf.}}$ for all values of $n$ qualitatively. For large values of $n$ they agree even quantitatively (less than 1\% disagreement for $n\geq21$). The disagreement for small values of $n$ can be explained by the validity of the semi-classical approximation used to derive Eq.~(\ref{eq:ConnectionLambdan}), as it only holds in the limit of large quantum numbers.

\subsection{Effective interactions}
\label{subsec:effective_interactions}

Another way to renormalize the potential is to incorporate not only the information about the ground-state energy, but also about the exact two-body solution into an effective two-body interaction. This approach is related to the ``Lee-Suzuki'' transformation in nuclear physics~\cite{suzuki1980, BOGNER2010} and its adaptations to cold-atom problems~\cite{Christensson2009, rotureau2013, Lindgren2014, Rammelmueller2023}.
The idea of this regularization scheme is to build a two-body potential in the truncated Hilbert space that reproduces the known two-body eigenenergies exactly.
To make the paper self-contained, we outline this idea below (this follows closely Refs.~\cite{Rammelmueller2023, rotureau2013, Stetcu2005}). The discussion shall also illustrate that
although the running coupling constant approach introduced in Sec.~\ref{sec:Runningcoupling} is easier to set up, the approaches introduced in this subsection provide additional benefits, e.g., they allow us to transform observables as well.

{\it General framework.} The Hamiltonian of the two-body system can always be written as:
\begin{equation}
    H^{(2)}=U^\dagger \widetilde{H}^{(2)} U
\end{equation}
where $\widetilde{H}^{(2)}=E^{(2)}$ is a diagonal matrix with the eigenenergies on the diagonal; $U$ is a unitary transformation matrix based upon the eigenvectors:
   $ U_{ij}=\braket{i|\Psi_j}$
with $\{{\ket{i}}\}$ being a two-body basis and $\{{\ket{\Psi_j}}\}$ eigenkets, $H^{(2)}\ket{\Psi_j}=E_j\ket{\Psi_j}$. To truncate the Hilbert space, we introduce the projection operator $P_{\mathcal{M}}$ which projects onto a finite Hilbert space, $\mathcal{M}$, in which we perform our calculations, $U_{\mathcal{M}\mathcal{M}}=P_{\mathcal{M}} U P_{\mathcal{M}}$, $E_{\mathcal{M}\mathcal{M}}=P_{\mathcal{M}} E^{(2)} P_{\mathcal{M}}$. 
In a finite Hilbert space, $U_{\mathcal{M}\mathcal{M}}$ is not unitary. The corresponding unitary matrix can be built as $Q_{\mathcal{M}\mathcal{M}}=U_{\mathcal{M}\mathcal{M}}\left(\sqrt{U_{\mathcal{M}\mathcal{M}}^\dagger U_{\mathcal{M}\mathcal{M}}}\right)^{-1}$.
With this, we can introduce an effective Hamiltonian in the truncated Hilbert space
\begin{equation}
    H^{\mathrm{eff}}_{\mathcal{M}\mathcal{M}}=Q_{\mathcal{M}\mathcal{M}}^\dagger E_{\mathcal{M}\mathcal{M}}Q_{\mathcal{M}\mathcal{M}}
\end{equation}
via a unitary transformation\footnote{Note that the normalization of the transformation operator, $\left(\sqrt{U_{\mathcal{M}\mathcal{M}}^\dagger U_{\mathcal{M}\mathcal{M}}}\right)^{-1}$, converges towards unity with increasing Hilbert space.}. 
Now, we extract the effective interaction as follows:
\begin{equation}
\label{eq:Veff}
    V^{\mathrm{eff}}_{\mathcal{M}\mathcal{M}}=H^{\mathrm{eff}}_{\mathcal{M}\mathcal{M}}-P_{\mathcal{M}} T P_{\mathcal{M}}
\end{equation}
with $T$ the kinetic energy operator.
This renormalization scheme could be interpreted as a unitary transformation of the Hamiltonian~\cite{BOGNER2010, Stetcu2005}, see also Sec. III in the supplementary material~\cite{SuppMat}. Therefore, operators should be transformed alongside the Hamiltonian transformation. 

Let us first consider a one-body operator:
\begin{equation}
    O^{(1)}=\sum\limits_{\alpha=1}^N O_\alpha,
\end{equation}
where $O_\alpha$ is an operator written in first quantization. In second quantization, $O^{(1)}=\sum_{ij} a_{ij}a_i^\dagger a_j$. For such a one-body operator many-body operators are induced via the unitary transformation. Because we built the effective interaction from the two-body solution, we only include the induced two-body operators and truncate higher order ones (see also Ref.~\cite{Stetcu2005}):
\begin{equation}
    O^{(1, 2)}_{\mathrm{eff}}=\sum\limits_{\alpha=1}^N O_\alpha+Q_{\mathcal{M}\mathcal{M}}^\dagger\widetilde{O}^{(2)}_{\mathcal{M}\mathcal{M}}Q_{\mathcal{M}\mathcal{M}}-O^{(2)}_{\mathcal{M}\mathcal{M}},
\label{eq:OneBodyOperatorTransformed}
\end{equation}
where
\begin{equation}
    O^{(2)}_{\mathcal{M}\mathcal{M}}=P_{\mathcal{M}}\sum\limits_{\alpha> \beta=1}^N \left(O_{\alpha}+O_{\beta}\right)P_{\mathcal{M}}.
    \label{eq:induced_two_body}
\end{equation}
Note that $\widetilde{O}^{(2)}_{\mathcal{M}\mathcal{M}}$ is represented in eigenstates of the two-body solution $\ket{\Psi_i}$ while $O^{(2)}_{\mathcal{M}\mathcal{M}}$ is written in the two-body basis functions $\ket{i}$ (cf. Sec. III B. in the supplementary material~\cite{SuppMat}).

For a two-body operator
\begin{equation}
    O^{(2)}=\sum\limits_{\alpha, \beta=1}^N O_{\alpha\beta},
\end{equation}
with $O_{\alpha\beta}$ an operator in first quantization\footnote{In second quantization such an operator can be written as $O^{(2)}=\sum_{ijkl} b_{ijkl}a_i^\dagger a_j^\dagger a_k a_l$.}   
the transformation reads:
\begin{equation}
    O^{(2)}_{\mathrm{eff}}=Q_{\mathcal{M}\mathcal{M}}^\dagger \widetilde{O}^{(2)}_{\mathcal{M}\mathcal{M}}Q_{\mathcal{M}\mathcal{M}},
    \label{eq:TwobodyOperatorTransformed}
\end{equation}
where the operator $\widetilde{O}^{(2)}_{\mathcal{M}\mathcal{M}}$ is represented in eigenstates of the two-body solution $\ket{\Psi_i}$.

Since we know the exact two-body solution of our system in terms of relative and center-of-mass coordinates (see below) the transformation of observables is straightforward if an observable can be factorized in these coordinates. However, if operators do not factorize, the transformation requires more involved calculations, for details see Secs. III C. and D. in the supplementary material~\cite{SuppMat}. 

Note that other renormalization methods exist, which also allow for a straightforward transformation of observables together with the potential like the similarity renormalization group method (SRG) used in nuclear physics~\cite{Bogner2007}. For a comparison of the SRG with methods similar to the running coupling constant approach in the context of nuclear physics see, e.g., Ref.~\cite{BOGNER2010}.

{\it Two-body problem}. Before we discuss effective potentials based upon the general framework above, let us present a solution to a two-body problem in a harmonic oscillator, which will be used to construct $U$ and $E^{(2)}$.
  The two-body solution of Eq.~(\ref{eq:Hamiltonian}) reads~\cite{avakian1987spectroscopy, busch1998}
\begin{equation}
\label{eq:ExactSolution}
    \Psi_i(x, X)=\Phi_{n_i}(X)\psi_{\nu_i}(x)
\end{equation}
with $x=(x_1-x_2)/\sqrt{2}$ and $X=(x_1+x_2)/\sqrt{2}$ being the relative and center-of-mass coordinates. Only the relative part of the wavefunction is influenced by the interaction and therefore $\Phi_{n_i}(X)$ is simply the harmonic oscillator eigenfunction; the corresponding energy is $E_{n_i}=n_i+1/2$.
For the eigenfunction in relative coordinates, odd states are not influenced by the interaction; they are the ones of the harmonic oscillator with the same energy. For even states one derives:
\begin{equation}
\label{eq:ExactWF}
    \psi_{\nu_i}(x)=\mathcal{N}_{\nu_i}e^{-x^2}U(-\nu_i, 1/2, x^2),
\end{equation}
where $U$ is the Tricomi function \cite{abramowitz1948handbook} and $\mathcal{N}_{\nu_i}=\sqrt{\frac{\Gamma(-\nu_i+1/2)\Gamma(-\nu_i)}{\pi (\psi(-\nu_i+1/2)-\psi(-\nu_i))}}$ with $\Gamma$ the gamma function and $\psi$ the digamma function. The eigenenergy of this state is given by $\epsilon_{\nu_i}=2\nu_i+1/2$ where $\nu_i$ solves the equation
\begin{align}
\label{eq:ExactEnergy}
    2\sqrt{2}\Gamma(-\nu_i+1/2)&=-g\Gamma(-\nu_i).
\end{align}

\subsubsection{Effective interaction from the energy spectrum in relative coordinates: \texorpdfstring{$V_{\mathrm{eff}}^{\mathrm{rel}}$}{V_eff^rel}}
\label{sec:Veff_rel}
We use the general framework and the two-body solution presented above to build the effective interaction in two different ways. First, we construct the effective interaction in relative coordinates (i.e., using $ \psi_{\nu_i}$ from above) and transform this interaction matrix to the laboratory frame (the frame in which Eq.~\eqref{eq:Hamiltonian} is defined) where our calculations are performed. This is possible as only the relative part of the wavefunction is affected by the interaction. This approach is the one used in Refs.~\cite{Rammelmueller2023, rotureau2013, Lindgren2014}. 
For the relevant equations we refer to Sec. I in the supplementary material~\cite{SuppMat} and Ref.~\cite{Rammelmueller2023}.

Note that $V_{\mathrm{eff}}^{\mathrm{rel}}$ does not lead immediately (i.e., for any cutoff parameter) to a correct full two-body spectrum in a truncated Hilbert space. Indeed, the Hamiltonian matrix built upon the truncated basis allows for coupling between relative and center-of-mass coordinates. As a result, the two-body spectrum is not exact when we construct the effective interaction in relative coordinates and then transform it to the laboratory frame. It is worth noting that for the low-energy part of the spectrum, the coupling between the relative and center-of-mass coordinates is in general weak; it decreases as the cutoff parameter increases.

\subsubsection{Effective interaction from the full two-body energy spectrum: \texorpdfstring{$V_{\mathrm{eff}}^{\mathrm{rel+cm}}$}{V_eff^rel+cm}} 
\label{sec:Veff_relcm}

We also build the effective interaction using the full two-body solution presented in Eq.~(\ref{eq:ExactSolution}), 
see the relevant equations in Sec. II of the supplementary material~\cite{SuppMat}.
This interaction explicitly accounts for the coupling between the relative and center-of-mass coordinates in a truncated Hilbert space, and fixes the spectrum of the $1+1$ system to the exact values for all one-body cutoffs. Note that $V_{\mathrm{eff}}^{\mathrm{rel+cm}}$ is harder to construct in comparison to $V_{\mathrm{eff}}^{\mathrm{rel}}$, 
in particular, for $V_{\mathrm{eff}}^{\mathrm{rel+cm}}$ the overlap between the laboratory frame and the exact two-body solution has to be calculated. 
However, its performance is in general better compared against our other renormalization methods as we illustrate in the next section. Furthermore, $V_{\mathrm{eff}}^{\mathrm{rel+cm}}$ is the only option for non-harmonic traps for which a factorization into relative and center-of-mass motions is not possible.

Note that effective interactions discussed above can be constructed in a similar manner for higher-dimensional systems and other (non-harmonic) traps~\footnote{When working with non-harmonic traps, one can employ harmonic oscillator eigenfunctions as one-body basis. In this approach, the effective interactions become independent of the trap and are constructed exactly as described in this paper. The information about the trap is then hidden in the one-body part of the Hamiltonian $A_{ij}$, which will be off-diagonal.}. The prerequisite is accurate knowledge of a solution to a two-body problem, which can be obtained either analytically or numerically. A numerical solution is typically the only option for non-contact interactions and non-harmonic traps for which an analytical solution is not attainable.

\section{Results}
\label{sec:Results}

In this section we compare the four different interactions - bare contact interaction $g\delta$, running coupling constant $g(n)\delta$ (see Sec.~\ref{sec:Runningcoupling}), $V_{\mathrm{eff}}^{\mathrm{rel}}$ (see Sec.~\ref{sec:Veff_rel}) and $V_{\mathrm{eff}}^{\mathrm{rel+cm}}$ (see Sec.~\ref{sec:Veff_relcm}) in terms of their convergence\footnote{Note that for higher dimensional systems the bare coupling constant, $g$, does not lead to a well-defined problem, making a benchmark against it useless.}.
We discuss the convergence properties for the energy, density, kinetic energy, $\langle x^2\rangle$, and transition matrix elements. For the latter three observables we will also demonstrate the effect of the unitary transformation presented in Eqs.~(\ref{eq:OneBodyOperatorTransformed}) and~(\ref{eq:TwobodyOperatorTransformed}). 

Note that although we use the $1+1$ system to renormalize our interactions, not all two-body observables, for example the density, are fixed by the renormalization scheme, i.e., a large number of basis states might still be needed to obtain accurate results. Therefore, already the simplest system with one spin-up and one spin-down fermion provides valuable insight into the convergence properties of the different renormalization schemes. In particular, since the $1+1$ system is exactly solvable, we can use its exact solution for accurate benchmarking. In practical calculations, the simulation of larger particle numbers is more important. We showcase the convergence of the renormalized interactions by studying a $1+3$ system with $N_{\uparrow}=1$ and $N_{\downarrow}=3$. [We observed similar convergence for other particle numbers, in particular for larger systems, see App.~\ref{App:MoreParticles}.]

We present our findings for two representative interaction strengths: One for repulsive and one for attractive interaction. However, we tested that our conclusions are valid for other interaction strengths as well. All results are shown for one-body basis cutoffs from $n=11$ up to $n=31$. With this choice, the calculations can be performed on a normal computer while at the same time the renormalized interactions enable converged results, see below.

\subsection{Energy}

\begin{figure}[!htb]
\centering
    \includegraphics[width=1\linewidth]{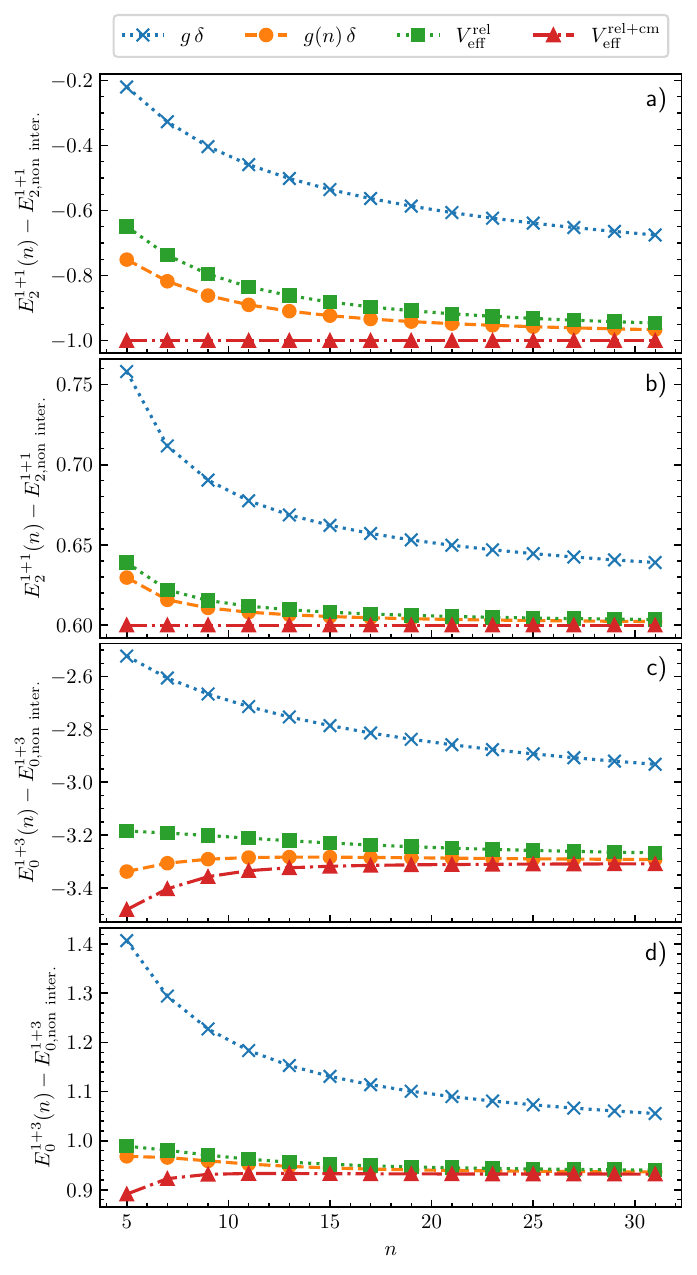}
    \caption{Energy of the system for different particle numbers and interactions. Panels a) and b) demonstrate the energy of the second excited state of the $1+1$ system; panels c) and d) show the ground state energy of $1+3$. Panels a) and c) are for attractive interaction with $g=-2.5067$ (which corresponds to $E_{1+1}=-1$); b) and d) illustrate the system with repulsive interactions, $g=3$ (which corresponds to $E_{1+1}=1.6$). Numerical data are shown with markers of different shape; lines are added to guide the eye. 
    The interested reader can see these data plotted as a function of $1/n$ and also using the log-log scale for $1+1$ in the supplementary material~\cite{SuppMat}, Figs.~S1 and~S6. 
    }
    \label{fig:Energy}
\end{figure}

The energies, $E(g)-E(g=0)$, obtained with the different renormalization schemes are shown in Figure~\ref{fig:Energy} as a function of the one-body cutoff parameter, $n$. As a relevant scale to judge the convergence we use the harmonic oscillator energy unit, $\hbar\omega=1$; all data are shown with respect to it. Panels a) and b) present results for the second excited state\footnote{We do not show the ground state energy of the $1+1$ system as this quantity is fixed for $g(n)\delta$ and $V_{\mathrm{eff}}^{\mathrm{rel+cm}}$.} of the $1+1$ system, for representative attractive and repulsive interactions, respectively. Panels c) and d) demonstrate the ground-state energy of the $1+3$ system, which does not have an analytical solution. Hence, its numerical solution is more relevant from the practical point of view.

We compare our numerical data for the $1+1$ system with the exact values, see Eq.~(\ref{eq:ExactEnergy}). First thing to notice is that the results obtained with the bare interaction always converge slower to the limit $n\to\infty$ than those for the effective interactions. For example, for the attractive interaction the energy calculated with $g\delta(x)$ has $E(n=31)-E(n=\infty)\approx 0.33$ for the highest presented cutoff; at the same time, the energy calculated with $g(n)\delta(x)$ deviates from the analytical result by $E(n=31)-E(n=\infty)\approx0.035$. For the effective interactions we see that $V_{\mathrm{eff}}^{\mathrm{rel}}$ converges similar to the running coupling constant.  For $V_{\mathrm{eff}}^{\mathrm{rel+cm}}$ the $1+1$ spectrum is immediately converged because the full energy spectrum is used for its construction. 

For the $1+3$ system, the effective potential $V_{\mathrm{eff}}^{\mathrm{rel+cm}}$ leads to the fastest approach to a reliable result. We observe that already for $n\simeq 21$ the energy changes by less than one percent of the relevant energy scale, $\hbar\omega=1$, when increasing the number of orbitals, e.g., for attractive interaction $E(n=21)-E(n=31)\lesssim 0.004$.  This observation motivates us to use results of $V_{\mathrm{eff}}^{\mathrm{rel+cm}}$ with a cutoff $n=31$ as a benchmark for the other interactions, i.e., we assume that $E(n\to\infty)\approx E(n=31)$. Running coupling constant and $V_{\mathrm{eff}}^{\mathrm{rel}}$ lead to similar convergence patterns. For these interactions and the largest $n$ considered, we have $E(n=31)-E(n\to\infty) \approx0.05$. At the same time, the energy calculated using the bare contact interaction deviates significantly from the $n\to\infty$ result: $E(n=31)-E(n\to\infty)\approx0.7$.

The slow convergence of the bare contact interaction is expected. Using Eqs.~(\ref{eq:large_lambda}) and~(\ref{eq:ConnectionLambdan}), we arrive at the following expression for $n\to\infty$ (for details of the calculation and a comparison to our data see App.~\ref{App:ConvergenceEnergy}):
\begin{equation}
\label{eq:ComparisonEnergy}
    E_{g\delta}(n)\simeq E_{g(n)\delta}(n)+\frac{C}{\pi\sqrt{2n}},
\end{equation}
where $E_{V}(n)$ is the energy calculated with the potential $V$ in the Hilbert space determined by the one-body cutoff $n$;
$C=g^2 \braket{\Psi|\sum\limits_{i<j}\delta(x_i-x_j)|\Psi}$ is the contact parameter~\cite{Barth2011}. As we have argued above, the running coupling constant approach leads to a fast convergence, and hence $E_{g(n)\delta}(n)$ can be considered constant with $n$ for sufficiently large cutoffs. We conclude that $E_{g\delta}(n)$ features a slow $1/\sqrt{n}$ convergence for any number of particles, in agreement with the numerical analysis of Ref.~\cite{grining2015many}. 

Motivated by Eq.~(\ref{eq:ComparisonEnergy}), we performed further analysis of the data by fitting them with the function (see also App.~\ref{App:Fits})
\begin{equation}
\label{eq:fit_function}
    E_V(n)=E_{V}(n\to\infty)+\frac{A_{V}}{n^{\sigma_{V}}}.
\end{equation}
Note that the parameters $E_{V}(n\to\infty), A_{V}$ and $\sigma_V$ depend on the potential $V$. For better readability, we will omit the subscript $V$ in the following and instead state explicitly to which potential these parameters belong.
The fit shows that renormalized interactions lead to a faster convergence rate (larger $\sigma$). It allows us to conclude that (i) the convergence of the energy is significantly faster for renormalized interactions in agreement with previous studies~\cite{Ernst2011, Rojo-Francas2022, Rammelmueller2023}; (ii) the effective interaction $V_{\mathrm{eff}}^{\mathrm{rel}}$ performs similar to the running coupling constant approach; (iii) the effective potential $V_{\mathrm{eff}}^{\mathrm{rel+cm}}$ outperforms other approaches.

Finally, we note that some energies for the $1+3$ system in Fig.~\ref{fig:Energy} can converge from below, i.e., the ground-state energy increases when we increase the cutoff parameter. This is at odds with the behavior expected for  CI methods, which are variational,
but it is in fact a typical feature of renormalized interactions (see also a non-monotonic behavior of the data calculated with $V_{\mathrm{eff}}^{\mathrm{rel+cm}}$ in Figs.~\ref{fig:kineticenergy},~\ref{fig:trap},~\ref{fig:transmatrix}). To rationalize this, let us consider the running coupling constant, see Fig.~\ref{fig:running}. When the Hilbert space is extended, the attractive interaction effectively weakens, see Eq.~\eqref{eq:gRG}. This can result in a competition between a decrease in energy due to a larger $n$ and an increase in energy due to a weaker potential\footnote{Note that we cannot rule out an oscillatory convergence behavior, even though it was not observed in our data. In general, we expect potential oscillations to be suppressed for large cutoffs.}.
The non-variational character of our calculations is not necessarily a disadvantage. Instead, this can be used to approximate the error imposed by the finite basis: Deviations between the different interactions, in particular if one converges from above and one from below, allow for an estimate of the `true' value that would result from an infinite Hilbert space.

\subsection{Density}

\begin{figure}[!htb]
   \centering
    \includegraphics[width=1\linewidth]{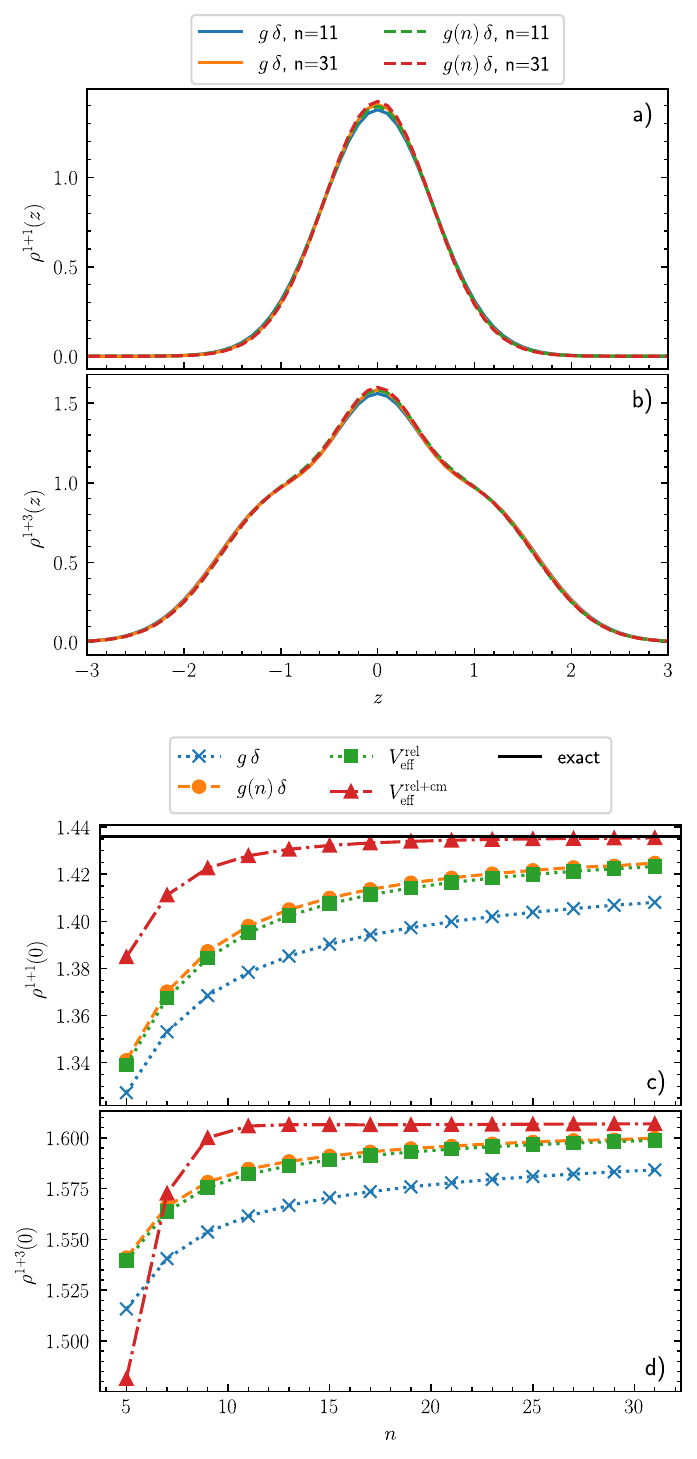}
    \caption{Panels a) and b): Density of the ground state calculated with the potentials $g\delta(x)$ and $g(n)\delta(x)$ and two cutoff parameters. Panels c) and d):  Density of the ground state in the center of the trap as a function of the cutoff parameter, $n$. Panels a) and c) show results for the $1+1$ system; panels b) and d) are for $1+3$. All panels are for attractive interactions: $g=-2.5067$, which corresponds to $E_{1+1}=-1$. Numerical data in panels c) and d) are shown with markers of different shapes. Lines are added to guide the eye.
    In the supplementary material~\cite{SuppMat}, we also present these data as a function of $1/n$, Fig. S2, and using the log-log scale for $1+1$, Fig. S7.}
    \label{fig:Density}
\end{figure}

In Fig.~\ref{fig:Density} we show the ground-state density for attractive interactions for the $1+1$ [panels a) and c)] and $1+3$ systems [panels b) and d)], whose operator is defined in first quantization picture as
\begin{equation}
    \rho(z)=\sum\limits_{i=1}^N\delta(z-x_i)
\end{equation}
while in second quantization it reads
\begin{equation}
    \rho(z)=\sum\limits_{i, j}\Phi_i(z)\Phi_j(z)a_i^\dagger a_j\,.
\end{equation}
The two upper panels show the density calculated with the bare contact interaction and with the running coupling constant approach. The two lower panels show the density at the trap center for all interactions. The plot demonstrates that the density converges fast to the limit $n\to\infty$ for $V_{\mathrm{eff}}^{\mathrm{rel+cm}}$ even without transforming 
the density operator alongside the interaction as discussed Sec.~\ref{subsec:effective_interactions}. Therefore, we leave the study of the effect of this transformation on the density to future works\footnote{Note that the induced two-body terms (see Eq.~\eqref{eq:OneBodyOperatorTransformed})
\begin{equation}
    \rho^{(2)}(z_\alpha, z_\beta)=\sum_{\alpha, \beta=1}^N \left(\rho(z_\alpha)+\rho(z_\beta)\right)
\end{equation}
do not factorize in relative $z_{\alpha,\beta}=\frac{1}{\sqrt{2}}(z_\alpha-z_\beta)$ and center-of-mass $Z_{\alpha,\beta}=\frac{1}{\sqrt{2}}(z_\alpha+z_\beta)$ coordinates, which complicates their computation. For more details see Sec. III in the supplementary material~\cite{SuppMat}.}.

While the overall shape of the density is reproduced with all interactions and even for a small cutoff, large values of $n$ are still required for an accurate estimate of the density. We illustrate this by considering the density at the center of the trap, $\rho(z=0)$, see panels c) and d). We have checked that the convergence of the density at the center is representative also of other values of $z$. To this end, we investigated the behavior of the relative density (not shown), defined as $(\rho_{n=11}(z) - \rho_{n=31}(z)) / \rho_{n=31}(z)$. 

For the $1+1$ system we see that results obtained with $V_{\mathrm{eff}}^{\mathrm{rel+cm}}$ approach the exact value quickly. Similar to the energy, the potentials $g(n)\delta(x)$ and $V_{\mathrm{eff}}^{\mathrm{rel}}$ 
enjoy similar convergence patterns, and the bare contact potential produces least accurate results for the density. The same conclusion holds also for larger systems, see Fig.~\ref{fig:Density}~d).

Similar to the energy, we analyzed the density with the fitting function
\begin{equation}
    O(n)=O(n\to\infty)+\frac{A}{n^\sigma}.
\end{equation}
The analysis allows us to quantify a quick convergence of the density calculated with $V_{\mathrm{eff}}^{\mathrm{rel+cm}}$ ($\sigma>1$). For a detailed description of the fitting analysis see App.~\ref{App:Fits}.
This shows that although the effective interactions are optimized using the energy spectrum, they also lead to better results for the density, in agreement with previous studies~\cite{Rammelmueller2023, Ernst2011, Rojo-Francas2022}.

\subsection{Transformation of observables}

In this subsection, we focus on the transformation of operators together with the potential (see Eq.~\eqref{eq:OneBodyOperatorTransformed}), which has received little attention in the corresponding cold-atom literature.
To this end, we calculate the expectation values of the one-body parts of the Hamiltonian, i.e., of the  kinetic energy operator\footnote{Note that strictly speaking the kinetic energy is not an observable quantity. In one spatial dimension it is well-defined, however, in higher dimensions its value diverges. This can be shown, e.g., using the Tan's contact parameter~\cite{zwerger2011bcs}. Nevertheless, we discuss the convergence of $\langle T\rangle$ as $T$ is one of the simplest operators that is sensitive to changes of the wave function on small length scales.}, $T$, and of the trapping potential, $W$, see Eq.~(\ref{eq:kinetic_trapped}).  The effective interactions $V_{\mathrm{eff}}^{\mathrm{rel}}$ and  $V_{\mathrm{eff}}^{\mathrm{rel+cm}}$ are constructed via a unitary transformation from which the transformation of operators alongside the construction of these potentials follows. The running coupling constant approach cannot be represented via a unitary transformation. Therefore, we cannot transform the operator together with this potential, and only present the expectation values of non-transformed operators in this case. Similar to the energy, the relevant scale to judge the convergence is the harmonic oscillator energy unit, $\hbar\omega=1$.

The transformation of the operators $T$ and $W$ alongside the effective interactions, Eq.~\eqref{eq:OneBodyOperatorTransformed}, yields the two-body operators [see Eq.~(\ref{eq:induced_two_body})]
\begin{equation}
    T^{(2)}=\sum\limits_{ijkl}t_{ijkl}a_i^\dagger a_j^\dagger a_k a_l, \qquad
    W^{(2)}=\sum\limits_{ijkl}w_{ijkl}a_i^\dagger a_j^\dagger a_k a_l,
    \nonumber
\end{equation}
where $t_{ijkl}=\delta_{j,l}t_{ik}+\delta_{i,k}t_{jl}$, and $w_{ijkl}=\delta_{j,l}w_{ik}+\delta_{i,k}w_{jl}$. Here, 
$t_{ik}=-\int dx \Phi_i(x)\Phi''_k(x)/2$ and $w_{ik}=\int dx \Phi_i(x)x^2\Phi_k(x_1)/2$. We can also straightforwardly calculate $\widetilde T^{(2)}$ and $\widetilde W^{(2)}$ as they factorize in center-of-mass and relative coordinates.

\begin{figure}[!htb]
\centering
    \includegraphics[width=1\linewidth]{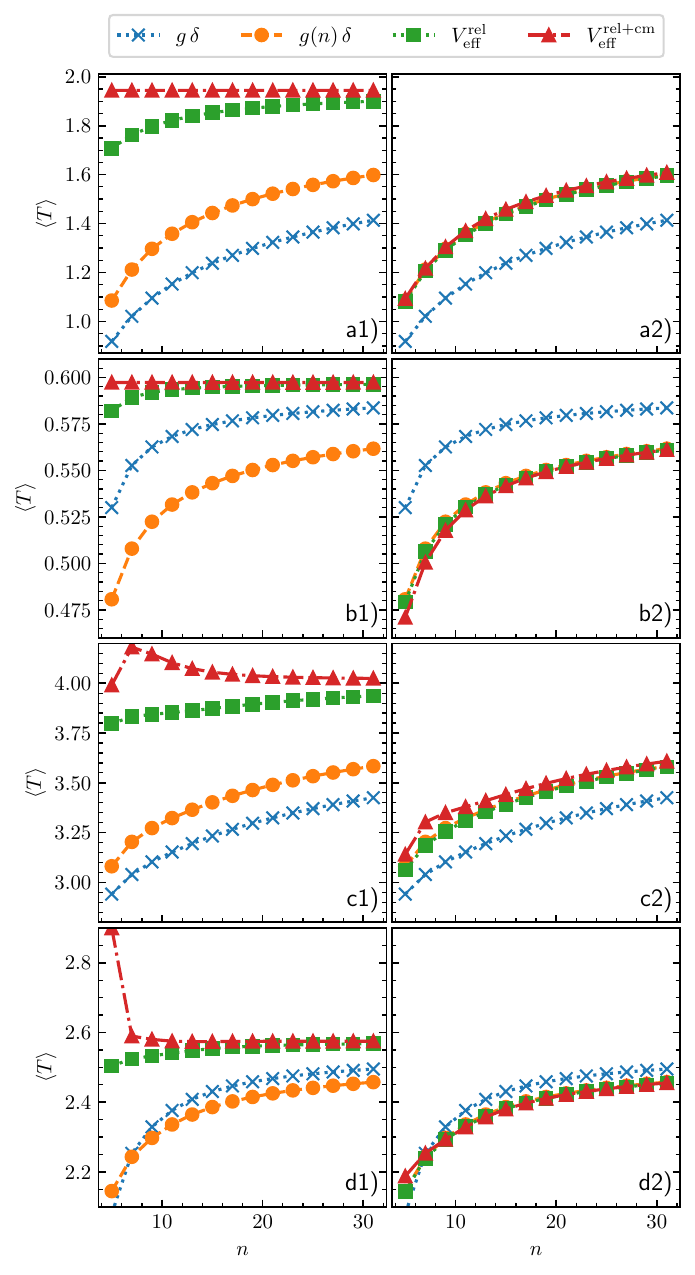}
    \caption{Expectation value of the kinetic energy operator, $\langle T\rangle$, as a function of the cutoff parameter, $n$. Left panels [*1)] show $\langle T\rangle$ for the operator $T$ that has been transformed together with $V_{\mathrm{eff}}^{\mathrm{rel}}$ and $V_{\mathrm{eff}}^{\mathrm{rel+cm}}$. Right panels [*2)] show what happens if the potential is renormalized but the operator $T$ is not transformed. Panels a) and b) are for the $1+1$ system, while panels c) and d) are for $1+3$. Panels a) and c) are for $g=-2.5067$ (corresponds to $E_{1+1}=-1$); panels b) and d) are for $g=3$ (corresponds to $E_{1+1}=1.6$).
    Our data are shown with markers; lines are added to guide the eye.
    We show these data in the supplementary material~\cite{SuppMat} also as a function of $1/n$, Fig. S3, and (for the $1+1$ system) using the log-log scale, Fig. S8.}
    \label{fig:kineticenergy}
\end{figure}

Figure~\ref{fig:kineticenergy} presents the expectation value of the kinetic energy operator for the $1+1$ [panels a), b)] and $1+3$ [c), d)] systems with attractive [a), c)] and repulsive [b), d)] interactions. The left panels [*1)] show $\langle T\rangle$ for the case when $T$ has been transformed together with the potential for $V_{\mathrm{eff.}}^{\mathrm{rel}}$ and $V_{\mathrm{eff.}}^{\mathrm{rel+cm}}$. The right panels [*2)] show $\langle T \rangle$ assuming that $T$ has not been transformed. A quantitative analysis of the data with a fitting procedure is given in App.~\ref{App:Fits}.

We observe that only renormalizing the interaction and not transforming $T$ does not necessarily improve the accuracy of results: For attractive interaction the convergence of $\langle T\rangle$ is improved slightly when compared to the bare contact interaction; for repulsive interaction the convergence rate is however slower\footnote{This behavior can be most easily explained with the running coupling constant approach: For repulsive interactions, renormalization leads to weaker interactions ($g(n)$ is smaller than $g$). This leads to a smaller value of the observable in question. However, increasing the cutoff increases the value of the observable and thus the renormalization works against the convergence. This is exactly opposite to the attractive-interaction case: There the renormalization leads to stronger interactions which result in a larger value of the observable. At the same time, increasing the cutoff also increases the value of the observable such that renormalization boosts the convergence.}. In contrast, if we transform the kinetic energy operator, Eq.~\eqref{eq:OneBodyOperatorTransformed}, alongside renormalization, the accuracy of our computation of $\langle T\rangle$ is significantly improved. As for the energy, the potential $V_{\mathrm{eff}}^{\mathrm{rel+cm}}$ leads to the fastest convergence.  
For the $1+1$ system, calculations with this potential 
produce an exact result for any cutoff if the observable is transformed together with the interaction. Further, we see that for the $1+3$ system this potential leads to a faster convergence with $n$ in comparison to all other potentials.

\begin{figure}[!htb]
\centering
    \includegraphics[width=1\linewidth]{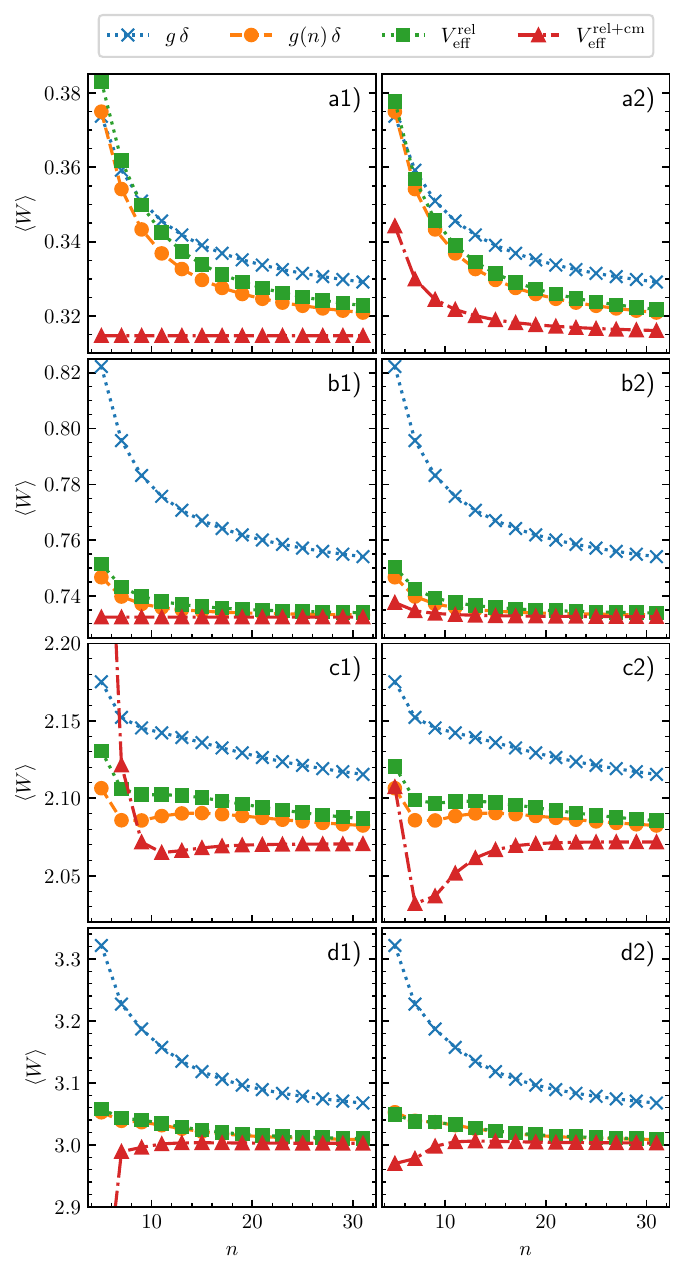}
    \caption{Expectation value of the trap operator, $\langle W \rangle$, as a function of the cutoff parameter, $n$. Left panels [*1)] show $\langle W\rangle$ for the operator $W$ that has been transformed together with   $V_{\mathrm{eff}}^{\mathrm{rel}}$ and $V_{\mathrm{eff}}^{\mathrm{rel+cm}}$. Right panels [*2)] show what happens if the potential is renormalized but the operator $W$ is not transformed. Panels a) and b) are for the $1+1$ system, while panels c) and d) are for $1+3$. Panels a) and c) are for $g=-2.5067$ (corresponds to $E_{1+1}=-1$); panels b) and d) are for $g=3$ (corresponds to $E_{1+1}=1.6$). Our data are shown with markers; lines are added to guide the eye. 
     In the supplementary material~\cite{SuppMat}, we show these data also as a function of $1/n$, Fig. S4 and (for the $1+1$ system) using the log-log scale, Fig. S9. }
    \label{fig:trap}
\end{figure}

Figure~\ref{fig:trap} shows the expectation value of the trap operator. We immediately notice that the transformation of this operator is not as important as of the kinetic energy operator. 
In particular, we see that the potentials
$V_{\mathrm{eff}}^{\mathrm{rel}}$ and  $g(n)\delta$ lead to similar convergence patterns. 
Similar to before, $V_{\mathrm{eff}}^{\mathrm{rel+cm}}$ leads to virtually converged (relative difference is $\simeq 0.01$) results for $n\simeq 25$ (for more details on the convergence see App.~\ref{App:Fits}).
 
To summarize the above discussion: The transformation of observables together with renormalizing interactions seems to be important only for observables that are sensitive to local changes of the wave function, such as the expectation value of the kinetic energy operator.

\subsection{Transition matrix elements}

Finally, we turn to an observable where (to the best of our knowledge) the improved convergence of renormalized interactions has not been demonstrated before: The transition rate between the ground and an excited states upon small periodic modulations of the interaction. There are two reasons to study this observable. First, the corresponding {\it two-body} operator is very sensitive to correlations between different particles and therefore it can further showcase the importance of transforming the operator alongside the interaction.
Second, it provides a useful probe in a cold-atom laboratory. For example, it has been used to study few-body precursors of a phase transition in systems of two-component fermions, theoretically~\cite{Bjerlin2016} and experimentally~\cite{bayha2020observing}. It has also been theoretically explored in the context of bosonic few-body droplets~\cite{chergui2023superfluid}. 

Let us consider the periodic modulation of the interaction strength,
\begin{equation}
    g(t)=g(0)+\eta\sin(\omega t),
\end{equation}
assuming that $|\eta|\; \ll |g(0)|$. The transition from the ground state $\ket{0}$ to an excited state $\ket{f}$ driven by this modulation can be estimated using
Fermi's golden rule  as
    $R_{0\to f}\propto\eta^2|M_{0\to f}|^2\delta(E_f-E_0-\omega)$. 
The focus of this subsection is on the transition matrix element, 
\begin{equation}
    M_{0\to f}=\braket{\Psi_f|\sum_{i<j}\delta(x_i-x_j)|\Psi_0}.
    \label{eq:transmatrix}
\end{equation}

\begin{figure}[!t]
\centering
    \includegraphics[width=0.95\linewidth]{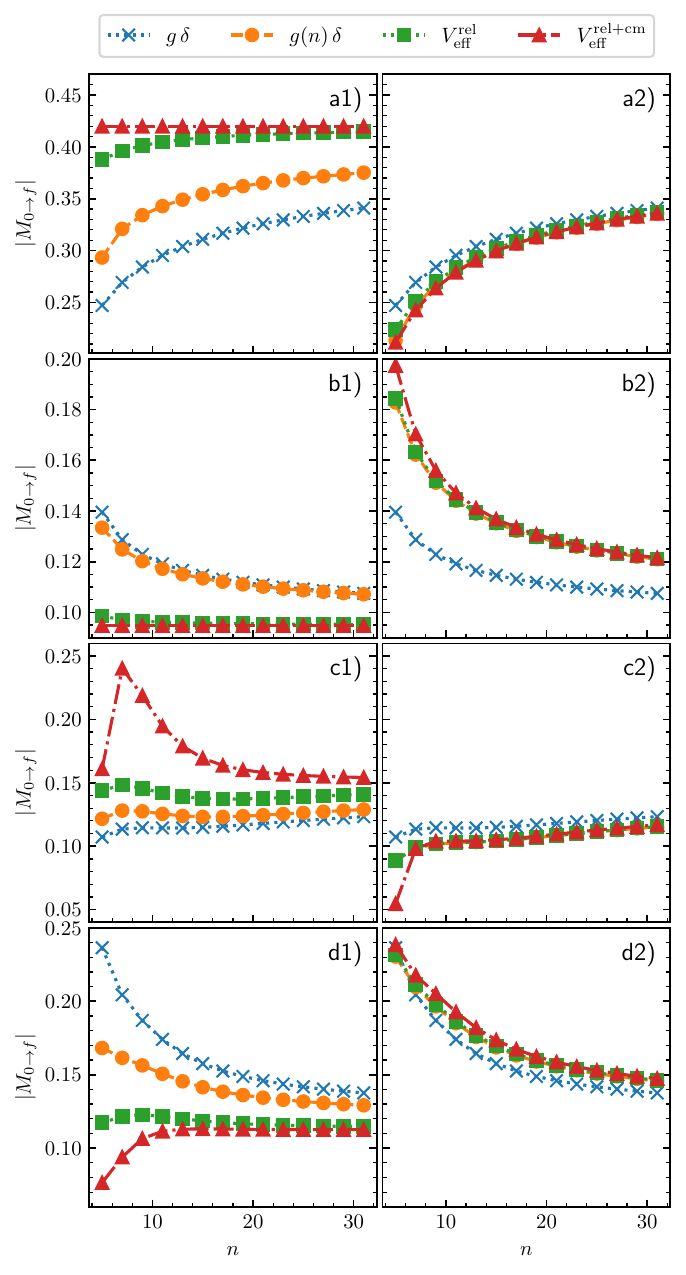}
    \caption{Transition matrix elements from the ground state to the lowest-energy state that can be excited by a periodic modulation of the interaction. Left panels [*1)] show $M_{0\to f}$ for the observable that has been transformed together with the potential  while the right ones [*2)] show what happens if the potential has been renormalized but the operator has not been modified. Panels a) and b) are for the $1+1$ system while panels c) and d) are for the $1+3$ system. Panels a) and c) are for $g=-2.5067$ (corresponds to $E_{1+1}=-1$); panels b) and d) are for $g=3$ (corresponds to $E_{1+1}=1.6$).
Our data are shown with markers; lines are added to guide the eye. 
    In the supplementary material~\cite{SuppMat}, we show these data also as a function of $1/n$, Fig. S5, and (for the $1+1$ system) using the log-log scale, Fig. S10. }
    \label{fig:transmatrix}
\end{figure}

It is straightforward to write  a transformed version of this observable for the effective interactions $V_{\mathrm{eff}}^{\mathrm{rel}}$ and $V_{\mathrm{eff}}^{\mathrm{rel+cm}}$ using Eq.~\eqref{eq:TwobodyOperatorTransformed}, because it only depends on the relative coordinates of the particles. Furthermore, we can also introduce a heuristic transformation of $M_{0\to f}$ for the $g(n)\delta$ potential. To this end, we note that the running coupling constant approach can be seen as a transformation of the delta-function potential:
\begin{align}
\label{eq:transformedTransmatrix}
\delta(x_i-x_j) \rightarrow \frac{g(n)}{g}\delta(x_i-x_j),
\end{align}
which implies the `natural' transformation of the transition matrix element.

Figure~\ref{fig:transmatrix}~a) and~b) presents $|M_{0\to f}|$ for the $1+1$ system while panels c) and d) demonstrate our results for $1+3$. We show the transitions from the ground state to the lowest state that can be excited via contact fermion-fermion interactions\footnote{Periodic modulations of the interaction can excite only certain states, e.g., only states with the same center-of-mass quantum number.}. Left panels ([*1)]) are for calculations where the observable has been transformed together with the potential while the right ones ([*2)]) are for a non-transformed observable.

In all cases presented in the figure, accurate results are obtained only by transforming the observable. For a non-transformed operator, our computations based upon $g\delta$ are actually more accurate than those based upon renormalized interactions. The reason is the same as for the one-body energy: The renormalization of the interaction works against the convergence of $M_{0\to f}$. Only by adjusting the observable, e.g., by using Eq.~\eqref{eq:transformedTransmatrix}, can the renormalization boost the convergence. We conclude that the transformation of $M_{0\to f}$ together with the potential has a significant effect on the accuracy of results. A detailed quantitative analysis of the data with a fitting procedure can be found in App.~\ref{App:Fits}. 

The highest accuracy is reached in calculations with the potential $V_{\mathrm{eff}}^{\mathrm{rel+cm}}$ and a transformed observable. As the corresponding results for the $1+1$ system are exact for any value of $n$, let us illustrate this statement by comparing convergence patterns for the $1+3$ system and attractive interactions, see Fig.~\ref{fig:transmatrix}~c1).  First of all, we notice that the standard fitting procedure with Eq.~\eqref{eq:fit_function},
gives meaningful results only for $V_{\mathrm{eff}}^{\mathrm{rel+cm}}$ and a transformed observable. In particular, the fit does not produce an accurate parameter $c$ for the data calculated with the bare contact interaction, at least for the considered truncated Hilbert spaces. 
Second, the data in Fig.~\ref{fig:transmatrix} defy the naive expectation that convergence patterns for systems with more than two particles should mimic those for the $1+1$ system. This expectation is based upon the idea that transition matrix elements are computed from a two-body operator, and hence their convergence can be studied with the 1+1 system. 
Instead, we observe that it is reasonable to fit the 1+1 data in Fig.~\ref{fig:transmatrix}~a1), whereas Eq.~(\ref{eq:fit_function}) yields poor results for the 1+3 data in Fig.~\ref{fig:transmatrix}~c1)
for all interactions except $V_{\mathrm{eff}}^{\mathrm{rel+cm}}$, see App.~\ref{App:Fits} for a more quantitative discussion. 

 In general, we observe that the use of $V_{\mathrm{eff}}^{\mathrm{rel+cm}}$ and $V_{\mathrm{eff}}^{\mathrm{rel}}$ is advantageous over the use of $g(n)\delta$ in computations of $M_{0\to f}$. This is expected as there is only a heuristic transformation~(\ref{eq:transformedTransmatrix}) of the latter -- not a unitary transformation as for $V_{\mathrm{eff}}^{\mathrm{rel}}$ and  $V_{\mathrm{eff}}^{\mathrm{rel+cm}}$. Still, the running coupling constant approach with a heuristically transformed observable leads to results that are more accurate than those for the bare interaction for all considered cases except for $1+1$ and repulsive interaction, where $g(n)\delta$ and $g\delta$ lead to similar results.  
To give some numbers: For the $1+3$ system and repulsive interactions [panel d1)], $M_{0\to f}$ changes  by 3\% when increasing the cutoff parameter from $n=25$ to $n=31$ for $g\delta$ while for $g(n)\delta$ this change is 2\%. The largest improvements in accuracy of results is seen for the effective interactions: For $V_{\mathrm{eff}}^{\mathrm{rel}}$ the corresponding difference is  0.6\% and for $V_{\mathrm{eff}}^{\mathrm{rel+cm}}$ it is only 0.05\%.

The discussion in this subsection highlights both the necessity of renormalizing interactions and the importance of transforming operators, such as the transition matrix elements, in numerical studies of few-fermion models. In particular, we are able to obtain accurate numerical data for attractive $1+3$ systems with only $n\simeq 10$, which leads to matrices that can be diagonalized on a standard computer.

\section{Conclusions}
\label{sec:Conclusion} 
 
\subsection{Summary}

We have contrasted zero-range one-dimensional interactions useful in studies of a few cold atoms in a trap.  In agreement with previous studies, we have demonstrated that numerically exact solutions to a few-body problem can be obtained by explicitly including information about converged two-body energies into the many-body calculations in a truncated Hilbert space.

A comparison between the renormalization schemes has demonstrated that a simple running coupling constant approach (used in particular in Refs.~\cite{Ernst2011, Rojo-Francas2022}) might be a go-to method for approximate calculations of the energy spectrum as it leads to a convergence pattern similar to $V_{\mathrm{eff}}^{\mathrm{rel}}$ also used in previous studies~\cite{Rammelmueller2023, rammelmuller2023magnetic, rotureau2013}. Indeed, running coupling constant approach is simple and transparent as it relies on a one-parameter renormalization of the interaction strength. 
At the same time, more accurate results at (almost) the same numerical cost can be obtained with $V_{\mathrm{eff}}^{\mathrm{rel+cm}}$. This effective potential fixes the two-body spectrum in a {\it truncated} Hilbert space to that in an {\it infinite} Hilbert space. The spectrum can be calculated either numerically or analytically. Although this potential has received little attention in the  cold-atom literature, we hope that our results will motivate its use in the future.

The advantage of numerical methods based upon $V_{\mathrm{eff}}^{\mathrm{rel+cm}}$ and $V_{\mathrm{eff}}^{\mathrm{rel}}$ becomes clear in calculations of observables, as these potentials are generated by a unitary transformation. We have observed that transformation of observables (in particular of the kinetic energy and transition matrix elements)
can improve convergence considerably. In particular, we have noticed that our data calculated with $V_{\mathrm{eff}}^{\mathrm{rel+cm}}$ can be considered numerically exact already for relatively small sizes of the Hilbert space.
One reason for that follows from the observation that the harmonic oscillator breaks translational invariance. In the oscillator relative and center-of-mass motion only decouple for infinite basis size. The effective interaction 
$V_{\mathrm{eff}}^{\mathrm{rel+cm}}$ corrects for this finite basis artefact and 
converges more quickly. For large basis sizes, the relative and center-of-mass motions
are effectively decoupled and $V_{\mathrm{eff}}^{\mathrm{rel}}$ gives similar results
as $V_{\mathrm{eff}}^{\mathrm{rel+cm}}$.

\subsection{Outlook}
\label{sec:Outlook}

\begin{figure}
\centering
    \includegraphics[width=1\linewidth]{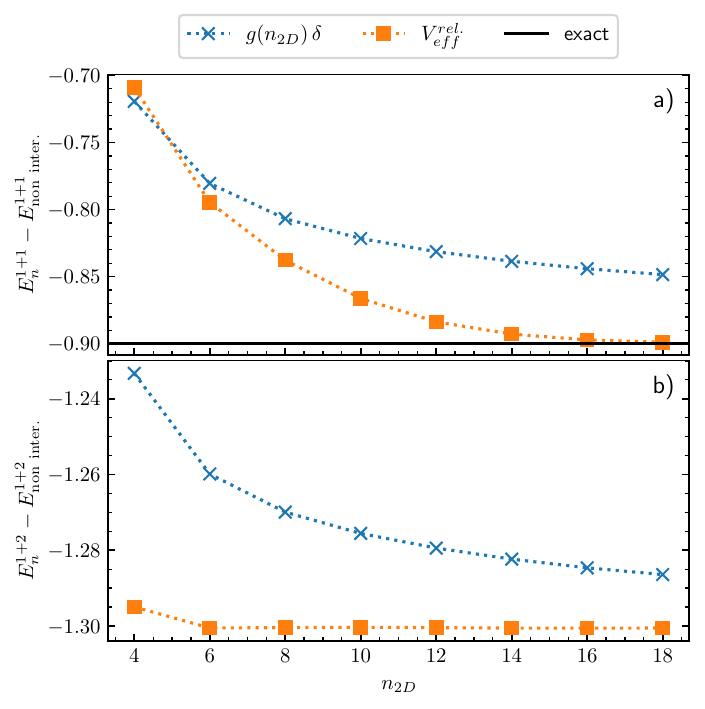}
    \caption{Energies in a two-dimensional harmonic oscillator for particles interacting via $g(n_{2D})\delta$ or $V_{\mathrm{eff}}^{\mathrm{rel}}$. Panel~a) is for the first excited state of the $1+1$ system with a vanishing total angular momentum. Panel~b) shows the energy of the ground state of the $1+2$ system.
    The parameters are chosen such that the ground state energy of two fermions is $E_{1+1}=1.1$. Our data are shown with markers; lines are added to guide the eye.}
    \label{fig:Energy2D}
\end{figure}

Our results have also demonstrated that the transformation of observables is not always necessary. For example, we have obtained accurate results for the density without it. Our current understanding is that observables influenced by ``global'' properties of the wave function are less susceptible to this transformation than ``local'' observables. The latter class is sensitive to changes of the wave function 
at short length scales, which are  determined mainly by particle-particle interactions. This boosts the importance of  induced two-body terms in Eq.~(\ref{eq:OneBodyOperatorTransformed}). We leave systematic investigations of this question to future studies. Such investigations are needed in particular to set the limits of validity of the simple $g(n)\delta$ potential, which does not allow for a unitary transformation of observables.

Our study motivates a comparative study of regularization methods employed in two-dimensional geometries~\cite{rontani2017renormalization,Christensson2009,Bjerlin2016}, which  are explored less than in one and three spatial dimensions. In Fig.~\ref{fig:Energy2D}, we showcase calculations for a two-dimensional harmonic oscillator confiment (see App.~\ref{App:2DHO} for the Hamiltonian and basis functions) that illustrate that already a simple potential $V_{\mathrm{eff}}^{\mathrm{rel}}$  can boost convergence of the energy. 
Here, we regularize the diverging two-dimensional $\delta$-potential by fixing the two-body ground state energy for each cutoff~\cite{rontani2017renormalization}. The cutoff is given by the highest one-body energy ($n_{2D}$) of the one-body basis used to construct the Hamiltonian. We show results for the first excited state of $1+1$ whose total angular momentum is zero and the ground-state energy for $1+2$\footnote{We have observed similar convergence patterns also for other particle numbers such as $3+3$ fermions.}. We expect that the use of the effective interaction will be beneficial also in calculations of observables, assuming that they are transformed together with the potential. 

We emphasize that our discussion on renormalized interaction is not limited to CI calculations and zero-range potentials. It can be extended to other numerical methods and to beyond `simple' delta-function potentials, e.g., dipole-dipole~\cite{DipolarRev1, DipolarRev2, DipolarRev3} or odd-channel interactions~\cite{Granger2004,Esslinger2005,mistakidis2022cold}. In an upcoming study, we will combine renormalized interactions and importance truncated CI (see e.g.~Refs~\cite{Bengtsson2020, chergui2023superfluid, Roth2009}) to calculate systems with larger particle numbers (mainly enabled by the importance truncation) and strong interactions (mainly enabled by renormalized interactions).

\section*{Acknowledgements}
We thank J. Cremon and J. Bjerlin for earlier contributions to the configuration interaction calculations used in this work (see Refs.~\cite{CremonPhDThesis, BjerlinPhdThesis}). F.B. and S.M.R. acknowledges helpful discussions with Carl Heintze, Sandra Brandstetter and Lila Chergui. We further want to thank Lila Chergui for helpful comments on the manuscript. This work was supported by a fellowship of the German Academic Exchange
Service (DAAD), the Swedish Research Council and the Knut and Alice Wallenberg foundation. 

\newpage
\bibliography{refs}

\newpage

\appendix

\counterwithin{table}{section} 
\renewcommand{\thetable}{\Alph{section} \arabic{table}} 

\section{Energy convergence for the bare contact interaction}
\label{App:ConvergenceEnergy}

In this appendix we discuss the convergence of the energy for a bare contact interaction. For this purpose we first
expand Eq.~\eqref{eq:gRG}, which
defines the running coupling constant in free space for attractive interaction, assuming large cutoffs $\lambda$

\begin{equation}
    \begin{split}
        -\frac{1}{g(\lambda)}&\simeq \frac{1}{2\sqrt{|E_{\mathrm{1+1}}^{\mathrm{unconf.}}}|}-\frac{\sqrt{2}}{\pi\lambda}\\
        \Leftrightarrow g&\simeq g(\lambda)+\frac{\sqrt{2}g^2}{\pi\lambda}.
        \label{eq:app:b1}
    \end{split}
\end{equation}
In the first step, we used the expression $E_{\mathrm{1+1}}^{\mathrm{unconf.}}=-g^2/4$. For the second step, we expanded the expression on the right-hand-side for large cutoffs $\lambda$ and used that $\frac{g(\lambda)^2}{\lambda}\simeq \frac{g^2}{\lambda}$ for $\lambda\to\infty$.

Next, we calculate $E_{\mathrm{1+1}}^{\mathrm{unconf.}}(\lambda)$. To this end, we insert our result for $g$ from Eq.~(\ref{eq:app:b1}) in the Schr{\"o}dinger equation at a finite cut-off $\lambda$: 
\begin{widetext}
\begin{equation}
\begin{split}
    E_{\mathrm{1+1}}^{\mathrm{unconf.}}(\lambda)\Psi_\lambda(x)&=\left(-\frac{1}{2}\frac{\partial^2}{\partial x^2}+\frac{g}{\sqrt{2}}\delta(x)\right)\Psi_\lambda(x)\\
    &=\left(-\frac{1}{2}\frac{\partial^2}{\partial x^2}+\left[g(\lambda)+\frac{\sqrt{2}g^2}{\pi\lambda}\right]{\sqrt{2}}\delta(x)\right)\Psi_\lambda(x)\\
    &=E_{\mathrm{1+1}}^{{\mathrm{unconf.}}(\lambda\to\infty)}\Psi_\lambda(x)+\frac{\sqrt{2}g^2}{\pi\lambda}\delta(x)\Psi_\lambda(x).
\end{split}
\end{equation}
\end{widetext}
In the last line, we utilized the fact that the use of $g(\lambda)$ leads (by construction) to the correct two-body energy. We treat the operator $\sqrt{2}g^2\delta(x)/(\pi\lambda)$ as perturbation, which leads to
\begin{equation}
    E_{\mathrm{1+1}}^{\mathrm{unconf.}}(\lambda)\simeq E_{\mathrm{1+1}}^{\mathrm{unconf.}(\lambda\to\infty)}+\frac{\sqrt{2}C}{\pi\lambda}
\end{equation}
with the contact parameter $C=g^2\braket{\Psi|\delta(x)|\Psi}$~\cite{Barth2011}; $\Psi$ is the wave function in the infinite Hilbert space. Here, we have used that $\ket{\Psi}_\lambda=\ket{\Psi}_{\lambda\to\infty}+o(1)$ ($o(1)$ includes small corrections to the wave function).  

Let us now perform the same calculation for the many-body system in a trap. To this end, we need to use the connection between $\lambda$ and $n$, Eq.~\eqref{eq:ConnectionLambdan}. We write the expectation value of the energy $E_{g\delta}(n)\equiv\prescript{}{n}{\bra{\Psi}}H\ket{\Psi}_n$:
\begin{widetext}
\begin{equation}
    \begin{split}
        E_{g\delta}(n)
        &=\prescript{}{n}{\bra{\Psi}}\sum\limits_{i=1}^N-\frac{1}{2}\frac{\partial^2}{\partial x_i^2}+\frac{1}{2}x_i^2+g\sum\limits_{i<j}\delta(x_i-x_j)\ket{\Psi}_n\\
        &=\prescript{}{n}{\bra{\Psi}}\sum\limits_{i=1}^N-\frac{1}{2}\frac{\partial^2}{\partial x_i^2}+\frac{1}{2}x_i^2+
        \left(g(n)+\frac{g^2}{\pi\sqrt{2n}}\right)\sum\limits_{i<j}\delta(x_i-x_j)\ket{\Psi}_n.\\
    \end{split}
\end{equation}
\end{widetext}

\begin{figure}[!htb]
\centering   
    \includegraphics[width=1\linewidth]{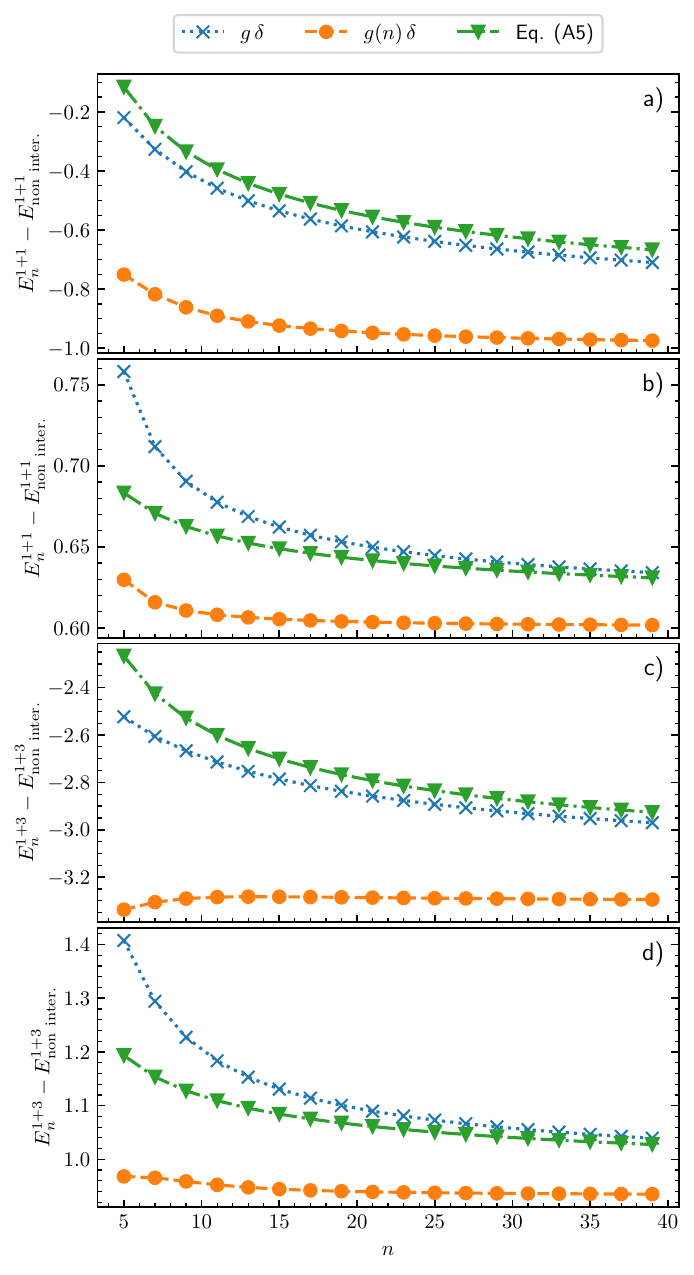}
    \caption{Energies of the $1+1$ and $1+3$ systems calculated with the bare contact interaction, running coupling constant and the predicted energy convergence of the bare contact interaction, Eq.~\eqref{eq:App:ComparisonEnergy}. Panels a) and b) show the energy of the second excited state of the $1+1$ system; c) and d) present the ground state energy of $1+3$. Panels a) and c) are for attractive interactions with $g=-2.5067$ (corresponds to $E_{1+1}=-1$); b) and d) are for repulsive interactions with $g=3$ (corresponds to $E_{1+1}=1.6$). Markers are results of our calculations. Curves in the panels are added to guide the eye. }
    \label{fig:App:ConvergenceEnergyComparisonToSqrtn}
\end{figure}

Assuming that $n$ is sufficiently large so that $\ket{\Psi}_n=\ket{\Psi}_{n\to\infty}+o(1)$, and that the energy from the running coupling constant approach, $E_{g(n)\delta}$, converges fast we arrive at a $1/\sqrt{n}$ convergence for the many-body energy for large cutoffs $n$,
\begin{equation}
\label{eq:App:ComparisonEnergy}
E_{g\delta}=E_{g(n)\delta}+\frac{1}{\pi\sqrt{2n}}C\simeq E_{n\to\infty}+\frac{1}{\pi\sqrt{2n}}C,
\end{equation}
where $C=g^2\braket{\Psi|\sum\limits_{i<j}\delta(x_i-x_j)|\Psi}$ is the contact parameter~\cite{Barth2011}.

In Fig.~\ref{fig:App:ConvergenceEnergyComparisonToSqrtn} we show a comparison between Eq.~\eqref{eq:App:ComparisonEnergy}\footnote{Here, we calculate the contact parameter $C$ using $V_{\mathrm{eff}}^{\mathrm{rel+cm}}$ with a transformed observable and a cutoff of $n=31$, which is sufficient for an accurate result. }, and the energy calculated using a bare contact interaction. For comparison, we also plot the outcome of the running coupling constant approach. As we can see, the running coupling approach converges fast so that the approximation, $E_{g(n)\delta}\simeq E_{n\to\infty}$,  in Eq.~\eqref{eq:App:ComparisonEnergy} is well justified. 
A good agreement between  the green and blue markers in Fig.~\ref{fig:App:ConvergenceEnergyComparisonToSqrtn} confirms that the bare contact interaction leads to a slow $1/\sqrt{n}$ convergence. Note that for repulsive interaction, the convergence in this range of the values of $n$ is somewhat faster than $1/\sqrt{n}$. At the same time, it is somewhat slower for attractive interaction, see also Tab.~\ref{Tab:Fit:Energy} where the fitting analysis shows this quantitatively. This observation is in agreement with the results of Ref.~\cite{grining2015many}.

\section{Analysis of convergence}
\label{App:Fits}

To analyze convergence patterns in our data, we perform fits with the function
\begin{equation}
\label{eq:fitfunc}
    f(n)=c+\frac{A}{n^\sigma},
\end{equation}
where $A,c$ and $\sigma>0$ are fit parameters.
We interpret the value of $\sigma$ as the rate of convergence: the larger $\sigma$, the faster the data approaches the result in an infinite Hilbert space. This is an additional measure of convergence compared to a simple energy gap from the converged value discussed in the main text. 
The parameter $c$ is the extrapolated expectation value of the observable of interest. Note that the values of $c$ can differ from one interaction to another if the maximal value of $n$ used to produce the data is not sufficiently large. This difference can be used as an additional estimate of the accuracy of the numerical method. 

The fitting function in Eq.~(\ref{eq:fitfunc}) is partly inspired by the discussion in App.~\ref{App:ConvergenceEnergy}, Eq.~\eqref{eq:App:ComparisonEnergy}, where we show the convergence behavior of the energy of many fermions in a one-dimensional harmonic trap interacting via bare contact interaction (see also Ref.~\cite{grining2015many}).
We therefore expect for a bare contact interaction $\sigma\approx 0.5$ for the convergence of the energy. In Ref.~\cite{Rammelmueller2023} a similar convergence rate was postulated and found for the density. In that reference, the authors studied the convergence of the energies and the density at $z=0$ using the effective interaction $V_{\mathrm{eff}}^{\mathrm{rel}}$ and the delta function potential. We extend their study by also considering the running coupling constant and $V_{\mathrm{eff}}^{\mathrm{rel+cm}}$. We further discuss the convergence rates of the transition matrix elements for a periodic modulation of the interaction strength, the expectation value of the kinetic energy as well as of the trapping potential. 

Note, that we do not know the convergence pattern for the data computed using the renormalized interactions. In fact, these data do not have to follow Eq.~(\ref{eq:fitfunc}). Therefore, in all fitted data there is a systematic error which we cannot estimate. Nevertheless, testing the validity of our fits with a $\chi^2$-test showed a chi-squared test statistic smaller than $10^{-8}$ and a p-value of one, which shows that Eq.~(\ref{eq:fitfunc}) models our data adequately. We conclude that fitting the data with Eq.~(\ref{eq:fitfunc}) helps one to quantify the convergence, e.g., larger values of $\sigma$ showcase faster convergence. 

\subsection{Energy}

\begin{table*}
\begin{tabular}{|l|lll|}
\hline
                            & \multicolumn{1}{l|}{$c$}                    & \multicolumn{1}{l|}{$A$}                  & $\sigma$           \\ \hline \hline
\multicolumn{4}{|l|}{$E^{1+1}, g=-2.5067$ (corresponds to $E_{1+1}=-1$)}                     \\ \hline
$g\delta$                   & \multicolumn{1}{l|}{$0.9954\pm 2\times 10^{-4}$}   & \multicolumn{1}{l|}{$1.7566\pm 4\times 10^{-4}$} & $0.4877\pm 3\times 10^{-4}$ \\ \hline
$g(n)\delta$                & \multicolumn{1}{l|}{$1.00151\pm 2\times 10^{-5}$} & \multicolumn{1}{l|}{$1.836\pm 0.002$}    & $1.1802\pm 5\times 10^{-4}$   \\ \hline
$V_{\mathrm{eff}}^{\mathrm{rel}}$                   & \multicolumn{1}{l|}{$0.9984\pm 2\times 10^{-4}$}    & \multicolumn{1}{l|}{$2.14 \pm 0.01$}    & $1.063\pm 0.003$    \\ \hline
$V_{\mathrm{eff}}^{\mathrm{rel+cm}}$                   & \multicolumn{3}{l|}{Constant by construction. Exact result: $c=1.00000$}                                                                    \\ \hline \hline

\multicolumn{4}{|l|}{$E^{1+1}, g=3$ (corresponds to $E_{1+1}=1.6$)}                  \\ \hline
$g\delta$                   & \multicolumn{1}{l|}{$2.6124\pm 5\times 10^{-4}$}   & \multicolumn{1}{l|}{$0.511\pm 0.008$}   & $0.86\pm 0.01$     \\ \hline
$g(n)\delta$                & \multicolumn{1}{l|}{$2.6004\pm 3\times 10^{-4}$}   & \multicolumn{1}{l|}{$0.226\pm 0.005$}   & $1.41\pm0.01$      \\ \hline
$V_{\mathrm{eff}}^{\mathrm{rel}}$                   & \multicolumn{1}{l|}{$2.6006\pm 4\times 10^{-4}$}   & \multicolumn{1}{l|}{$0.2636\pm 5\times 10^{-4}$}  & $1.319\pm 0.009$   \\ \hline
$V_{\mathrm{eff}}^{\mathrm{rel+cm}}$                   & \multicolumn{3}{l|}{Constant by construction. Exact result: $c=2.6000$}                                                                    \\ \hline \hline

\multicolumn{4}{|l|}{$E^{1+3}, g=-2.5067$ (corresponds to $E_{1+1}=-1$)}            \\ \hline
$g\delta$ $^*$                  & \multicolumn{1}{l|}{$2.119\pm 0.003$}       & \multicolumn{1}{l|}{$1.715\pm 0.003$}   & $0.390\pm 0.003$     \\ \hline

$g(n)\delta$ $^*$                 & \multicolumn{1}{l|}{$2.155\pm 0.009$}     & \multicolumn{1}{l|}{$0.122\pm 0.004$}   & $0.25\pm 0.04$     \\ \hline
$V_{\mathrm{eff}}^{\mathrm{rel}}$ $^*$            & \multicolumn{1}{l|}{$2.183\pm 0.007$}       & \multicolumn{1}{l|}{$0.74\pm 0.01$}   & $0.783\pm 0.009$     \\ \hline
$V_{\mathrm{eff}}^{\mathrm{rel+cm}}$ $^*$                   & \multicolumn{1}{l|}{$2.19341\pm 2\times 10^{-5}$}   & \multicolumn{1}{l|}{$-17.9\pm0.1$}          & $2.76\pm 0.02$    \\ \hline \hline

\multicolumn{4}{|l|}{$E^{1+3}, g=3$ (corresponds to $E_{1+1}=1.6$)}                  \\ \hline
$g\delta$                   & \multicolumn{1}{l|}{$6.481\pm 0.002$}      & \multicolumn{1}{l|}{$2.02\pm 0.05$}     & $0.96\pm 0.01$    \\ \hline
$g(n)\delta$                & \multicolumn{1}{l|}{$6.4331\pm 1\times 10^{-4}$}     & \multicolumn{1}{l|}{$1.18\pm 0.05$}    & $1.71\pm 0.02$      \\ \hline
$V_{\mathrm{eff}}^{\mathrm{rel}}$                   & \multicolumn{1}{l|}{$6.4338\pm 1\times10^{-4}$}   & \multicolumn{1}{l|}{$0.93\pm 0.02$}     & $1.5\pm0.1$       \\ \hline
$V_{\mathrm{eff}}^{\mathrm{rel+cm}}$                   & \multicolumn{1}{l|}{$6.4313\pm5\times 10^{-4}$}      & \multicolumn{1}{l|}{$0.02\pm0.02$}     & $1.0\pm0.4$     \\ \hline
\end{tabular}
\label{Tab:Fit:Energy}
\caption{Fits of the energies from Fig.~\ref{fig:Energy} (the second excited state for $1+1$ and the ground state for $1+3$) with the function $f(n)=c+\frac{A}{n^\sigma}$. The values from $n=11$ to $n=31$ were used for all fits except for the ones marked with $^*$. There a fitting range of $n=21-31$ was used. The uncertainties given are the uncertainties from the fitting procedure.}
\end{table*}

In Tab.~\ref{Tab:Fit:Energy} we show the results of fitting the energies from Fig.~\ref{fig:Energy}. For the $1+1$ system and attractive interactions, $\sigma\approx0.5$ for the bare contact interaction, as expected. For repulsive interactions the value of $\sigma$ is larger suggesting that we do not yet have enough states to suppress higher orders, e.g., $\sim 1/n$, in Eq.~(\ref{eq:App:ComparisonEnergy}). For $1+1$, all renormalized interactions yield values of $\sigma$ that are significantly larger than those for the bare contact interaction. This shows that not only are the data closer to the infinite limit, as discussed in Sec.~\ref{sec:Results}, they also converge faster. We further see, that the results for $c$ -- the extrapolated values in infinite Hilbert space -- are somewhat closer to the exact values for renormalized interactions in comparison to a bare contact potential. Only small differences between the running coupling constant approach and $V_{\mathrm{eff}}^{\mathrm{rel}}$ are visible. Since $V_{\mathrm{eff}}^{\mathrm{rel+cm}}$ is constant by construction, we do not fit these data points.

 Let us now consider the convergence of the energy for the $1+3$ system computed with the bare contact interaction, $g\delta$. For attractive interactions, we observe that $\sigma<0.5$. For the repulsive $1+3$ system, we see that $\sigma\approx 1$, indicating again the importance of higher order terms in Eq.~(\ref{eq:App:ComparisonEnergy}).

For the energies of the $1+3$ system calculated using the effective interaction in relative coordinates and the running coupling constant approach, we find $\sigma\approx0.78$ for attractive interactions. Note that the non-monotonic behavior shown in Fig.~\ref{fig:Energy} for $g(n)\delta$ cannot be reproduced with Eq.~(\ref{eq:fitfunc}). Therefore, we only fit the data for a few large values of $n$, $n=21-31$, in this case. To make the comparison with the other interactions faithful, we also use the restricted fitting regime for them. Nevertheless, the resulting values of $c$ are in good agreement with that for $V_{\mathrm{eff}}^{\mathrm{rel+cm}}$, whose use leads to by far the largest value of $\sigma$ and to the most stable fit. For repulsive interactions, we observe that the data based upon running coupling constant and $V_{\mathrm{eff}}^{\mathrm{rel}}$ converge faster than the bare contact interaction. For $V_{\mathrm{eff}}^{\mathrm{rel+cm}}$, the value of $\sigma$ has a high uncertainty: This is because the results for the energy are already nearly converged, which means that there is only little variation in the fitted data. This can be also seen from the fact that $A\approx 0$.

\subsection{Density}

\begin{table*}
\begin{tabular}{|llll|}
\hline
\multicolumn{1}{|l|}{}             & \multicolumn{1}{l|}{$c$}                        & \multicolumn{1}{l|}{$A$}                  & $\sigma$                      \\ \hline \hline
\multicolumn{4}{|l|}{$\rho(0)^{1+1}, g=-2.5067$ (corresponds to $E_{1+1}=-1$)}                                                                                                      \\ \hline
\multicolumn{1}{|l|}{$g\delta$}    & \multicolumn{1}{l|}{$1.4276\pm 3\times 10^{-4}$}        & \multicolumn{1}{l|}{$-0.4137\pm 7\times 10^{-4}$} & $0.887\pm 0.001$              \\ \hline
\multicolumn{1}{|l|}{$g(n)\delta$} & \multicolumn{1}{l|}{$1.4338\pm 4\times 10^{-4}$}        & \multicolumn{1}{l|}{$-0.83\pm 0.05$}    & $1.31\pm 0.03$                \\ \hline
\multicolumn{1}{|l|}{$V_{\mathrm{eff}}^{\mathrm{rel}}$}    & \multicolumn{1}{l|}{$1.43441\pm 6\times10^{-5}$} & \multicolumn{1}{l|}{$-0.72\pm 0.05$}    & $1.215\pm 0.003$              \\ \hline
\multicolumn{1}{|l|}{$V_{\mathrm{eff}}^{\mathrm{rel+cm}}$}    & \multicolumn{1}{l|}{$1.43607\pm 2\times10^{-5}$} & \multicolumn{1}{l|}{$-3.1\pm0.1$}       & $2.47\pm 0.02$                \\ \hline \hline
\multicolumn{4}{|l|}{$\rho(0)^{1+3}, g=-2.5067$ (corresponds to $E_{1+1}=-1$)}                                                                                                      \\ \hline
\multicolumn{1}{|l|}{$g\delta$}    & \multicolumn{1}{l|}{$1.6001\pm 3\times 10^{-4}$}       & \multicolumn{1}{l|}{$-0.297\pm 0.005$}  & $0.85\pm 0.01$                \\ \hline
\multicolumn{1}{|l|}{$g(n)\delta$} & \multicolumn{1}{l|}{$1.6069\pm 5\times 10^{-4}$}        & \multicolumn{1}{l|}{$-0.31\pm 0.03$}    & $1.10\pm 0.05$                \\ \hline
\multicolumn{1}{|l|}{$V_{\mathrm{eff}}^{\mathrm{rel}}$}    & \multicolumn{1}{l|}{$1.6076\pm 2\times 10^{-4}$}        & \multicolumn{1}{l|}{$-0.295\pm 0.006$}   & $1.02\pm$ 0.01 \\ \hline
\multicolumn{1}{|l|}{$V_{\mathrm{eff}}^{\mathrm{rel+cm}}$}    & \multicolumn{1}{l|}{$1.6071\pm 4\times 10^{-4}$}        & \multicolumn{1}{l|}{$-0.04\pm 0.08$}    & $2\pm 1$                 \\ \hline
\end{tabular}
\label{Tab:Fit:Density}
\caption{Fitting parameters (fitting function $f(n)=c+\frac{A}{n^\sigma}$) for the density at the origin, see Fig.~\ref{fig:Density}. The data in the range $n=11$ to $n=31$ are used for the fit. The uncertainties given are the uncertainties from the fitting procedure.}
\end{table*}

As we have discussed in the main text, the behavior of the energy as a function of $n$  is not a herald of convergence properties for other observables. Therefore, we continue our analysis of convergence patterns by considering the density at $z=0$ in this subsection, see Tab.~\ref{Tab:Fit:Density}. The conclusions we can draw are similar to the ones above. The results computed with renormalized interactions have larger values of $\sigma$, which demonstrates that not only are the results closer to the limit $n\to\infty$, they also converge faster. The results based upon $V_{\mathrm{eff}}^{\mathrm{rel}}$ and the running coupling approach are similar. The potential $V_{\mathrm{eff}}^{\mathrm{rel+cm}}$ has the strongest performance. The corresponding value of $\sigma$ is the largest of all. Further, the data are already nearly constant as can be seen by the fact that $A\approx 0$ within the error bars.

\subsection{One-body Hamiltonian}

\begin{table*}
\begin{tabular}{|llll|}
\hline
\multicolumn{1}{|l|}{}                   & \multicolumn{1}{l|}{$c$}                   & \multicolumn{1}{l|}{$A$}                     & $\sigma$                      \\ \hline \hline
\multicolumn{4}{|l|}{$\langle T\rangle , N_\uparrow=1, N_\downarrow=1, g=-2.5067$ (corresponds to $E_{1+1}=-1$)}                                                                                                       \\ \hline
\multicolumn{1}{|l|}{$g\delta$}          & \multicolumn{1}{l|}{$2.31\pm 0.02$}  & \multicolumn{1}{l|}{$-2.10\pm 0.01$}    & $0.245\pm 0.004$  \\ \hline
\multicolumn{1}{|l|}{$g(n)\delta$}       & \multicolumn{1}{l|}{$1.966\pm 0.002$}  & \multicolumn{1}{l|}{$-1.945\pm 0.003$}    & $0.484\pm 0.002$              \\ \hline
\multicolumn{1}{|l|}{$V_{\mathrm{eff}}^{\mathrm{rel}}$ obs trans.}          & \multicolumn{1}{l|}{$1.9527\pm 7\times 10^{-4}$}  & \multicolumn{1}{l|}{$-1.07\pm 0.01$} & $0.879\pm 0.007$            \\ \hline
\multicolumn{1}{|l|}{$V_{\mathrm{eff}}^{\mathrm{rel+cm}}$ obs trans.}            & \multicolumn{3}{l|}{Constant by construction. Exact result: $c=1.9442$}                                                                    \\ \hline
\multicolumn{1}{|l|}{$V_{\mathrm{eff}}^{\mathrm{rel}}$ obs non-trans.} & \multicolumn{1}{l|}{$1.975\pm 0.003$}   & \multicolumn{1}{l|}{$ -1.964\pm 0.004$}      & $0.479\pm 0.003$              \\ \hline
\multicolumn{1}{|l|}{$V_{\mathrm{eff}}^{\mathrm{rel+cm}}$ obs non-trans.} & \multicolumn{1}{l|}{$1.884\pm 0.005$}   & \multicolumn{1}{l|}{$ -2.16\pm 0.02$}      & $0.599\pm 0.007$              \\ 
\hline \hline
\multicolumn{4}{|l|}{$\langle T\rangle , N_\uparrow=1, N_\downarrow=1, g=3$ (corresponds to $E_{1+1}=1.6$)}                                                                                                             \\ \hline
\multicolumn{1}{|l|}{$g\delta$}          & \multicolumn{1}{l|}{$0.5919\pm 2\times 10^{-4}$}  & \multicolumn{1}{l|}{$-0.261\pm 0.006$}       & $1.00\pm0.01$               \\ \hline
\multicolumn{1}{|l|}{$g(n)\delta$}       & \multicolumn{1}{l|}{$0.5883\pm 4\times 10^{-4}$}   & \multicolumn{1}{l|}{$ -0.323\pm0.004$}       & $0.726\pm0.007$               \\ \hline
\multicolumn{1}{|l|}{$V_{\mathrm{eff}}^{\mathrm{rel}}$ obs trans.}          & \multicolumn{1}{l|}{$0.59698\pm 2\times 10^{-5}$}  & \multicolumn{1}{l|}{$-0.150\pm0.006$}       & $1.56\pm0.02$                \\ \hline
\multicolumn{1}{|l|}{$V_{\mathrm{eff}}^{\mathrm{rel+cm}}$ obs trans.}            & \multicolumn{3}{l|}{Constant by construction. Exact result: $c=0.59731$}                                                                    \\ \hline

\multicolumn{1}{|l|}{$V_{\mathrm{eff}}^{\mathrm{rel}}$ obs non-trans.}          & \multicolumn{1}{l|}{$0.5887
\pm 4\times 10^{-4}$}  & \multicolumn{1}{l|}{$-0.325\pm0.003$}       & $0.716\pm0.007$                 \\ \hline
\multicolumn{1}{|l|}{$V_{\mathrm{eff}}^{\mathrm{rel+cm}}$ obs non-trans.}          & \multicolumn{1}{l|}{$0.5834\pm 7\times 10^{-4}$}  & \multicolumn{1}{l|}{$-0.43\pm0.01$}       & $0.86\pm0.01$                 \\ \hline \hline
\multicolumn{4}{|l|}{$\langle T\rangle, N_\uparrow=1, N_\downarrow=3 , g=-2.5067$ (corresponds to $E_{1+1}=-1$)}                                                                                                       \\ \hline
\multicolumn{1}{|l|}{$g\delta$}          & \multicolumn{3}{l|}{Data could not be fitted by Eq.~\eqref{eq:fitfunc}.}                                                                                                   \\ \hline
\multicolumn{1}{|l|}{$g(n)\delta$}       & \multicolumn{3}{l|}{Data could not be fitted by Eq.~\eqref{eq:fitfunc}.}                                                                                  \\ \hline
\multicolumn{1}{|l|}{$V_{\mathrm{eff}}^{\mathrm{rel}}$ obs trans.}          & \multicolumn{3}{l|}{Data could not be fitted by Eq.~\eqref{eq:fitfunc}.}                                                                                     \\ \hline
\multicolumn{1}{|l|}{$V_{\mathrm{eff}}^{\mathrm{rel+cm}}$ obs trans.}          & \multicolumn{1}{l|}{$4.0187\pm4\times 10^{-4}$}     & \multicolumn{1}{l|}{$56\pm 6$}        & $2.71\pm 0.04$ \\ \hline
\multicolumn{1}{|l|}{$V_{\mathrm{eff}}^{\mathrm{rel}}$ obs non-trans.} & \multicolumn{3}{l|}{Data could not be fitted by Eq.~\eqref{eq:fitfunc}.}                         \\ \hline
\multicolumn{1}{|l|}{$V_{\mathrm{eff}}^{\mathrm{rel+cm}}$ obs non-trans.} & \multicolumn{3}{l|}{Data could not be fitted by Eq.~\eqref{eq:fitfunc}.}                                                                                  \\ \hline
\multicolumn{4}{|l|}{$\langle T\rangle, N_\uparrow=1, N_\downarrow=3 , g=3$ (corresponds to $E_{1+1}=1.6$)}                                                                                                             \\ \hline
\multicolumn{1}{|l|}{$g\delta$ $^*$}         & \multicolumn{1}{l|}{$2.539\pm 0.001$}   & \multicolumn{1}{l|}{$2.9\pm 0.1$}        & $1.22\pm 0.02$                \\ \hline
\multicolumn{1}{|l|}{$g(n)\delta$ $^*$}       & \multicolumn{1}{l|}{$2.531\pm 0.001$} & \multicolumn{1}{l|}{$-1.77\pm 0.05$}        & $0.93\pm 0.01$ \\ \hline
\multicolumn{1}{|l|}{$V_{\mathrm{eff}}^{\mathrm{rel}}$ obs trans.  $^*$}         & \multicolumn{1}{l|}{$2.5723\pm 1\times 10^{-4}$}  & \multicolumn{1}{l|}{$-3.4\pm 0.2$}        & $1.92\pm 0.02$                \\ \hline
\multicolumn{1}{|l|}{$V_{\mathrm{eff}}^{\mathrm{rel+cm}}$ obs trans. $^*$}          & \multicolumn{1}{l|}{$ 2.575092\pm9\times 10^{-6}$}     & \multicolumn{1}{l|}{$-60\pm 20$}        & $3.9\pm 0.1$ \\ \hline
\multicolumn{1}{|l|}{$V_{\mathrm{eff}}^{\mathrm{rel}}$ obs non-trans. $^*$} & \multicolumn{1}{l|}{$2.532\pm 0.002$} & \multicolumn{1}{l|}{$-1.75\pm 0.05$}        & $0.90\pm 0.01$ \\ \hline
\multicolumn{1}{|l|}{$V_{\mathrm{eff}}^{\mathrm{rel+cm}}$ obs non-trans. $^*$} & \multicolumn{1}{l|}{$2.525\pm 0.001$} & \multicolumn{1}{l|}{$-2.12\pm 0.06$}        & $0.99\pm 0.01$ \\ \hline
\end{tabular}
\label{Tab:Fit:kineticenergy}
\caption{Fitting parameters (fitting function $f(n)=c+\frac{A}{n^\sigma}$) for the kinetic energy, see Fig.~\ref{fig:kineticenergy}. The data in the range from $n=11$ to $n=31$ are used for all fits except the ones marked with  $^*$, where $n=17-31$. The uncertainties given are the uncertainties from the fitting procedure.}
\end{table*}

{\it Kinetic energy.}  
Table~\ref{Tab:Fit:kineticenergy} summarizes our fitting procedure for the kinetic energy, which is demonstrated in Fig.~\ref{fig:kineticenergy}. For the attractive $1+1$ system, we see that the bare contact interaction features the slowest convergence rate, i.e., the smallest value of $\sigma$. Transforming the observable together with the interaction speeds up convergence: For $V_{\mathrm{eff}}^{\mathrm{rel}}$ the convergence rate is nearly doubled while for $V_{\mathrm{eff}}^{\mathrm{rel+cm}}$ the calculated data are constant by construction of the interaction matrix and of the induced two-body elements, see Eq.~\eqref{eq:OneBodyOperatorTransformed}. For the repulsive $1+1$ system, the advantage of transforming the operator is even more pronounced: In these cases the renormalized interactions without any transformation of the operator perform worse than $g\delta(x)$, in particular, the corresponding values of  $\sigma$ are smaller. 

For the attractive $1+3$ system, our fitting procedure features large systematic uncertainties, making it impossible to fit all of the data except for $V_{\mathrm{eff}}^{\mathrm{rel+cm}}$ with a transformed kinetic energy operator. In that case a large value of $\sigma$ implies that the data are nearly converged for relatively small values of $n$. For the repulsive $1+3$ system, the fitting procedure works similarly to the $1+1$ case discussed above. The use of the renormalized interactions does not lead to any improvement over the bare contact interaction if the observable is not transformed. By contrast, if we transform the observable together with the potential, the renormalized interactions is clearly more useful than the bare interaction.

\begin{table*}
\begin{tabular}{|llll|}
\hline
\multicolumn{1}{|l|}{}                   & \multicolumn{1}{l|}{$c$}                   & \multicolumn{1}{l|}{$A$}                     & $\sigma$                      \\ \hline \hline
\multicolumn{4}{|l|}{$\langle W\rangle, N_\uparrow=1, N_\downarrow=1, g=-2.5067$ (corresponds to $E_{1+1}=-1$)}                                                                                                       \\ \hline
\multicolumn{1}{|l|}{$g\delta$}          & \multicolumn{1}{l|}{$0.31782\pm 9\times 10^{-5}$}  & \multicolumn{1}{l|}{$0.219\pm 0.002$}    & $0.863\pm 0.004$  \\ \hline
\multicolumn{1}{|l|}{$g(n)\delta$}       & \multicolumn{1}{l|}{$0.3153\pm 2\times 10^{-4}$}  & \multicolumn{1}{l|}{$0.480\pm 0.003$}    & $1.295\pm 0.003$              \\ \hline
\multicolumn{1}{|l|}{$V_{\mathrm{eff}}^{\mathrm{rel}}$ obs trans.}          & \multicolumn{1}{l|}{$0.31474\pm 2\times 10^{-5}$}  & \multicolumn{1}{l|}{$0.480\pm 0.001$} & $1.191\pm 0.003$            \\ \hline
\multicolumn{1}{|l|}{$V_{\mathrm{eff}}^{\mathrm{rel+cm}}$ obs trans.}            & \multicolumn{3}{l|}{Constant by construction. Exact result: $c=0.31472$}                                                                    \\ \hline
\multicolumn{1}{|l|}{$V_{\mathrm{eff}}^{\mathrm{rel}}$ obs non-trans.} & \multicolumn{1}{l|}{$0.31515\pm 1\times 10^{-5}$}   & \multicolumn{1}{l|}{$ 0.459\pm 0.003$}      & $1.233\pm 0.001$              \\ \hline
\multicolumn{1}{|l|}{$V_{\mathrm{eff}}^{\mathrm{rel+cm}}$ obs non-trans.} & \multicolumn{1}{l|}{$0.31492\pm 2\times 10^{-5}$}   & \multicolumn{1}{l|}{$ 0.377\pm 0.007$}      & $1.672\pm 0.009$              \\ 
\hline \hline
\multicolumn{4}{|l|}{$\langle W\rangle, N_\uparrow=1, N_\downarrow=1, g=3$ (corresponds to $E_{1+1}=1.6$)}                                                                                                             \\ \hline
\multicolumn{1}{|l|}{$g\delta$}          & \multicolumn{1}{l|}{$0.7399\pm 3\times 10^{-4}$}  & \multicolumn{1}{l|}{$0.302\pm 0.006$}       & $0.89\pm0.01$               \\ \hline
\multicolumn{1}{|l|}{$g(n)\delta$}       & \multicolumn{1}{l|}{$0.73265\pm 1\times 10^{-5}$}   & \multicolumn{1}{l|}{$ 0.168\pm0.005$}       & $1.66\pm0.01$               \\ \hline
\multicolumn{1}{|l|}{$V_{\mathrm{eff}}^{\mathrm{rel}}$ obs trans.}          & \multicolumn{1}{l|}{$0.73277\pm 2\times 10^{-5}$}  & \multicolumn{1}{l|}{$0.153\pm0.004$}       & $1.41\pm0.01$                \\ \hline
\multicolumn{1}{|l|}{$V_{\mathrm{eff}}^{\mathrm{rel+cm}}$ obs trans.}            & \multicolumn{3}{l|}{Constant by construction. Exact result: $c=0.732404$}                                                                    \\ \hline

\multicolumn{1}{|l|}{$V_{\mathrm{eff}}^{\mathrm{rel}}$ obs non-trans.}          & \multicolumn{1}{l|}{$0.73278
\pm 2\times 10^{-5}$}  & \multicolumn{1}{l|}{$0.143\pm0.004$}       & $1.42\pm0.01$                 \\ \hline
\multicolumn{1}{|l|}{$V_{\mathrm{eff}}^{\mathrm{rel+cm}}$ obs non-trans.}          & \multicolumn{1}{l|}{$0.732442\pm 5\times 10^{-6}$}  & \multicolumn{1}{l|}{$0.088\pm0.005$}       & $1.93\pm0.03$                 \\ \hline \hline
\multicolumn{4}{|l|}{$\langle W\rangle, N_\uparrow=1, N_\downarrow=3, g=-2.5067$ (corresponds to $E_{1+1}=-1$)}                                                                                                       \\ \hline
\multicolumn{1}{|l|}{$g\delta$ $^*$}          & \multicolumn{1}{l|}{$ 1.99
\pm 0.02$}  & \multicolumn{1}{l|}{$0.26\pm0.01$}       & $0.24\pm0.04$                                                                                   \\ \hline
\multicolumn{1}{|l|}{$g(n)\delta$}       & \multicolumn{3}{l|}{Data could not be fitted by Eq.~\eqref{eq:fitfunc}.}                                                                                  \\ \hline
\multicolumn{1}{|l|}{$V_{\mathrm{eff}}^{\mathrm{rel}}$ obs trans. $^*$}          & \multicolumn{1}{l|}{$ 2.01
\pm 0.03$}  & \multicolumn{1}{l|}{$0.17\pm0.01$}       & $0.24\pm0.08$                                                                     \\ \hline
\multicolumn{1}{|l|}{$V_{\mathrm{eff}}^{\mathrm{rel+cm}}$ obs trans. $^*$}          & \multicolumn{1}{l|}{$2.070624\pm7\times 10^{-6}$}     & \multicolumn{1}{l|}{$-3000\pm 800$}        & $5.13\pm 0.09$ \\ \hline
\multicolumn{1}{|l|}{$V_{\mathrm{eff}}^{\mathrm{rel}}$ obs non-trans. $^*$}  & \multicolumn{1}{l|}{$1.9\pm 0.1$}     & \multicolumn{1}{l|}{$0.2\pm 0.1$}        & $0.12\pm 0.09$                                                                   \\ \hline
\multicolumn{1}{|l|}{$V_{\mathrm{eff}}^{\mathrm{rel+cm}}$ obs non-trans. $^*$} & \multicolumn{1}{l|}{$2.07192\pm 3\times10^{-5}$}     & \multicolumn{1}{l|}{$8\times10^{5}\pm 2\times10^{5}$}        & $6.9\pm 0.3$                                                                   \\ \hline
\multicolumn{4}{|l|}{$\langle W\rangle, N_\uparrow=1, N_\downarrow=3, g=3$ (corresponds to $E_{1+1}=1.6$)}                                                                                                             \\ \hline
\multicolumn{1}{|l|}{$g\delta$ $^*$}         & \multicolumn{1}{l|}{$3.0307\pm 8\times 10^{-4}$}   & \multicolumn{1}{l|}{$2.16\pm 0.09$}        & $1.18\pm 0.02$                \\ \hline
\multicolumn{1}{|l|}{$g(n)\delta$ $^*$}      & \multicolumn{1}{l|}{$3.00419\pm 4\times 10^{-5}$} & \multicolumn{1}{l|}{$3.09\pm 0.07$}        & $1.918\pm 0.009$ \\ \hline
\multicolumn{1}{|l|}{$V_{\mathrm{eff}}^{\mathrm{rel}}$ obs trans. $^*$}         & \multicolumn{1}{l|}{$3.00227\pm 1\times 10^{-5}$}  & \multicolumn{1}{l|}{$2.21\pm 0.08$}        & $1.77\pm 0.02$                \\ \hline
\multicolumn{1}{|l|}{$V_{\mathrm{eff}}^{\mathrm{rel+cm}}$ obs trans. $^*$}          & \multicolumn{1}{l|}{$3.00227\pm1\times 10^{-5}$}     & \multicolumn{1}{l|}{$1.6\pm 0.4$}        & $2.66\pm 0.09$ \\ \hline
\multicolumn{1}{|l|}{$V_{\mathrm{eff}}^{\mathrm{rel}}$ obs non-trans. $^*$} & \multicolumn{1}{l|}{$3.00460\pm 8\times 10^{-5}$} & \multicolumn{1}{l|}{$2.21\pm 0.09$}        & $1.78\pm 0.02$ \\ \hline
\multicolumn{1}{|l|}{$V_{\mathrm{eff}}^{\mathrm{rel+cm}}$ obs non-trans. $^*$} & \multicolumn{1}{l|}{$3.00241\pm 2\times10^{-5}$} & \multicolumn{1}{l|}{$2.7\pm 0.3$}        & $2.46\pm 0.04$ \\ \hline
\end{tabular}
\caption{Fitting parameters (fitting function $f(n)=c+\frac{A}{n^\sigma}$) for the harmonic trapping potential, see Fig.~\ref{fig:trap}. The data in the range from $n=11$ to $n=31$ are used for all fits except the ones marked with  $^*$, where $n=17-31$. The uncertainties given are the uncertainties from the fitting procedure.}
\label{Tab:Fit:trap}
\end{table*}

{\it Harmonic trap.} Next, we discuss the second part of the one-body Hamiltonian, the harmonic trapping confiment\footnote{Note that in general  the confiment does not define a useful observable. However, for a harmonic oscillator, the expectation value of the trapping confiment corresponds to the mean-square radius -- a standard observable in cold-atom systems.}. The plots with the data are presented in the main text in Fig.~\ref{fig:trap} and the outcome of our fitting analysis is summarized in Tab.~\ref{Tab:Fit:trap}. As discussed in the main text, the trapping confiment is less sensitive to the transformation of observables than the kinetic energy.
We can further see that for $1+1$ the convergence rates for attractive and repulsive interactions are similar and that the renormalized interactions provide a slight boost in performance. 

Even without a transformation of the harmonic trap, the use of the operator $V_{\mathrm{eff}}^{\mathrm{rel+cm}}$ leads to a fast convergence. As in all considered cases, $V_{\mathrm{eff}}^{\mathrm{rel+cm}}$ clearly outperforms other potentials. At the same time, the running coupling constant demonstrates the weakest performance. 
In particular, for the $1+3$ system with attractive interactions, it does not produce the data that can be fit with Eq.~(\ref{eq:fitfunc}) for the used values of $n$. Note that the convergence of the data calculated with  $V_{\mathrm{eff}}^{\mathrm{rel}}$ is slow ($\sigma=0.24\pm0.08$, comparable to the rate of the bare contact interaction), making this potential suboptimal for the study of this observable.  
For the repulsive $1+3$ system, all potentials can in principle be used: Their convergence rates are comparable. Still, $V_{\mathrm{eff}}^{\mathrm{rel+cm}}$ yields the most accurate result.

\subsection{Transition matrix elements}

\begin{table*}
\begin{tabular}{|llll|}
\hline
\multicolumn{1}{|l|}{}                   & \multicolumn{1}{l|}{$c$}                   & \multicolumn{1}{l|}{$A$}                     & $\sigma$                      \\ \hline \hline
\multicolumn{4}{|l|}{$M^{1+1}_{0\to f}, g=-2.5067$ (corresponds to $E_{1+1}=-1$)}                                                                                                       \\ \hline
\multicolumn{1}{|l|}{$g\delta$}          & \multicolumn{1}{l|}{$0.4367\pm 5\times 10^{-4}$}   & \multicolumn{1}{l|}{$-0.3490\pm 1\times 10^{-4}$}    & $0.377 \pm 0.002$            \\ \hline
\multicolumn{1}{|l|}{$g(n)\delta$}       & \multicolumn{1}{l|}{$0.4283\pm 5\times 10^{-4}$}  & \multicolumn{1}{l|}{$-0.4348\pm 6\times 10^{-4}$}    & $0.449\pm 0.002$              \\ \hline
\multicolumn{1}{|l|}{$V_{\mathrm{eff}}^{\mathrm{rel}}$ obs trans.}          & \multicolumn{1}{l|}{$0.4202\pm 2\times 10^{-4}$}  & \multicolumn{1}{l|}{$-0.1642\pm 8\times 10^{-4}$} & $0.983\pm 0.003$            \\ \hline
\multicolumn{1}{|l|}{$V_{\mathrm{eff}}^{\mathrm{rel+cm}}$ obs trans.}            & \multicolumn{3}{l|}{Constant by construction. Exact result: $c=0.4199$}                                                                    \\ \hline
\multicolumn{1}{|l|}{$g(n)\delta$ obs non-trans.} & \multicolumn{1}{l|}{$0.4181\pm 5\times 10^{-4}$}   & \multicolumn{1}{l|}{$-0.279\pm 0.001$}      & $0.544\pm 0.005$              \\ \hline
\multicolumn{1}{|l|}{$V_{\mathrm{eff}}^{\mathrm{rel}}$ obs non-trans.} & \multicolumn{1}{l|}{$0.4271\pm 4\times 10^{-4}$}   & \multicolumn{1}{l|}{$-0.42149\pm 4\times 10^{-4}$}      & $0.448\pm 0.001$              \\ \hline
\multicolumn{1}{|l|}{$V_{\mathrm{eff}}^{\mathrm{rel+cm}}$ obs non-trans.} & \multicolumn{1}{l|}{$0.411\pm 0.001$}   & \multicolumn{1}{l|}{$-0.485\pm 0.003$}      & $0.5421\pm 0.005$              \\ 
\hline \hline
\multicolumn{4}{|l|}{$M^{1+1}_{0\to f}, g=3$ (corresponds to $E_{1+1}=1.6$)}                                                                                                             \\ \hline
\multicolumn{1}{|l|}{$g\delta$}          & \multicolumn{1}{l|}{$0.0982\pm 1\times 10^{-4}$}  & \multicolumn{1}{l|}{$0.137\pm0.002$}       & $0.785\pm0.007$               \\ \hline
\multicolumn{1}{|l|}{$g(n)\delta$}       & \multicolumn{1}{l|}{$0.101\pm 0.002$}   & \multicolumn{1}{l|}{$0.255\pm0.002$}       & $0.744\pm0.006$               \\ \hline
\multicolumn{1}{|l|}{$V_{\mathrm{eff}}^{\mathrm{rel}}$ obs trans.}          & \multicolumn{1}{l|}{$0.094920\pm 6\times 10^{-6}$}  & \multicolumn{1}{l|}{$0.0209\pm4\times 10^{-4}$}       & $1.19\pm0.01$                \\ \hline
\multicolumn{1}{|l|}{$V_{\mathrm{eff}}^{\mathrm{rel+cm}}$ obs trans.}            & \multicolumn{3}{l|}{Constant by construction. Exact result: $0.094829$}                                                                    \\ \hline
\multicolumn{1}{|l|}{$g(n)\delta$ obs non-trans.} & \multicolumn{1}{l|}{$0.0977\pm 1\times 10^{-4}$}  & \multicolumn{1}{l|}{$0.1034\pm9\times 10^{-4}$}     & $0.695\pm0.006$               \\ \hline

\multicolumn{1}{|l|}{$V_{\mathrm{eff}}^{\mathrm{rel}}$ obs non-trans.}          & \multicolumn{1}{l|}{$0.1014\pm 3\times 10^{-4}$}  & \multicolumn{1}{l|}{$0.263\pm0.003$}       & $0.754\pm0.006$                 \\ \hline
\multicolumn{1}{|l|}{$V_{\mathrm{eff}}^{\mathrm{rel+cm}}$ obs non-trans.}          & \multicolumn{1}{l|}{$0.1046\pm 4\times 10^{-4}$}  & \multicolumn{1}{l|}{$0.355\pm0.009$}       & $0.89\pm0.01$                 \\ \hline \hline
\multicolumn{4}{|l|}{$M^{1+3}_{0\to f}, g=-2.5067$ (corresponds to $E_{1+1}=-1$)}                                                                                                       \\ \hline
\multicolumn{1}{|l|}{$g\delta$}          & \multicolumn{3}{l|}{Data could not be fitted by Eq.~\eqref{eq:fitfunc}.}                                                                                  \\ \hline
\multicolumn{1}{|l|}{$g(n)\delta$}       & \multicolumn{3}{l|}{Data could not be fitted by Eq.~\eqref{eq:fitfunc}.}                                                                                  \\ \hline
\multicolumn{1}{|l|}{$V_{\mathrm{eff}}^{\mathrm{rel}}$ obs trans.}          & \multicolumn{3}{l|}{Data could not be fitted by Eq.~\eqref{eq:fitfunc}.}                                                                                  \\ \hline
\multicolumn{1}{|l|}{$V_{\mathrm{eff}}^{\mathrm{rel+cm}}$ obs trans.}          & \multicolumn{1}{l|}{$0.1518\pm3\times 10^{-4}$}     & \multicolumn{1}{l|}{$47\pm 7$}        & $2.92\pm 0.07$ \\ \hline
\multicolumn{1}{|l|}{$g(n)\delta$ obs non-trans.} & \multicolumn{3}{l|}{Data could not be fitted by Eq.~\eqref{eq:fitfunc}.}                                                                                  \\ \hline
\multicolumn{1}{|l|}{$V_{\mathrm{eff}}^{\mathrm{rel}}$ obs non-trans.} & \multicolumn{3}{l|}{Data could not be fitted by Eq.~\eqref{eq:fitfunc}.}                                                                                  \\ \hline
\multicolumn{1}{|l|}{$V_{\mathrm{eff}}^{\mathrm{rel+cm}}$ obs non-trans.} & \multicolumn{3}{l|}{Data could not be fitted by Eq.~\eqref{eq:fitfunc}.}                                                                                  \\ \hline \hline
\multicolumn{4}{|l|}{$M^{1+3}_{0\to f}, g=3$ (corresponds to $E_{1+1}=1.6$)}                                                                                                             \\ \hline
\multicolumn{1}{|l|}{$g\delta$}          & \multicolumn{1}{l|}{$0.1256\pm 2\times 10^{-4}$}   & \multicolumn{1}{l|}{$1.22\pm 0.03$}        & $1.34\pm 0.01$                \\ \hline
\multicolumn{1}{|l|}{$g(n)\delta$}       & \multicolumn{1}{l|}{$0.1265\pm 6\times 10^{-4}$} & \multicolumn{1}{l|}{$0.77\pm 0.03$}        & $1.08\pm 0.02$ \\ \hline
\multicolumn{1}{|l|}{$V_{\mathrm{eff}}^{\mathrm{rel}}$ obs trans.}          & \multicolumn{1}{l|}{$0.1124\pm 3\times 10^{-4}$}  & \multicolumn{1}{l|}{$0.30\pm 0.07$}        & $1.4\pm 0.1$                \\ \hline
\multicolumn{1}{|l|}{$V_{\mathrm{eff}}^{\mathrm{rel+cm}}$ obs trans.}          & \multicolumn{1}{l|}{$0.11239\pm7\times 10^{-5}$}     & \multicolumn{1}{l|}{$0.2\pm 0.2$}        & $2.1\pm 0.4$ \\ \hline
\multicolumn{1}{|l|}{$g(n)\delta$ obs non-trans.} & \multicolumn{1}{l|}{$0.1198\pm 5\times 10^{-4}$} & \multicolumn{1}{l|}{$0.50\pm 0.03$}        & $1.16\pm 0.03$ \\ \hline
\multicolumn{1}{|l|}{$V_{\mathrm{eff}}^{\mathrm{rel}}$ obs non-trans.} & \multicolumn{1}{l|}{$0.2624\pm 6\times 10^{-4}$} & \multicolumn{1}{l|}{$0.77\pm 0.03$}        & $1.06\pm 0.02$ \\ \hline
\multicolumn{1}{|l|}{$V_{\mathrm{eff}}^{\mathrm{rel+cm}}$ obs non-trans.} & \multicolumn{1}{l|}{$0.126\pm 0.001$} & \multicolumn{1}{l|}{$0.95\pm 0.05$}        & $1.11\pm 0.03$ \\ \hline
\end{tabular}
\caption{Fitting parameters (fitting function $f(n)=c+\frac{A}{n^\sigma}$) for the transition matrix elements, see Fig.~\ref{fig:transmatrix}. The data in the range from $n=11$ to $n=31$ are used for all fits. For $M^{1+3}_{0\to f}, g=-2.5067$ the data could not be fitted except for the $V_{\mathrm{eff}}^{\mathrm{rel+cm}}$ interaction with a transformed observable. The uncertainties given are the uncertainties from the fitting procedure.}
\label{Tab:Fit:TransMatrix}
\end{table*}

Finally, we analyze the data for the transition matrix elements, see Fig~\ref{fig:transmatrix} and Table~\ref{Tab:Fit:TransMatrix}. Already for the $1+1$ system, it is beneficial to transform this observable together with the interaction. For $V_{\mathrm{eff}}^{\mathrm{rel+cm}}$ the result is exact by construction  if we transform the operator with the potential, while for $V_{\mathrm{eff}}^{\mathrm{rel}}$, the data converge significantly faster (i.e., larger $\sigma$) than the bare interaction. For repulsive interactions, we even see that the value of $A$ is small, signaling that the expectation value of the observable in the limit $n\to\infty$ can be estimated in a reliable manner using our data with relatively small values of $n$.

For the repulsive $1+3$ system, the importance to transform the operator becomes even more clear. 
The effective interactions with a transformed operator feature the best performance: the use $V_{\mathrm{eff}}^{\mathrm{rel+cm}}$ leads to the largest convergence rate and also to a small value of $A$, signaling that the data are already nearly converged for the used values of $n$; the transformation of the observable together with $V_{\mathrm{eff}}^{\mathrm{rel}}$ improves the convergence rate slightly and also reduces the value of $A$.
For attractive interaction, the fitting procedure failed for all cases except for $V_{\mathrm{eff}}^{\mathrm{rel+cm}}$ and a transformed observable. The reason for this is most likely that the convergence rate is so slow that the function we use for the fitting procedure has in this case a large systematic error (see also main text). However, for $V_{\mathrm{eff}}^{\mathrm{rel+cm}}$, we see that the convergence rate is reasonably fast ($\sigma=2.92\pm0.07$), allowing us to attain accurate results.

\section{Two-dimensional harmonic oscillator}
\label{App:2DHO}

In this appendix, we first introduce the few-fermion problem in a two-dimensional harmonic oscillator; then, we explain the truncation and subsequent renormalization schemes used in our numerical calculations. Finally,
 we calculate the interaction matrix for the effective interaction approach, $V_{\mathrm{eff}}^{\mathrm{rel}}$.

{\it Hamiltonian.} The Hamiltonian of the system reads:
\begin{equation}
    H=\sum\limits_{i=1}^N\left(-\frac{1}{2}\Vec{\nabla_i}^2+\frac{1}{2}\Vec{x}_i^2\right)+g\sum\limits_{i<j}\delta(\Vec{x_i}-\Vec{x_j}).
    \label{eq:Hamiltonian2D}
\end{equation}
Like in the main text, we consider a system of $N=N_\uparrow+N_\downarrow$ fermions. The interaction cannot flip the polarization of the spins and due to the Pauli exclusion principle only fermions of different polarization interact. 

The eigenfunctions of the one-body part of the Hamiltonian are the eigenfunctions of the two-dimensional harmonic oscillator  $\Phi_i(\Vec{x})=\frac{1}{\sqrt{2\pi}}e^{im_i\phi}F_{n_i, m_i}(\rho)$ with
\begin{equation}
   F_{n_i, m_i}(\rho)=(-1)^{n_i}\sqrt{\frac{2\Gamma(n_i+1)}{\Gamma(|m_i|+n_i+1)}}e^{-\frac{\rho^2}{2}}\rho^{|m_i|}L_{n_i}^{|m_i|}(\rho^2),
   \nonumber
\end{equation}
where $i$ is the index of the basis state; $m_i$ is the angular component of this index and $n_i$ is the radial one. We define the basis in polar coordinates $\rho$ and $\phi$. $L$ is the generalized Laguerre polynomial. The energy of the state $\Phi_i(\Vec{x})$ is 
\begin{equation}
\label{eq:App:OneBodyEnergy2DHO}
    E_{m_i, n_i}=2n_i+|m_i|+1\,.
\end{equation}
In the CI calculations, we build the Hamiltonian matrix using the functions $\Phi_i(\Vec{x})$ as one-body basis states. The size of the basis is defined by a cutoff $n_{2D}$, which fixes the maximal one-body energy via Eq.~\eqref{eq:App:OneBodyEnergy2DHO}.  

{\it Fermion-fermion interactions.} Note that in two dimensions the contact interaction requires regularization (see, e.g., Refs.~\cite{rontani2017renormalization, nyeo2000regularization} and references therein). In the CI method, it is naturally regularized by the use of a finite one-body basis. In practice, we calculate the two-body energy for each cutoff and use it to parameterize the strength of fermion-fermion interactions in the calculations of other quantities, e.g., the ground-state energy of systems with more particles. This renormalization scheme is somewhat similar to the running coupling constant approach discussed in Sec.~\ref{sec:Runningcoupling}. Our implementation of this approach in two dimensions follows Ref.~\cite{rontani2017renormalization}.

To build an operator of effective interactions using ideas of Sec.~\ref{subsec:effective_interactions}, we need to know the exact solution to the two-body problem. This solution can be obtained analytically (see Refs.~\cite{busch1998, avakian1987spectroscopy} and below) for a system with a regularized pseudopotential~\cite{Olshanii2001}:
\begin{equation}
    V_{\mathrm{Pseudo}}=-\frac{2\pi \delta(\Vec{x})}{\mathrm{ln}(\mathcal{A} a\Lambda)}\left[1-\mathrm{ln}(\mathcal{A}\Lambda \rho)\rho\frac{\partial}{\partial \rho}\right]_{\rho\to 0^+},
\end{equation}
with $\Lambda$ being an arbitrary constant, $\mathcal{A}=e^\gamma/2$ where $\gamma$ is the Euler-Mascheroni constant, and $a$ is the scattering length in two dimensions. 

{\it Two-body solution.}  To write down the two-body solution, we first notice that the Schr{\"o}dinger equation factorizes in the center-of-mass $\Vec{X}=\frac{1}{\sqrt{2}}(\Vec{x_1}+\Vec{x_2})$ and relative coordinates $\Vec{x}=\frac{1}{\sqrt{2}}(\Vec{x_1}-\Vec{x_2})$. The interaction enters only via the latter. Furthermore, because we work with a delta-function potential, only relative states with angular momentum that equal to zero are influenced by it. All other states are the eigenstates of the harmonic oscillator. 

To find the states with vanishing angular momentum, we solve the relative part of the Schr{\"o}dinger equation in polar coordinates: 
\begin{equation}
    \left(-\frac{\nabla_\rho^2}{2}+\frac{\rho^2}{2}+V_{\mathrm{Pseudo}}\right)\Psi_i(\rho)=\epsilon_i\Psi(\rho).
\end{equation}
The solution depends only on the radial coordinate:
\begin{equation}
    \Psi_i(\rho)=\frac{1}{\sqrt{2\pi}}\mathcal{N}_{\nu_i}\Gamma(-\nu_i)e^{-\rho^2/2}U(-\nu_i, 1, \rho^2),
\end{equation}
where $U$ is the Tricomi function \cite{abramowitz1948handbook} and $\Gamma$ is the gamma function. $\nu_i$ is determined from the equation
\begin{equation}
    \psi(-\nu)=\mathrm{ln}(2/ a^2)-2\gamma\,
\end{equation}
where $\psi$ is the digamma function.
The energy of this system is given by $\epsilon_i=2\nu_i+1$. The normalization constant $\mathcal{N}_\nu$ is
\begin{equation}
    \mathcal{N}_\nu=\frac{\sqrt{2}}{\sqrt{\psi'(-\nu)}}.
\end{equation}
Here, $\psi'$ is the trigamma function.

{\it Effective interaction.} Before we start building the effective interaction matrix we  note two things. First, values of the Talmi-Moshinsky-Smirnov brackets,
\begin{equation}
    \alpha_{ij, ab}=\int d\Vec{x}_1\int d \Vec{x}_2 \Phi_i(\Vec{x}_1)\Phi_j(\Vec{x}_2)\Phi_a(\Vec{x})\Phi_b(\Vec{X}),
\end{equation}
are known in two dimensions~\cite{Wenying_1997, CremonPhDThesis}. Second, the overlap integral,
\begin{equation}
    U_{ij}=\int d \Vec{x} \Phi_i(\Vec{x})\Psi_j(\Vec{x}),
\end{equation}
can be calculated analytically. Indeed, let use the fact that $\Psi_i(\Vec{x})$ solves the Schr{\"o}dinger equation with the pseudopotential and rewrite the expression above as:
\begin{equation}
    U_{ij}= \frac{\int d \Vec{x} V_{\mathrm{Pseudo}} \Phi_i(\Vec{x})\Psi_j(\Vec{x})}{\epsilon_{\nu_j}-T_i}
\end{equation}
with $T_i=2i+1$ being the energy of the non-interacting system. Next, by using that $U(-\nu, 1, \rho^2)\to-(2\mathrm{ln}(\rho)+2\gamma+\psi(-\nu))/\Gamma(-\nu)$ in the limit $\rho\to0$, we arrive at:
\begin{equation}
    U_{ij}=\sqrt{2}\pi\frac{\Phi_i(0)\mathcal{N}_{\nu_j}}{\epsilon_{\nu_j}-T_i}.
\end{equation}

Using the Talmi-Moshinsky-Smirnov brackets, orthogonality of the harmonic oscillator eigenfunctions and that the interaction is of $s$-wave character, we calculate $V_{ijkl}$ (cf. Eq.~(\ref{eq:hamiltonian_second}))
\begin{widetext}
\begin{equation}
    \begin{split}
    V_{ijkl}&=\sum\limits_{a, b, a_1, b_1}\alpha^\dagger_{ij, ab}\alpha_{kl, a_1b_1}\int d\Vec{x}\int d\Vec{X}\Phi_a^\dagger(\Vec{x})\Phi_b^\dagger(\Vec{X})V(\Vec{x})\Phi_{a_1}(\Vec{x})\Phi_{b_1}(\Vec{X})\\
    &=\sum\limits_{a, b, a_1}\alpha^\dagger_{ij, ab}\alpha_{kl, a_1 b}\delta_{m_a, 0}\delta_{m_{a_1}, 0}v_{aa_{1}},
    \end{split}
\end{equation}
\end{widetext}
where $V(\Vec{x})$ describes the fermion-fermion interaction. We express it in terms of $v_{aa_{1}}$, which describes the effective interaction in relative coordinates. 

To write $v_{aa_{1}}$, we use Eq.~(\ref{eq:Veff})
\begin{equation}
    v^{\mathrm{eff}}=Q^\dagger E Q-T\,, ~~\text{with}~~ Q=\frac{U}{\sqrt{U^\dagger U}}.
\end{equation}
Here, the matrix $T_{ij}=\delta_{ij} (2n_i+|m_i|+1)$ is a diagonal matrix that contains the energies of a non-interacting system, whereas $E$ is a diagonal matrix with the energies of an interacting system. Because only matrix elements with vanishing angular momentum contribute to the interaction, it is sufficient to consider only these. This means that $m_i=0$, and that the matrices $U$, $E$ and $T$ are restricted only by a cutoff in the $n$-quantum number. 
Because center-of-mass excitations can be distributed between two quantum numbers, there is an ambiguity in the cutoffs in the laboratory and center-of-mass reference frames. We found empirically that the choice $n_{\mathrm{rel}}=n_{2D}/2+1$\footnote{Remember that $n_{2D}$ defines the maximal one-body energy; $n_{rel}$ is the highest radial quantum number used to construct the effective interaction in relative coordinates.} is adequate for our calculations. However, because of this ambiguity, we did not find a useful relation for cutoffs for the $V_{\mathrm{eff}}^{\mathrm{rel+cm}}$ interaction. Therefore, we leave a construction of this interaction in 2D to future studies.

\section{Energy for larger particle number}
\label{App:MoreParticles}

Here, we show that the conclusions of the main text also hold for larger particle numbers. To this end, we present in this Appendix additional data for the $1+N$ system, with $N$ up six particles, and the $2+N$ system, with $N$ up to four particles. Due to the increased particle number, we are no longer able to use a maximal cutoff of $n=31$ but instead, we use $n=21$. For the largest particle number considered, this corresponds to more than one million many-body basis states, making an even larger cutoff impossible on a standard computer. As we will show below, this cutoff is insufficient in producing converged results for large particle numbers, however, the main characteristics of the convergence behavior of renormalized interactions becomes clear.  

\begin{figure}
    \centering
    \includegraphics[width=\linewidth]{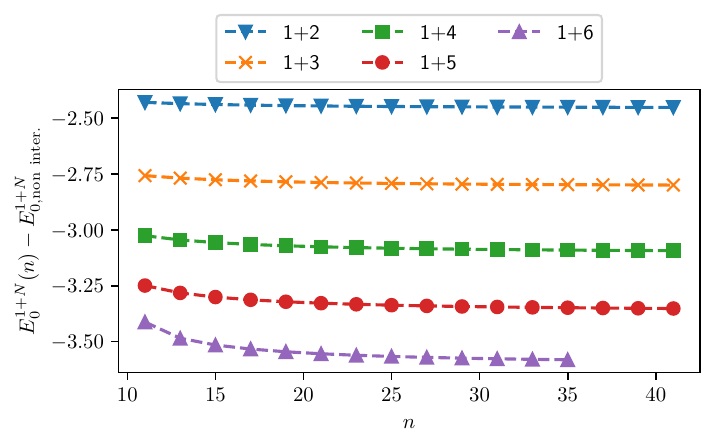}
    \caption{Interaction energy of the $1+N$ system for different particle numbers as a function of the one-body cutoff $n$ calculated by diagonalizing Eq.~\eqref{eq:LLPrel}. We consider attractive interactions with $g=-2.5067$ (which corresponds to $E_{1+1}=-1$). Numerical data are shown with markers of different shape for the different particle numbers; lines are added to guide the eye. 
    }
    \label{fig:LLPonly}
\end{figure}

\subsection{Calculation in relative coordinates}

To judge the performance of the $1+N$ calculations, we perform an additional calculation which uses the symmetries of our system by transforming the Hamiltonian to frame of reference centered at the impurity (similar to the Lee-Low-Pines transformation in real space), so that instead of the $1+N$ system with contact interaction, we solve a $N$-body Schrödinger equation. The corresponding Hamiltonian in the center-of-mass and relative coordinates reads:
\begin{widetext}
\begin{align} 
H^{\mathrm{cm}}&=-\frac{\hbar}{2 N}\frac{\partial^2}{\partial y^2}+\frac{N}{2}y^2,\\ 
H^{\mathrm{rel}}&=-\sum\limits_{i=1}^{N}\frac{\partial^2}{\partial x_i^2}+\frac{1}{2}\frac{N-1}{N}\sum\limits_{i=1}^{N} x_i^2+\sum\limits_{i=1}^{N}g\delta(x_i)+\sum\limits_{i, j}V(x_i, x_j),
\label{eq:LLPrel}
\end{align}
\end{widetext}
where
$V(x_i, x_j)=-\frac{1}{2}\partial_{x_i}\partial_{x_j}-\frac{1}{N+1}x_ix_j$. The solutions of the center-of-mass part are simply the eigenstates of the harmonic oscillator. To construct a Hamiltonian matrix from $H^{\mathrm{rel}}$, we use as one-body basis states the exact solution of the one-body part of the Hamiltonian from Eq.~\eqref{eq:ExactSolution}. For more details on this approach we refer the interested reader to Ref.~\cite{Wlodzinsky2022}.

Note that as we have effectively decreased the particle number by one, this approach allows us to use a one-body basis cutoff of $n=37$ for the $1+6$ system and $n=41$ for lower particle numbers. Furthermore, the basis already includes the information about the delta-function interaction, allowing us to reach nearly converged results as we show in Fig.~\ref{fig:LLPonly}. We extrapolate the data to $n\to\infty$ using the fitting function described above, Eq.~\eqref{eq:fit_function}. This produces a benchmark for our calculations with effective interactions.

\subsection{Energy convergence $1+N$ system}

\begin{figure}
    \centering
    \includegraphics[width=\linewidth]{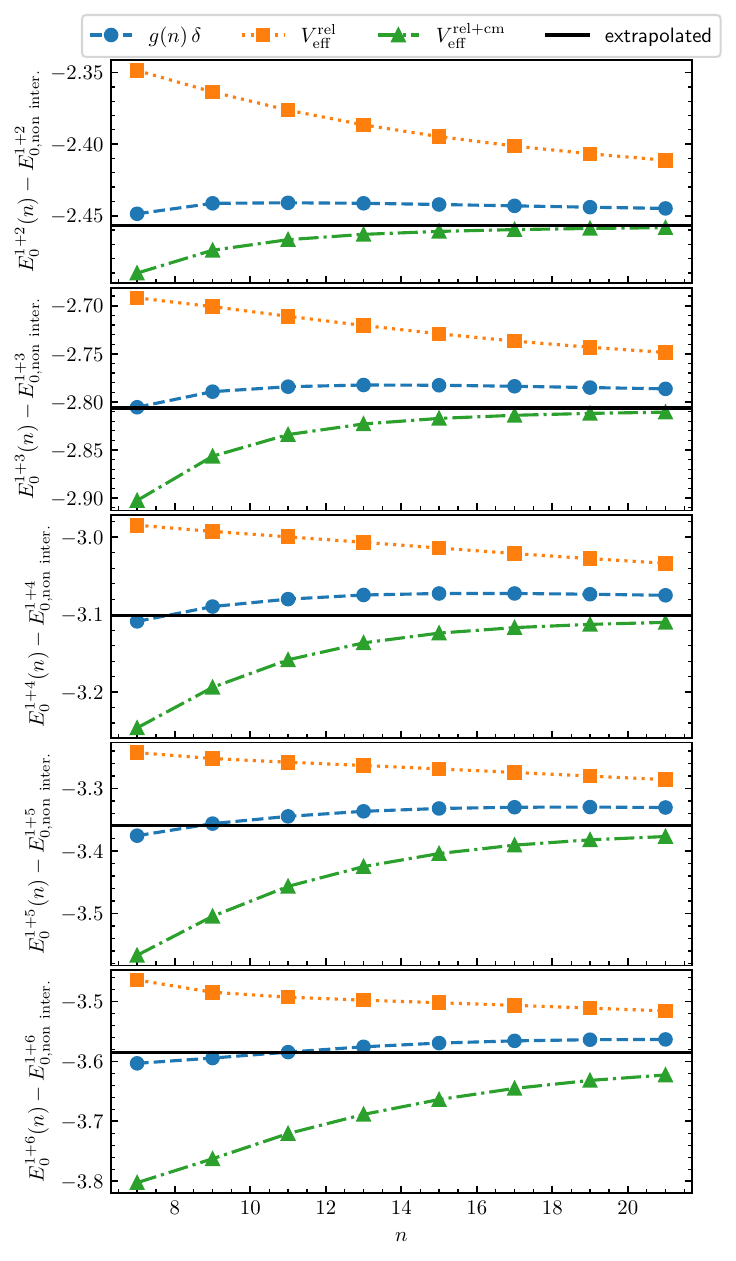}
    \caption{Interaction energy of the $1+N$ system for different particle numbers as a function of the one-body cutoff $n$ using the renormalized interactions. We consider attractive interactions with $g=-2.5067$ (which corresponds to $E_{1+1}=-1$). Numerical data are shown with markers of different shape for the different particle numbers; dashed lines are added to guide the eye. The black solid line corresponds to an extrapolation of the data shown in Fig.~\ref{fig:LLPonly} using the fitting function from Eq.~\eqref{eq:fit_function}. Different panels correspond to different particle numbers. 
    }
    \label{fig:Energy_1+N}
\end{figure}

In Fig.~\ref{fig:Energy_1+N}, we show the interaction energy, $E(g)-E(g=0)$, as a function of the one-body cutoff, $n$, for $1+N$ particles. We only present the data for the renormalized interactions as the energy computed with the bare contact interaction converges considerably slower. Additionally, we also present the energies obtained by extrapolating the data in Fig.~\ref{fig:LLPonly} to $n\to\infty$. These energies are used as a benchmark. 

We see that the convergence behavior discussed in the main text also holds for larger systems. In particular, for the $1+2$, $1+3$, $1+4$ and $1+5$ systems the $V_{\mathrm{eff}}^{\mathrm{rel+cm}}$ interaction leads to the results that are the closest to the benchmark values. For the $1+6$ system, one can conclude naively that  
the running coupling constant approach, $g(n)$, outperforms other methods. However, it is worthwhile noting that the corresponding energies are increasing as a function of $n$ for the considered cutoff parameters. We conclude that for the $1+6$ system more one-body basis states are needed before the energies calculated with the running coupling constant approach reach maximum and start to decrease to the correct value. 

\subsection{Energy convergence $2+N$ system}

\begin{figure}
    \centering
    \includegraphics[width=\linewidth]{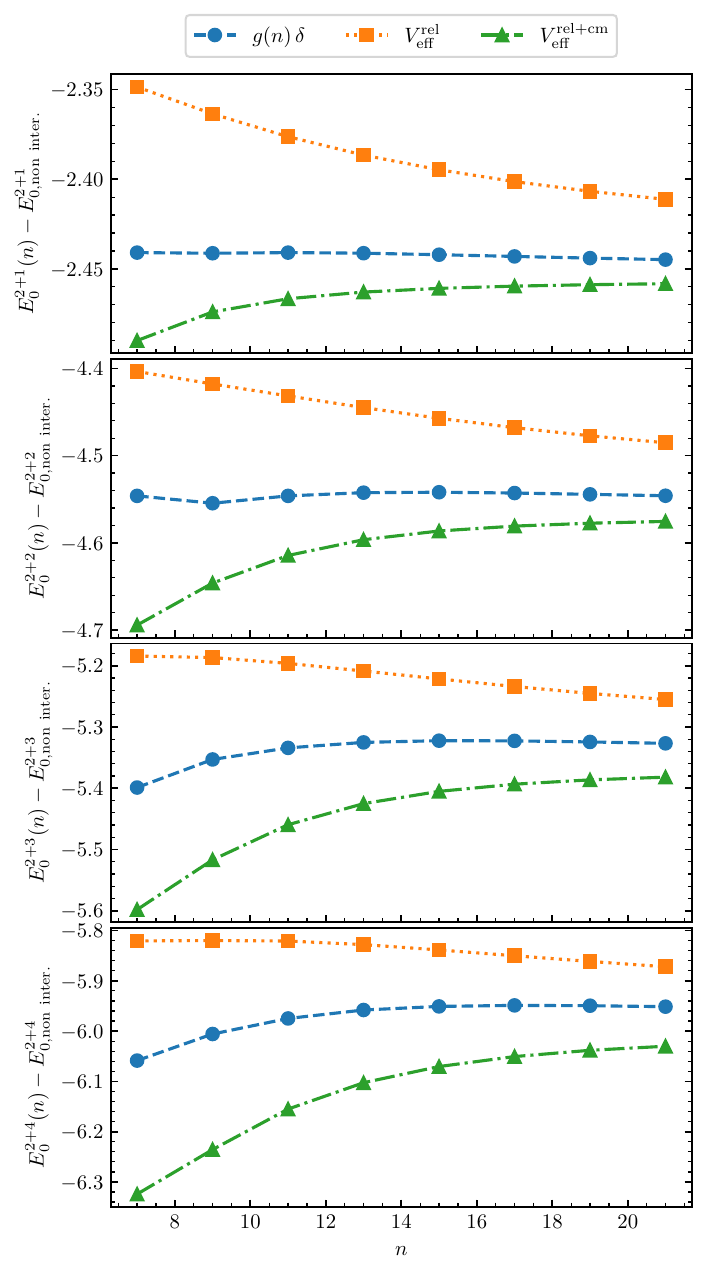}
    \caption{Interaction energy of the $2+N$ system for different particle numbers as a function of the one-body cutoff $n$ using the renormalized interactions. We consider attractive interactions with $g=-2.5067$ (which corresponds to $E_{1+1}=-1$). Numerical data are shown with markers of different shape for the different particle numbers; dashed lines are added to guide the eye. Different panels correspond to different particle numbers. 
    }
    \label{fig:Energy_2+N}
\end{figure}

Finally, we present the ground-state energies for the $2+N$ system in Fig.~\ref{fig:Energy_2+N}. These data showcases that the our conclusions also hold for systems that cannot be represented as an impurity in a Fermi gas. In this case, we have no comparison data for infinite cutoff. Nevertheless, we can see that the general convergence behavior is similar to the one for $1+N$ systems.

\end{document}


\title{Supplementary material for: Comparison of renormalized interactions using one-dimensional few-body systems as a testbed
}

	\author{Fabian Brauneis}
	\affiliation{Technische Universit\"{a}t Darmstadt$,$ Department of Physics$,$ 64289 Darmstadt$,$ Germany}
	
	\author{Hans-Werner Hammer}
	\affiliation{Technische Universit\"{a}t Darmstadt$,$ Department of Physics$,$ 64289 Darmstadt$,$ Germany}
	\affiliation{ExtreMe Matter Institute EMMI and Helmholtz Forschungsakademie
  Hessen f\"ur FAIR (HFHF)$,$ GSI Helmholtzzentrum f\"ur Schwerionenforschung GmbH$,$ 64291 Darmstadt$,$ Germany}
	
	\author{Stephanie M. Reimann}
	\affiliation{Division of Mathematical Physics and NanoLund$,$ Lund University$,$ SE-221 00 Lund$,$ Sweden}
	
	\author{Artem G. Volosniev}
	\affiliation{Institute of Science and Technology Austria (ISTA)$,$ Am Campus 1$,$ 3400 Klosterneuburg$,$ Austria} 
   \affiliation{Department of Physics and Astronomy$,$ Aarhus University$ ,$ Ny Munkegade 120$,$ DK-8000 Aarhus C$,$ Denmark}

\begin{abstract}
    This supplementary material contains technical details and additional information that support the discussion in the main text. In Sec.~\ref{sec:App:EffRel}, we showcase how to implement the $V_{\mathrm{eff}}^{\mathrm{rel}}$ interaction, while in Sec.~\ref{sec:App:EffTot}, we explain how $V_{\mathrm{eff}}^{\mathrm{rel+cm}}$ is built. We outline a transformation of operators alongside renormalization of interactions in Sec.~\ref{App:TrafoOperators}. We illustrate the running coupling constant approach in the unconfined system in Sec.~\ref{sec:App:Derivation_momentum_spaceRG}. In Sec.~\ref{sec:App:Derivation_connection_qn_mom}, we connect the momentum space cutoff with the cutoff for a system in a polynomial trapping potential. Finally, in Secs.~\ref{sec:Convergence1_over_n} and~\ref{sec:Convergenceloglog}, we showcase convergence of our data as a function of $1/n$, and using a log-log scale for $1+1$ particles. 
\end{abstract}

\maketitle

\section{Implementation of \texorpdfstring{$V_{\mathrm{eff}}^{\mathrm{rel}}$}{V_eff^rel}}
\label{sec:App:EffRel}

To make the paper self-contained, in this section, we explain the necessary steps to build the interaction matrix, $V_{\mathrm{eff}}^{\mathrm{rel}}$. This discussion follows Ref.~\cite{Rammelmueller2023}. 

We start with the interaction matrix needed for our CI-simulation
\begin{equation}
    V_{ijkl}=\int dx_1\int dx_2 \Phi_i(x_1)\Phi_j(x_2)V(x_1-x_2)\Phi_k(x_1)\Phi_l(x_2).
\end{equation}
Note that the interaction depends only on the relative distance between two particles.  
Further, as the system is in a harmonic oscillator, 
the two-body wavefunction factorizes in relative $x=\frac{1}{\sqrt{2}}(x_1-x_2)$ and center-of-mass $X=\frac{1}{\sqrt{2}}(x_1+x_2)$ coordinates:
\begin{equation}
    \Phi_i(x_1)\Phi_j(x_2)=\sum\limits_{a, b}\alpha_{ij, ab}\Phi_a(x)\Phi_b(X),
\end{equation}
where the expansion coefficients (Talmi-Moshinsky-Smirnov brackets),
\begin{equation}
    \alpha_{ij, ab}=\int dx_1\int dx_2 \Phi_i(x_1)\Phi_j(x_2)\Phi_a(x)\Phi_b(X),
\end{equation}
can be calculated analytically \cite{Rammelmueller2023, moshinsky1969harmonic}.
Using these coefficients, we rewrite the interaction matrix as
\begin{equation}
\label{eq:RelToLabVijkl}
    V_{ijkl}=\sum\limits_{a, a', b}\alpha_{ij, ab}\alpha_{kl, a'b}v_{aa'} \qquad \mathrm{with} \qquad    v_{aa'}=\int dx \Phi_a(x)V(\sqrt{2}x)\Phi_{a'}.
\end{equation}

To build an effective interaction matrix for $v_{aa'}$ utilized in the main text, we employ the Schr{\"o}dinger equation in relative coordinates:
\begin{equation}
    \left(-\frac{1}{2}\frac{\partial^2}{\partial x^2}+\frac{x^2}{2}+\frac{g}{\sqrt{2}}\delta(x)\right)\psi_i(x)=\epsilon_i\psi_i(x),
\end{equation}
solving this equation allows us now to build the effective interaction. 
First, we use its solution (see Refs.~\cite{avakian1987spectroscopy,busch1998} and the main text) to build the diagonal matrix $E_{ij}=\delta_{ij}\epsilon_i$ and to calculate the overlap matrix $U_{ij}=\int dx \psi_i(x)\Phi_j(x)$:
\begin{equation}
\begin{split}
  U_{ij}  =
    \begin{cases}
        -g\frac{\psi_{i}(0)\Phi_j(0)}{j+1/2-\epsilon_{i}} &\text{for $i$ and $j$ even},\\
        \delta_{ij}& \text{for } i \text{ and } j \text { odd },\\
        0& \text{otherwise}.
	\end{cases}
\end{split}
\end{equation}
We use this information to build the effective interaction matrix according to:
\begin{equation}
\label{eq:Veffrel}
    v^{\mathrm{eff}}=Q^\dagger E Q-T\,, ~~\text{with}~~ Q=\frac{U}{\sqrt{U^\dagger U}}
\end{equation}
where we omitted the indices of the matrices for better readability. The matrix $T$ is given by $T_{ij}=\delta_{ij} (i+1/2)$. All matrices $U$, $E$ and $T$ are of dimension $n_{\mathrm{rel}}\times n_{\mathrm{rel}}$ with $n_{\mathrm{rel}}=2n$. $n$ is the cutoff introduced in the main text for the maximal number of one-body basis states in the laboratory frame. The relation between $n$ and $n_{\mathrm{rel}}$ stems from the condition that the maximal two-body energy in the laboratory frame ($2n+1$) for a given one-body cutoff $n$ has to be equal to the maximal two-body energy in relative and center-of-mass coordinates. The matrix $v^{\mathrm{eff}}$ is then transformed via Eq.~\eqref{eq:RelToLabVijkl} to $V_{ijkl}$. 

For the transformation of observables that corresponds to this interaction see App.~\ref{App:ImplementationObsRel}.

\section{Implementation of \texorpdfstring{$V_{\mathrm{eff}}^{\mathrm{rel+cm}}$}{V_eff^rel+cm}}
\label{sec:App:EffTot}

Instead of going first to relative coordinates and building the effective interaction matrix there as in Eq.~(\ref{eq:RelToLabVijkl}), one can engineer the effective interaction matrix using both relative and center-of-mass excitations: 

\begin{equation}
\label{eq:Veffrelcm}
\begin{split}
    V_{ijkl}&=\left[QEQ^\dagger  \right]_{p(i, j), t(k,l)}-(i+l+1)\delta_{i,k}\delta_{j,l}\\
    &=\sum\limits_{r, m} Q_{p(i,j), r}E_{r, m}Q_{m, t(k, l)}^\dagger - (i+l+1)\delta_{i,k}\delta_{j,l}\\
    &=\sum\limits_{r} Q_{p(i,j), r}E_{r, r}Q_{r, t(k, l)}^\dagger - (i+l+1)\delta_{i,k}\delta_{j,l}
\end{split}
\end{equation}
with $E_{r, m}=\delta_{rm}(\epsilon_{a(r)}+b(r)+1/2)$ being the diagonal eigenenergy matrix, here $\epsilon_{a(r)}$ is the eigenenergy of the Hamiltonian in relative coordinates and $b(r)+1/2$ is the eigenenergy of the corresponding center-of-mass part. $p, t, r, m$ are two-body indices build from one-body indices $i, j, k, l, a, b$, e.g. $p=i n + j$ with $n$ the singe-body cutoff introduced in the main text. Note that $i, j, k, l, p, t$ are defined in the laboratory frame; $a, b, m, r$ operate in the space defined by the relative and center-of-mass coordinates. 

To construct the effective interaction, we should calculate the unitary (overlap) matrix 
\begin{equation}
    Q_{pr}=\sum_d\frac{U_{p d}}{\sqrt{\sum_{b}U_{d b}^\dagger U_{b r}}} \qquad  \mathrm{where }   \qquad  U_{r(a, b), t(k,l)}=\int dx_1\int dx_2 \Psi_{r(a,b)}(x_1, x_2)\Phi_k(x_1)\Phi_l(x_2),
\end{equation}
where $\Psi_{r(a,b)}(x_1, x_2)=\psi_{a(r)}(x)\Phi_{b(r)}(X)$; $\psi_{a(r)}(x)$ is the solution of the Schr{\"o}dinger equation in relative coordinates and $\Phi_{b(r)}(X)$ is the harmonic oscillator wave function (see the main text for the corresponding expressions).  
The matrix $U$ can be calculated analytically. Indeed, if $a(r)$ is odd,  then $\psi_{a(r)}(x)=\Phi_{a(r)}(x)$ is an eigenfunction of a non-interacting problem and the overlap is
\begin{equation}
\label{eq:ExactSolutionApp}
    U_{r(a, b), t(k,l)}=\alpha_{ab, kl}.
\end{equation}
If $a(r)$ is even, we use that $\Psi_r(x_1, x_2)$ is a solution of the two-body Schr{\"o}dinger equation to rewrite the integral as follows
\begin{equation}
     U_{r(a, b), t(k,l)}=g\frac{\int dx \Psi_{r(a, b)}(x, x)\Phi_k(x)\Phi_l(x)}{E_{r(a, b), r(a, b)}-(k+l+1)}.
\end{equation}
Now, we use $\Psi_{r(a, b)}(x, x)=\psi_{a(r)}(0)\Phi_{b(r)}(\sqrt{2}x)$ and  $\Phi_k(x)\Phi_l(x)=\sum\limits_{a_1, b_1}\alpha_{kl, a_1 b_1}\Phi_{a_1}(0)\Phi_{b_1}(\sqrt{2}x)$ to arrive at
\begin{equation}
    U_{r(a, b), t(k,l)}=g\psi_{a(r)}(0)\sum\limits_{b_1}\alpha_{kl, (k+l-b_1)b_1}\Phi_{k+l-b_1}(0)\frac{\int dx \Phi_{b(r)}(\sqrt{2}x)\Phi_{b_1}(\sqrt{2}x)}{E_{r(a, b), r(a, b)}-(k+l+1)},
\end{equation}
where we used energy conservation in the transformation from the laboratory to center-of-mass frames, $2a_1+2b_1+1=2k+2l+1$. Using the orthogonality of the harmonic oscillator eigenfunctions, this expression can be further simplified
\begin{equation}
    U_{r(a, b), t(k,l)}=\frac{g\psi_{a(r)}(0)}{\sqrt{2}}\alpha_{kl, (k+l-b(r))b(r)}\frac{\Phi_{k+l-b(r)}(0)}{E_{r(a, b), r(a, b)}-(k+l+1)}.
\end{equation}

Now everything can be combined to calculate the interaction matrix $V_{ijkl}$.
All one-body indices $i, j, k, l, a, b$ are restricted by the cutoff $n$ from our one-body basis and the two-body indices by $n^2$. Therefore the matrices $E, U, Q$ are of dimension $n^2\times n^2$.

For the transformation of observables that corresponds to this interaction see App.~\ref{App:ImplementationObsRelCM}.

\section{Transformation of operators}
\label{App:TrafoOperators}

In this section we derive Eqs.~\eqref{eq:OneBodyOperatorTransformed} and~\eqref{eq:TwobodyOperatorTransformed} that describe the transformation of operators discussed in the main text. 

\subsection{Two-body system}
First, we show that the transformation of the effective interaction is indeed a unitary transformation for the two-body system described by the Hamiltonian $H^{(2)}$\footnote{The subscript $(2)$ denotes that we work with a two-particle system.}. To this end, we first extend the effective two-body Hamiltonian to an infinite Hilbert space

\begin{equation}
H^{\mathrm{eff}, (2)}=  
\begin{pmatrix}
Q_{\mathcal{M}\mathcal{M}}^T & \vline & 0 \\
\hline
0 & \vline & \mathbb{I}
\end{pmatrix}
\begin{pmatrix}
E_{\mathcal{M}\mathcal{M}}^{(2)} & \vline & 0 \\
\hline
0 & \vline & E_{high}^{(2)}
\end{pmatrix}
\begin{pmatrix}
Q_{\mathcal{M}\mathcal{M}} & \vline & 0 \\
\hline
0 & \vline & \mathbb{I}
\end{pmatrix}.
\end{equation}
Here, $E_{\mathcal{M}\mathcal{M}}^{(2)}$ is the low-energy spectrum that we use in our calculations. We add the high-energy spectrum $E_{high}^{(2)}$ here explicitly to introduce a unitary matrix in an infinite Hilbert space, $Q=\begin{pmatrix}
Q_{\mathcal{M}\mathcal{M}} & \vline & 0 \\
\hline
0 & \vline & \mathbb{I}
\end{pmatrix}$.

Next, we define the diagonal matrix that contains eigenvalues of the Hamiltonian $H^{(2)}$
\begin{equation}
    E^{(2)} \equiv UH^{(2)}U^T = \widetilde{H}^{(2)}\footnote{Remember that the tilde shows that the matrix is expanded in the eigenbasis of the interacting system.},
\end{equation}
where as before $U_{ij}=\braket{\Phi_i|\Psi_j}$, here $\Phi_i^{(2)}$ is a two-body basis state, and $\Psi_i$ is the exact two-body solution. 

Equations from above allow us to connect $ H^{\mathrm{eff}, (2)}$ and $H^{(2)}$ via a unitary transformation $\mathcal{W}=Q U^T$. Indeed,
\begin{equation}
    H^{\mathrm{eff}, (2)}=Q^T E^{(2)} Q = Q^T U H^{(2)} U^T Q = \mathcal{W}^T H^{(2)} \mathcal{W}.
\end{equation}
 We see that the effective interaction can be connected to a unitary transformation for the two-body system. Therefore, the matrix $\mathcal{W}$ should be also used to transform any operator of interest.

\subsection{Many-body system}
Let us now consider the many-body case. Using the discussion from the previous subsection, we write the interaction in the form
\begin{equation}
    V^{\mathrm{eff}}_{ij}=\left(\mathcal{W}^T H^{(2)} \mathcal{W}\right)_{ij}-T_i-T_j.
\end{equation}
This expression assumes that the operator only acts on the particles $i$ and $j$. 
The corresponding effective many-body Hamiltonian reads
\begin{equation}
    H^{\mathrm{eff}}=\sum\limits_i T_i+\sum\limits_{i>j}V^{\mathrm{eff}}_{ij}.
\end{equation}
If we increase the cutoff, $Q$ converges to $U$ and therefore we can approximate
\begin{equation}
    \mathcal{W}\approx \mathbb{I}+\Delta
\end{equation}
with $\Delta$ a small perturbation for sufficiently large Hilbert spaces. Because $\mathcal{W}$ is a unitary matrix, it follows that $\Delta^T=-\Delta$. Inserting this approximation into the expression for the effective Hamiltonian, we arrive at
\begin{equation}
    \begin{split}
        H^{\mathrm{eff}}&\approx\sum\limits_i T_i + \sum\limits_{i>j}\left\{ \left(\left(\mathbb{I}+\Delta\right)^T H^{(2)}\left(\mathbb{I}+\Delta\right)\right)_{ij}-T_i-T_j\right\}\\
        &\approx\sum\limits_i T_i+\sum\limits_{i>j}\left\{T_i+T_j+V_{ij}- \left(\Delta H^{(2)}\mathbb{I}\right)_{ij}+\left(\mathbb{I} H^{(2)}\Delta\right)_{ij}-T_i-T_j\right\}\\
        &\approx\sum\limits_i T_i+\sum\limits_{i>j}V_{ij}-\sum\limits_{i>j}\left(\Delta H^{(2)}\mathbb{I}\right)_{ij}+\sum\limits_{i>j}\left(\mathbb{I} H^{(2)}\Delta\right)_{ij}\\
        &\approx H - \sum\limits_{i>j} \left(\Delta H^{(2)}\mathbb{I}\right)_{ij}+\sum\limits_{i>j}\left(\mathbb{I} H^{(2)}\Delta\right)_{ij},
    \end{split}
\end{equation}
where we have used that: $H=\sum\limits_i T_i+\sum\limits_{i>j}V_{ij}.$
Next, we extend $\Delta$ such that $\Delta H = \Delta H^{(2)}$, i.e. $\Delta_{ij}T_k=0$ if $k\neq i, j$ and $\Delta_{ij}V_{ik}=0$ if $k\neq j$. With this we can now write:
\begin{equation}
    H^{\mathrm{eff}}\approx \left(\mathbb{I}-\sum\limits_{i>j}\Delta\right)H\left(\mathbb{I}+\sum\limits_{i>j}\Delta\right)=\mathcal{W}^T H \mathcal{W},
\end{equation}
We have shown that for large cutoffs the transformation for the effective Hamiltonian of the many-body system is approximately unitary.

Now, we apply the same transformation to operators:
\begin{equation}
    O^{\mathrm{eff}}=\mathcal{W}^T O \mathcal{W}.
\end{equation}
For a one-body operator $O^{(1)}=\sum\limits_i O_i$ we derive:
\begin{equation}
\label{eq:OneBodyOperatorTransformed}
\begin{split}
    O^{\mathrm{eff}, (1, 2)}&=\left(\mathbb{I}-\sum\limits_{i>j}\Delta\right)O^{(1)}\left(\mathbb{I}+\sum\limits_{i>j}\Delta\right)\\
    &=\sum\limits_i O_i - \sum\limits_{ij}\left(\Delta O^{(2)}\right)_{ij}+\sum\limits_{ij}\left(O^{(2)}\Delta\right)_{ij}\\
    &=\sum\limits_i O_i+ \sum\limits_{ij} \left(\left(\mathbb{I}+\Delta\right)^T O^{(2)}\left(\mathbb{I}+\Delta\right)\right)_{ij}-O^{(2)}\\
    &=\sum\limits_i O_i +\sum\limits_{ij} \left(Q^T \widetilde{O}^{(2)}Q\right)_{ij}-O^{(2)}.
\end{split}
\end{equation}

For a two-body operator $O^{(2)}=\sum\limits_{i>j} O_{ij}$ we get:
\begin{equation}
\label{eq:TwobodyOperatorTransformed}
\begin{split}
    O^{\mathrm{eff}, (2)}&=\left(\mathbb{I}-\sum\limits_{i>j}\Delta\right)O^{(2)}\left(\mathbb{I}+\sum\limits_{i>j}\Delta\right)\\
    &=\sum\limits_{i>j} O_{ij} - \sum\limits_{ij}\left(\Delta O^{(2)}\right)_{ij}+\sum\limits_{ij}\left(O^{(2)}\Delta\right)_{ij}\\
    &=\sum\limits_{i>j} O_{ij}+ \sum\limits_{ij} \left(\left(\mathbb{I}+\Delta\right)^T O^{(2)}\left(\mathbb{I}+\Delta\right)\right)_{ij}-O^{(2)}\\
    &=\sum\limits_{ij} \left(Q^T \widetilde{O}^{(2)}Q\right)_{ij}.
\end{split}
\end{equation}
For our CI-calculations, we only use the restricted Hilbert space provided by the projection operator $P_{\mathcal{M}}$. Therefore, in a final step, one has to apply $P_{\mathcal{M}}$ such all matrices of operators are of finite dimension. 

\subsection{Implementation for \texorpdfstring{$V_{\mathrm{eff}}^{\mathrm{rel}}$}{V_eff^rel}}
\label{App:ImplementationObsRel}

To implement the transformation of observables for $V_{\mathrm{eff}}^{\mathrm{rel}}$, we need to transform the (induced) two-body operator into relative and center-of-mass coordinates $O(x_1, x_2)\rightarrow O(x, X)$ with $x=(x_1-x_2)/\sqrt{2}$ and $X=(x_1+x_2)/\sqrt{2}$. If the operator factorizes in these coordinates, $O(x, X)=O(x)O(X)$, the calculation is straightforward: We evaluate the expectation value of the relative part of the operator, $O(x)$, with the exact solution in relative coordinates to build a matrix equivalent to $E$ in Eq.~\eqref{eq:Veffrel}. Then we plug in the same matrices $Q$ and (depending on whether we calculate a one- or two-body operator according to Eqs.~\eqref{eq:OneBodyOperatorTransformed},~\eqref{eq:TwobodyOperatorTransformed}) subtract the matrix stemming from the expectation value of the operator $O(x)$ with the basis states ($T$ in Eq.~\eqref{eq:Veffrel}). If the operator does not factorize, e.g. for the density, the computation becomes more involved and requires an additional calculation of the expectation value of $O(x_1, x_2)$.

\subsection{Implementation for \texorpdfstring{$V_{\mathrm{eff}}^{\mathrm{rel+cm}}$}{V_eff^rel+cm}}
\label{App:ImplementationObsRelCM}

Here, we explain how to evaluate the expectation value of the (induced) $O(x_1, x_2)$ 
if we use $V_{\mathrm{eff}}^{\mathrm{rel+cm}}$. In this case, we should first build a matrix similar to $E$ in Eq.~\eqref{eq:Veffrelcm}. Next, we transform the operator with the overlap matrices $Q$ used to construct the effective potential and, depending on whether we calculate a one- or two-body operator according to Eqs.~\eqref{eq:OneBodyOperatorTransformed},~\eqref{eq:TwobodyOperatorTransformed}, we also need to calculate and subtract the matrix given by the expansion of this operator in the basis states. 

If the operator  $O(x_1, x_2)$ factorizes in relative and center-of-mass coordinates, $O(x, X)=O(x)O(X)$, the run time of the calculation is reduced significantly because only one-dimensional integrals need to be evaluated. For general operators the calculation of the expectation values leads to two-dimensional integrals, which can require a long run time if calculated numerically. Nevertheless, the evaluation of these integrals might be more straightforward than an implementation of the transformation needed for $V_{\mathrm{eff}}^{\mathrm{rel}}$, see Sec.~\ref{App:ImplementationObsRel}.

\section{Derivation of the running coupling constant \texorpdfstring{$g_{\mathrm{unconf.}}(\lambda)$}{g_unconf.(lambda)} in free space}
\label{sec:App:Derivation_momentum_spaceRG}

In this section, we show in more detail how to derive the expression
\begin{equation}
\label{eq:gRG}
    -\frac{1}{g_{\mathrm{unconf.}}(\lambda)}=\frac{1}{\pi\sqrt{|E_{1+1}^{\mathrm{unconf.}}|}}\mathrm{arctan}\left[\frac{\lambda}{\sqrt{|2E_{1+1}^{\mathrm{unconf.}}|}}\right].
\end{equation}
Note that the discussion below adapts the approach for 2D systems presented in Ref.~\cite{nyeo2000regularization}.

We start by considering the ground state of the relative part of the Schr\"odinger equation for two particles interacting via an attractive contact interaction in free space
\begin{equation}
    \left(-\frac{1}{2}\frac{\partial^2}{\partial x^2}+\frac{g}{\sqrt{2}}\delta(x)\right)\Psi(x)=E_{1+1}^{\mathrm{unconf.}}\Psi(x),
\end{equation}
where  $x=\frac{1}{\sqrt{2}}(x_1-x_2)$ is the relative coordinate, and $g<0$.
Note that $E_{1+1}^{\mathrm{unconf.}}<0$ -- there is a bound state in this problem~\cite{griffiths2018introduction}. We perform the Fourier transform
\begin{align}
    \Psi(x)&=\int\frac{dk}{2\pi}e^{ikz}\phi(k).
\end{align}
The corresponding Schr\"odinger equation in momentum space reads
\begin{equation}
    \frac{1}{2}k^2\phi(k)+\frac{g}{\sqrt{2}}\Psi(0)=E_{1+1}^{\mathrm{unconf.}}\phi(k).
\end{equation}
Its solution is
\begin{equation}
    \phi(k)=-\frac{\sqrt{2}g\Psi(0)}{k^2-2E_{1+1}^{\mathrm{unconf.}}}.
\end{equation}
Now, we note that $\Psi(0)=\int\frac{dk}{2\pi}\phi(k)$, which leads to the expression:
\begin{equation}
    \int \phi(k)dk=-\int\frac{\sqrt{2}g}{k^2-2E_{1+1}^{\mathrm{unconf.}}}dk\int \phi(k')\frac{dk'}{2\pi}.
\end{equation}
We divide by $\int \phi(k)dk$ 
\begin{equation}
    1=-\frac{\sqrt{2}g}{2\pi}\int\frac{1}{k^2-2E_{1+1}^{\mathrm{unconf.}}}dk,
\end{equation}
 introduce a momentum space cutoff $\lambda$ and change $g$ to $g_{\mathrm{unconf.}}(\lambda)$:
\begin{equation}
    \frac{1}{g_{\mathrm{unconf.}}(\lambda)}=-\frac{\sqrt{2}}{2\pi}\int_{-\lambda}^\lambda\frac{dk}{k^2-2E_{1+1}^{\mathrm{unconf.}}}=-\frac{\mathrm{Arctan}(\lambda/\sqrt{|2E_{1+1}^{\mathrm{unconf.}}|})}{\pi\sqrt{|E_{1+1}^{\mathrm{unconf.}}|}}
\end{equation}

\section{Connecting quantum numbers to momenta}
\label{sec:App:Derivation_connection_qn_mom}

To find the connection between the quantum number $n$ for a general potential $x^k$ and the cutoff in relative momentum space $\lambda$ in $g_{\mathrm{unconf.}}(\lambda)$, Eq.~\eqref{eq:gRG}, we first use the Wilson–Sommerfeld rule to connect $n$ to the energy
\begin{equation}
    \oint_{H(p, x)=E} p\,dx=2\pi n,
\end{equation}
with $H=\frac{1}{2}p^2+\frac{1}{2^{k-1}}x^k$. This leads to:
\begin{align}
    2\pi n&=2\int_{x=-2^{1-1/k}E^{1/k}}^{x=2^{1-1/k}E^{1/k}} \sqrt{2E-\frac{1}{2^{k-2}}x^k}\,dx\\
    &=2\sqrt{2E}x ~_2F_1(-1/2, 1/k, 1+1/k, 2^{1-k}x^k/E)|_{x=-2^{1-1/k}E^{1/k}}^{x=2^{1-1/k}E^{1/k}}\\
    &= 2^{7/2-1/k} E^{1/2+1/k} \frac{\sqrt{\pi}\Gamma(1/k)}{(2+k)\Gamma(1/2+1/k)},
\end{align}
where $_2F_1$ is the hypergeometric function~\cite{abramowitz1948handbook}.

Then, we note that the cutoff $n$ corresponds to an energy cutoff ($E$) in the single particle Hilbert space while $\lambda$ restricts the relative momentum of two particles. To combine both, we use the energy relation
\begin{equation}
    2E=\frac{1}{2}\lambda^2, 
\end{equation}
i.e., a one-body cutoff of $E$ allows for a maximal energy of $2E$ in the two-body Hilbert space which has to be equal to the maximal energy that can be obtained with the momentum cutoff $\lambda$. Note that this is only true in either a harmonic oscillator or in a system on ring where the center of mass and relative motion factorize. However, for other potentials and high enough cutoffs, we can still assume this relation to hold approximately. Using this expression, we derive
\begin{equation}
    2\pi n= 2^{5/2-3/k} \lambda^{1+2/k} \frac{\sqrt{\pi}\Gamma(1/k)}{(2+k)\Gamma(1/2+1/k)}.
\end{equation}
For the harmonic oscillator, $k=2$, this expression can be simplified
\begin{align}
    2\pi n&=\frac{\pi}{2} \lambda^2
    \Leftrightarrow \lambda=2\sqrt{n}.
\end{align}
We also note the expression for hard wall boundary conditions, which can be approximated with $k\to\infty$,
\begin{align}
    2\pi n&=2^{5/2} \lambda
    \Leftrightarrow \lambda=\frac{\pi n}{\sqrt{8}}.
\end{align}
\newpage

\newpage
\section{Convergence as a function of \texorpdfstring{$1/n$}{1/n}}
\label{sec:Convergence1_over_n}

\begin{figure}
\centering
    \includegraphics[width=0.5\linewidth]{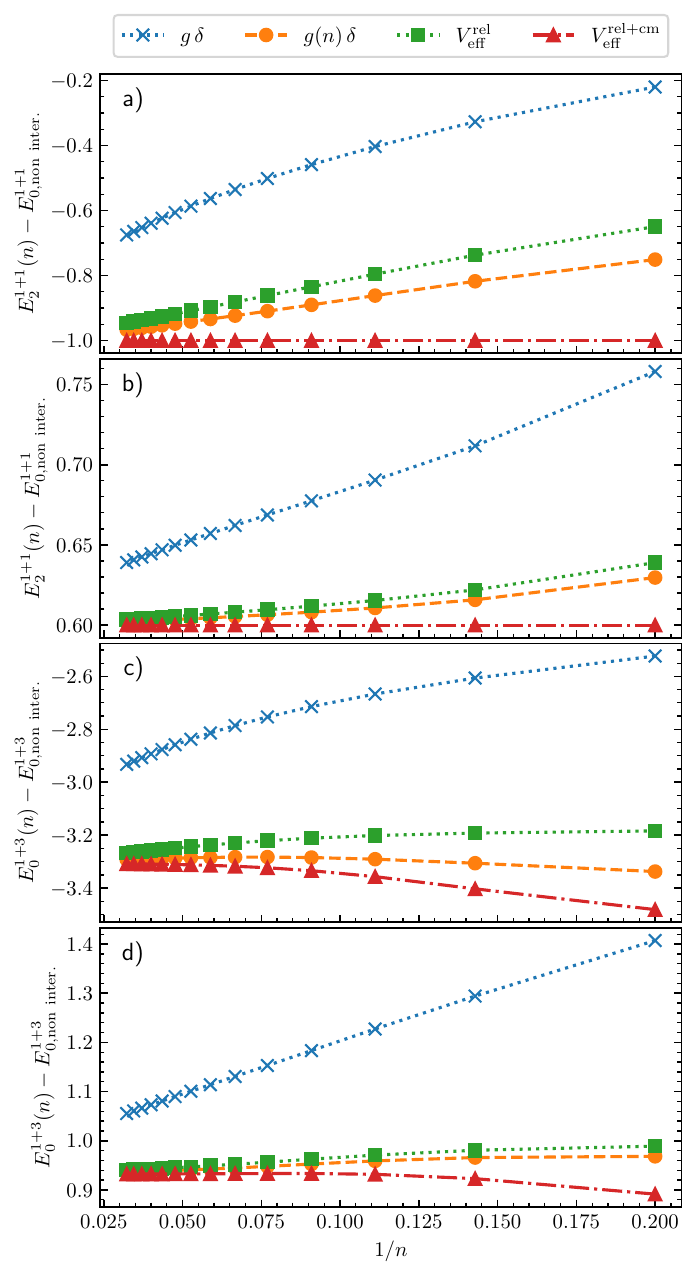}
    \caption{Energy of the system for different particle numbers and interactions calculated with the bare contact interaction, running coupling constant, $V_{\mathrm{eff}}^{\mathrm{rel}}$ and $V_{\mathrm{eff}}^{\mathrm{rel+cm}}$ as a function of $1/n$. Panels a) and b) show the result for the second excited state of the $1+1$ system; c) and d) present the ground state energy of $1+3$. Panels a) and c) are for attractive interaction with $g=-2.5067$ (corresponds to $E_{1+1}=-1$); panels b) and d) are for repulsive interactions with $g=3$ (corresponds to $E_{1+1}=1.6$). Our numerical results are presented with markers; lines are added to guide the eye.}
    \label{fig:Energy_1_over_n}
\end{figure}

\begin{figure}
    \centering
    \includegraphics[width=0.5\linewidth]{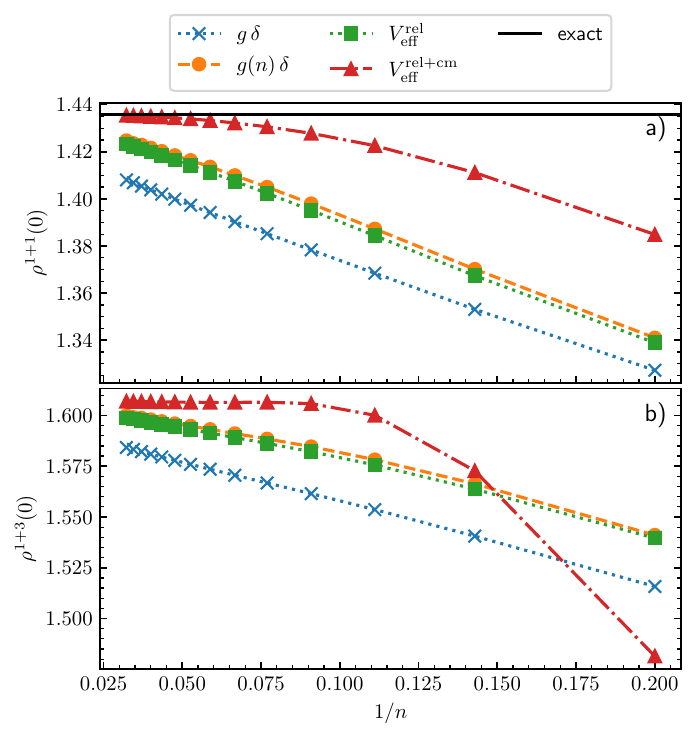}
    \caption{Ground state density at the center of the trap calculated with bare contact interaction, running coupling constant, $V_{\mathrm{eff}}^{\mathrm{rel}}$ and $V_{\mathrm{eff}}^{\mathrm{rel+cm}}$ as a function of $1/n$. Panel a) shows results for the $1+1$ system; panel b) is for $1+3$. All panels are for attractive interactions with $g=-2.5067$ (corresponds to $E_{1+1}=-1$).  Our numerical results are presented with markers; lines are added to guide the eye. The exact value of the density for the $1+1$ system is shown as a horizontal solid line.}
    \label{fig:Density_1_over_n}
\end{figure}

\begin{figure}
\centering
    \includegraphics[width=0.5\linewidth]{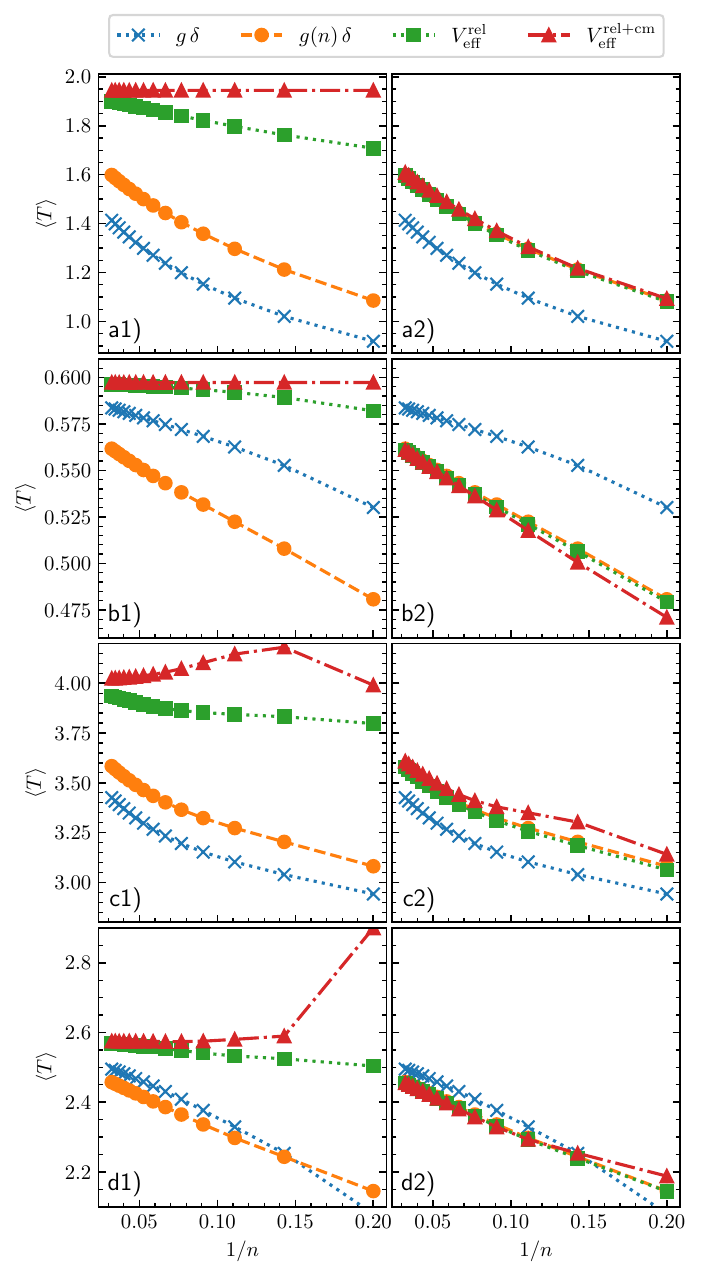}
    \caption{Kinetic energy for different particle numbers and interactions as a function of $1/n$. Left panels [*1)] show the results where the observable has been transformed together with the interaction potential for $V_{\mathrm{eff}}^{\mathrm{rel}}$ and $V_{\mathrm{eff}}^{\mathrm{rel+cm}}$ while the right ones [*2)] show results where only the interaction potential but not the observable has been transformed. Panels a) and b) are for $1+1$; panels c) and d) show results for the $1+3$ system. Panels a) and c) are for $g=-2.5067$ (corresponds to $E_{1+1}=-1$); panels b) and d) are for $g=3$ (corresponds to $E_{1+1}=1.6$). Our numerical results are presented with markers; lines are added to guide the eye.}
    \label{fig:kinetic_1_over_n}
\end{figure}

\begin{figure}
\centering
    \includegraphics[width=0.5\linewidth]{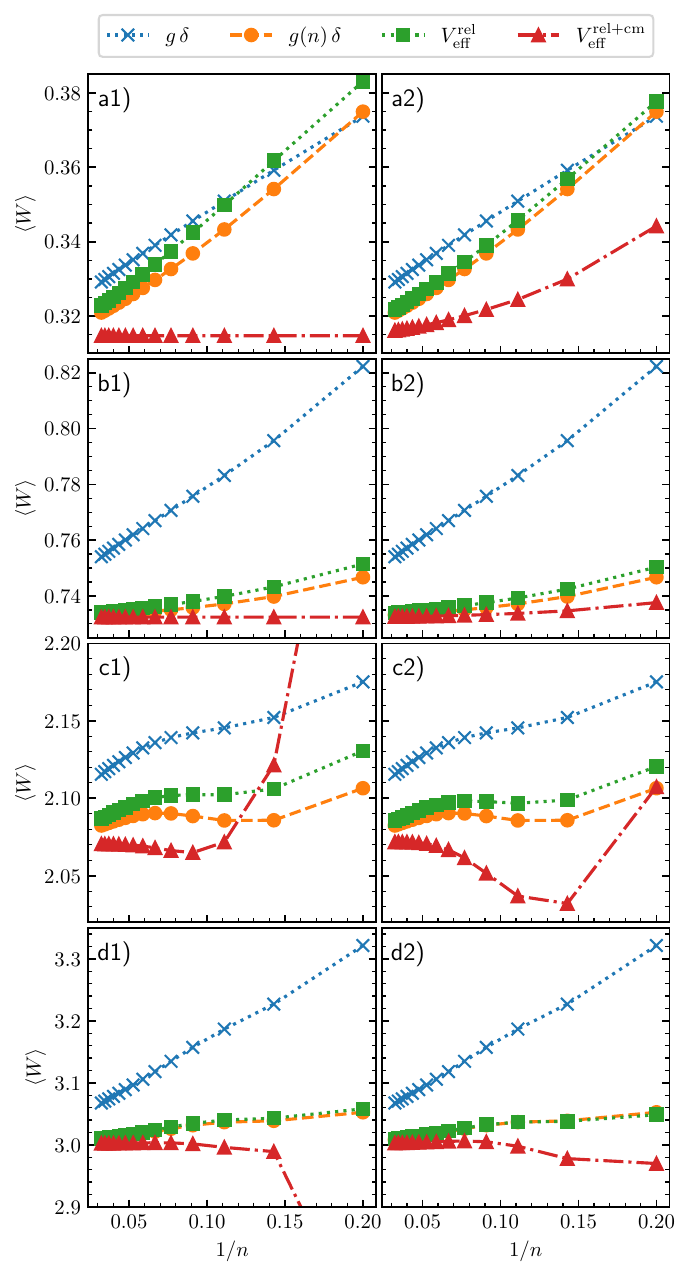}
    \caption{Expectation value of the trapping potential for different particle numbers and interactions as a function of $1/n$. Left panels [*1)] show the results where the observable has been transformed together with the interaction potential for $V_{\mathrm{eff}}^{\mathrm{rel}}$ and $V_{\mathrm{eff}}^{\mathrm{rel+cm}}$ while the right ones [*2)] show results where only the potential but not the observable has been transformed. Panels a) and b) are for $1+1$; panels c) and d) show results for the $1+3$ system. Panels a) and c) are for $g=-2.5067$ (corresponds to $E_{1+1}=-1$); panels b) and d) are for $g=3$ (corresponds to $E_{1+1}=1.6$). Our numerical results are presented with markers; lines are added to guide the eye.}
    \label{fig:trap_1_over_n}
\end{figure}

\begin{figure}
\centering
    \includegraphics[width=0.5\linewidth]{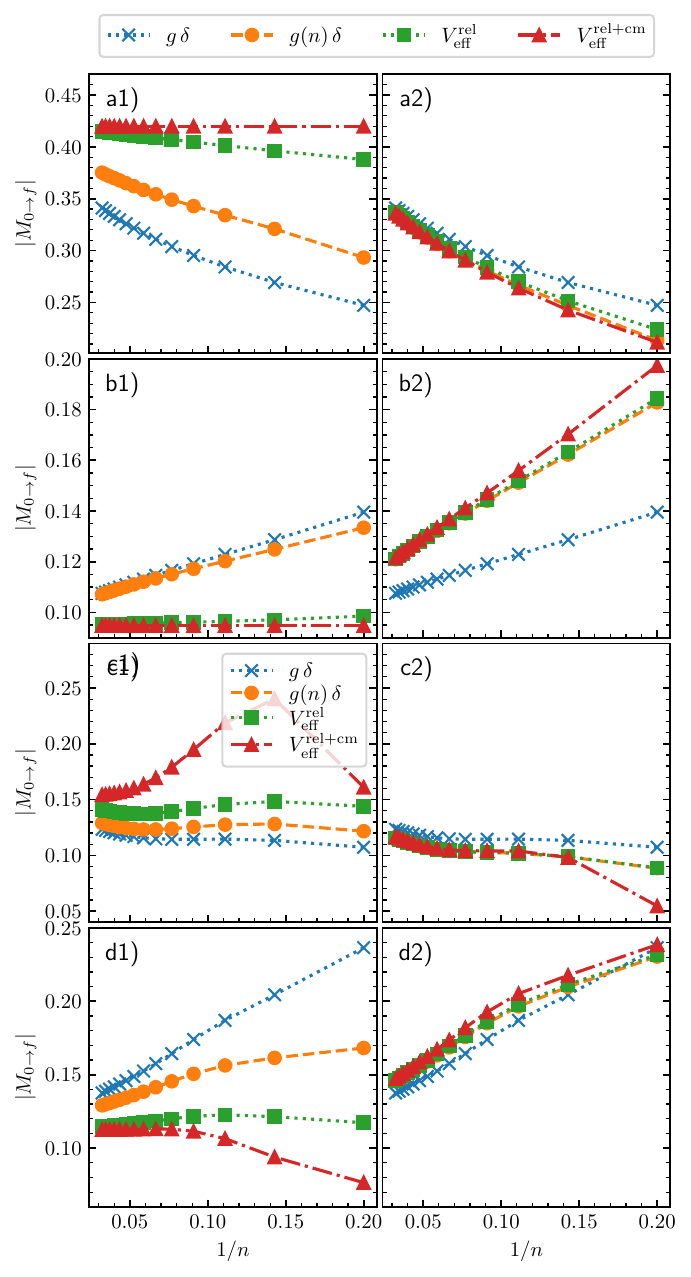}
    \caption{Transition matrix elements for different particle numbers and interactions as a function of $1/n$. Left panels [*1)] show the results where the observable has been transformed together with the interaction potential for the renormalized interactions $g(n)\delta$, $V_{\mathrm{eff}}^{\mathrm{rel}}$ and $V_{\mathrm{eff}}^{\mathrm{rel+cm}}$ while the right ones [*2)] show results where only the potential but not the observable has been transformed. Panels a) and b) are for $1+1$ and the transition from the ground to the fourth excited state. Panels c) and d) are for $1+3$ and the transition c) from the ground state to the fifth excited state, d) from the ground to the sixths excited state. Panels a) and c) are for $g=-2.5067$ (corresponds to $E_{1+1}=-1$); panels  b) and d) are for $g=3$ (corresponds to $E_{1+1}=1.6$). Our numerical results are presented with markers; lines are added to guide the eye.}
    \label{fig:transmatrix_1_over_n}
\end{figure}

For the interested reader, we provide in this section the convergence plots as a function of $1/n$ with $n$ the one-body basis cutoff. Fig.~\ref{fig:Energy_1_over_n} shows the convergence of the energy, Fig.~\ref{fig:Density_1_over_n} of the density at the trap center, Fig~\ref{fig:kinetic_1_over_n} of the kinetic energy, Fig~\ref{fig:trap_1_over_n} of the trapping confiment and Fig.~\ref{fig:transmatrix_1_over_n} of the transition matrix elements.

\newpage
\section{Double logarithmic plots}
\label{sec:Convergenceloglog}

\begin{figure}
\centering
    \includegraphics[width=0.5\linewidth]{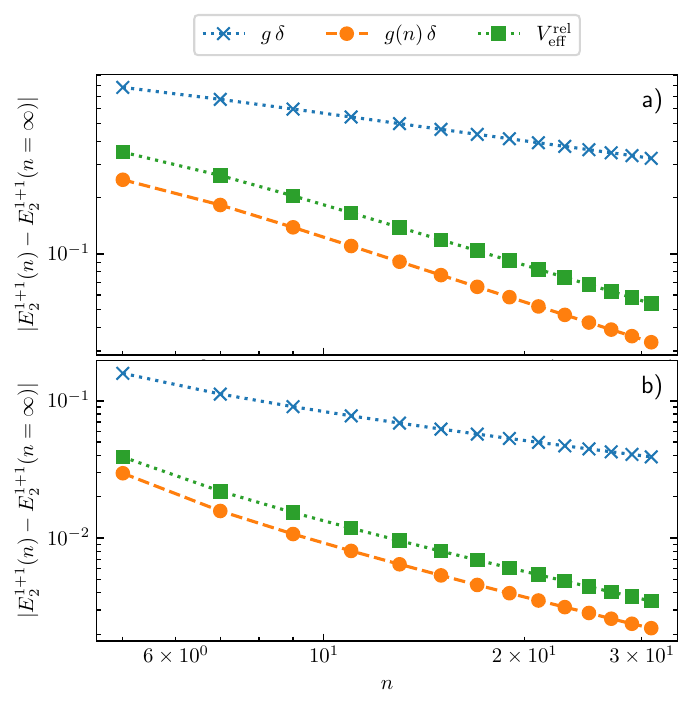}
    \caption{Energy of the second excited state of the $1+1$ system calculated with the bare contact interaction, running coupling constant and $V_{\mathrm{eff}}^{\mathrm{rel}}$. We subtract the exact result given by the solution of Refs.~\cite{avakian1987spectroscopy, busch1998} and plot the data on a double logarithmic graph. Panel a) is for attractive interactions with $g=-2.5067$ (corresponds to $E_{1+1}=-1$); panel b) is for repulsive interaction with $g=3$ (corresponds to $E_{1+1}=1.6$).  Our numerical results are presented with markers; lines are added to guide the eye. }
    \label{fig:Energy_loglog}
\end{figure}

\begin{figure}
\centering
    \includegraphics[width=0.5\linewidth]{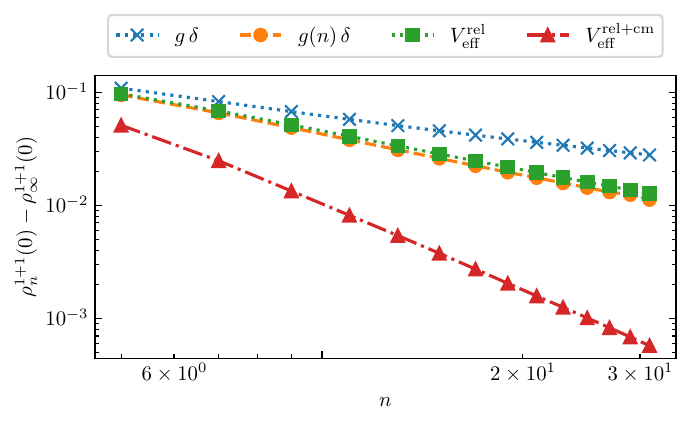}
    \caption{Density of the ground state in the middle of the trap as a function of the cutoff parameter for the four potentials considered in this study for the $1+1$ system. We subtract the exact result given by the solution of Refs.~\cite{avakian1987spectroscopy, busch1998} and plot the data as a double logarithmic graph. We use attractive interaction $g=-2.5067$, which corresponds to $E_{1+1}=-1$.  Our numerical results are presented with markers; lines are added to guide the eye.}
    \label{fig:Density_loglog}
\end{figure}

\begin{figure}
\centering
    \includegraphics[width=0.5\linewidth]{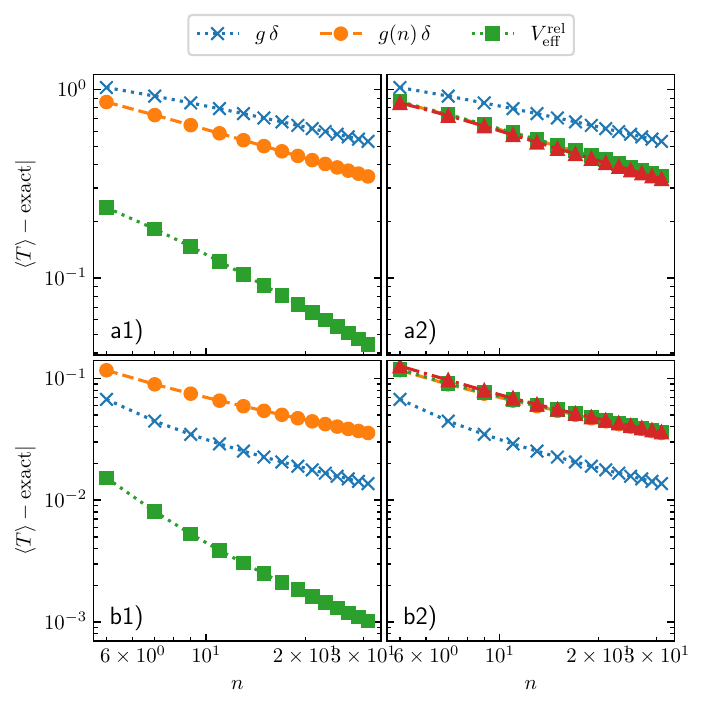}
    \caption{Kinetic energy for the $1+1$ system. Left panels [*1)] show the results where the observable has been transformed together with the interaction potential for $V_{\mathrm{eff}}^{\mathrm{rel}}$ and $V_{\mathrm{eff}}^{\mathrm{rel+cm}}$ while the right ones [*2)] show results where only the potential but not the observable has been transformed. Panel a) is for $g=-2.5067$ (corresponds to $E_{1+1}=-1$); panel b) shows results for $g=3$ (corresponds to $E_{1+1}=1.6$). Note that we have subtracted the exact result calculated using the solution of Refs.~\cite{avakian1987spectroscopy, busch1998}.  Our numerical results are presented with markers; lines are added to guide the eye.}
    \label{fig:kinetic_loglog}
\end{figure}

\begin{figure}
\centering
    \includegraphics[width=0.5\linewidth]{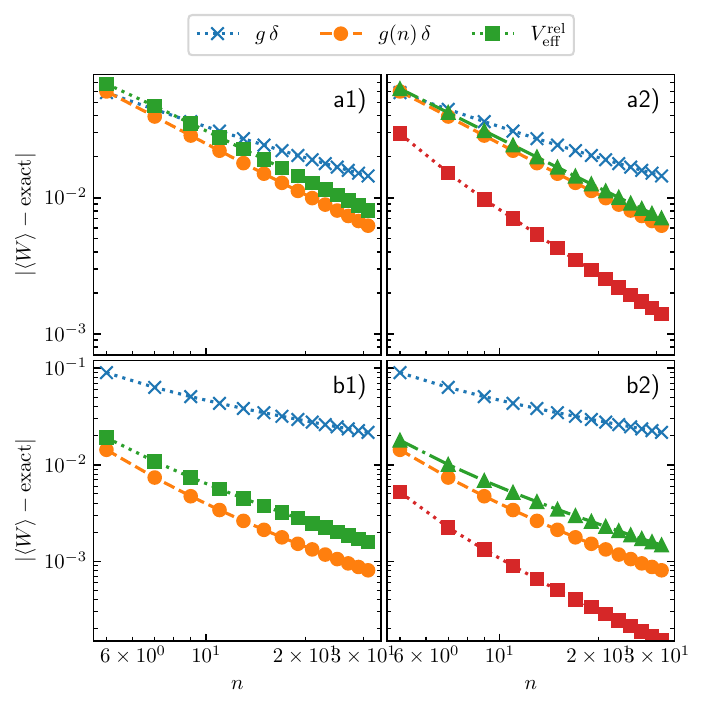}
    \caption{Expectation value of the trapping potential for the $1+1$. Left panels [*1)] show the results where the observable has been transformed together with the interaction potential for $V_{\mathrm{eff}}^{\mathrm{rel}}$ and $V_{\mathrm{eff}}^{\mathrm{rel+cm}}$ while the right ones [*2)] show results where only the potential but not the observable has been transformed. Panel a) is for $g=-2.5067$ (corresponds to $E_{1+1}=-1$); panel b) shows results for $g=3$ (corresponds to $E_{1+1}=1.6$). Note that we have subtracted the exact result calculated using the solution in Refs.~\cite{avakian1987spectroscopy, busch1998}.  Our numerical results are presented with markers; lines are added to guide the eye.}
    \label{fig:trap_loglog}
\end{figure}

\begin{figure}
\centering
    \includegraphics[width=0.5\linewidth]{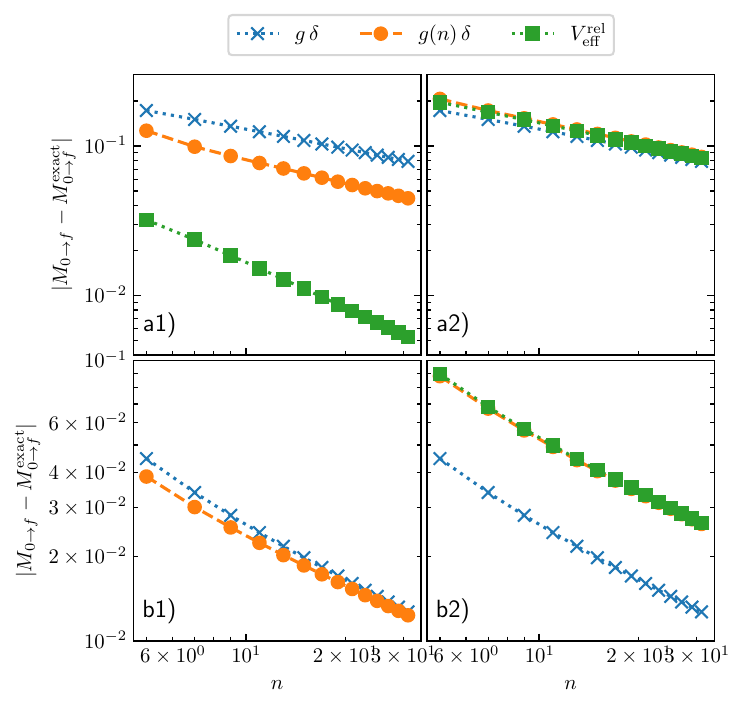}
    \caption{Transition matrix elements from the ground to the fourth excited states. Left panels [*1)] show the results where the observable has been transformed together with the interaction potential for $g(n)\delta$, $V_{\mathrm{eff}}^{\mathrm{rel}}$ and $V_{\mathrm{eff}}^{\mathrm{rel+cm}}$ while the right ones [*2)] show results where only the potential but not the observable has been transformed. Panel a) is for $g=-2.5067$ (corresponds to $E_{1+1}=-1$); panel b) shows results for $g=3$ (corresponds to $E_{1+1}=1.6$). Note that we have subtracted the exact result calculated using the solution of Refs.~\cite{avakian1987spectroscopy, busch1998}.  Our numerical results are presented with markers; lines are added to guide the eye.}
    \label{fig:transmatrix_loglog}
\end{figure}

We also provide double logarithmic plots for the $1+1$ system where we have the exact result: The energy is shown in Fig.~\ref{fig:Energy_loglog}, the density at the middle of the trap is presented in Fig.~\ref{fig:Density_loglog} , the kinetic energy is demonstrated in  Fig.~\ref{fig:kinetic_loglog}, the expectation value of the harmonic trapping confiment is in  Fig.~\ref{fig:trap_loglog} and the transition matrix elements are in Fig.~\ref{fig:transmatrix_loglog}. Note that the results for $V_{\mathrm{eff}}^{\mathrm{rel+cm}}$ and transformed operators are not shown -- these results are exact by construction. 

The double logarithmic plots showcase (by the steeper slope) that the use of renormalized interactions leads to a faster convergence of the results in comparison to a bare contact interaction.

\clearpage

\bibliography{refs}